\documentclass[]{report}

\usepackage{a4}
\usepackage{german}
\usepackage{epsfig}
\usepackage{cite}
\usepackage{amsmath}
\usepackage{amsfonts}
\usepackage{amssymb}

\newcommand{\f}[2]{\frac{#1}{#2}}
\newcommand{\eqnanf}{\begin{eqnarray}}
\newcommand{\eqnend}{\end{eqnarray}}
\newcommand{\mathanf}{\begin{displaymath}}
\newcommand{\mathend}{\end{displaymath}}

\newcommand{\schwups}{\\[2.6ex]}
\newcommand{\wups}{\nonumber \\[2.6ex]}
\newcommand{\ok}{\nonumber}
\newcommand{\glok}{}

\def\buildrel #1 \under #2 {\mathrel{\mathop{\kern0pt #2}\limits_{#1}}}

\newcommand{\ep}{\epsilon}
\newcommand{\strich}{\biggr|}

\newcommand{\demand}{\mathrel{\mathop{\kern0pt =}\limits^{!}}}

\newcommand{\del}{\partial}

\newcommand{\schwinger}{\mbox{\Large{\sf{S}}}}
\newcommand{\swsum}[1]{\buildrel #1 \under \schwinger}

\newcommand{\dbyd}[2]{\f{\partial \, #1}{\partial \, #2}}

\newcommand{\bra}[1]{\mbox{$ \langle #1 | $}}
\newcommand{\ket}[1]{\mbox{$ | #1 \rangle $}}

\newcommand{\bracket}[2]{\mbox{$ \langle #1 | #2 \rangle $}}

\newcommand{\scalar}[2]{\mbox{$ \langle #1\,,\,#2 \rangle $}}

\newcommand{\com}[2]{[ #1 \,,\, #2 ]}
\newcommand{\anticom}[2]{\{#1\,,\,#2\}}

\newcommand{\slf}[1]{\not{\!#1}}
\newcommand{\relscp}[2]{#1 \, \cdot \, #2}
\newcommand{\scp}[2]{\vec{#1} \cdot \vec{#2}}

\newcommand{\quabla}{\square}

\newcommand{\SFV}{\frac{1}{\slf{p} - \slf{k} - m - \go V}}

\newcommand{\psq}{{\vec{p}\,}^2}
\newcommand{\lquer}{\lambda_{\rm e}}
\newcommand{\Os}{{\cal O}}
\newcommand{\Es}{{\cal E}} 

\newcommand{\modw}{|\omega|}
\newcommand{\modk}{|\vec{k}|}
\newcommand{\intk}{\int \frac{d^4 k}{(2 \pi)^4}}
\newcommand{\intclk}{\int_{C_L} \frac{d^4 k}{(2 \pi)^4}}

\newcommand{\id}{{\rm Id}}
\newcommand{\go}{\gamma^0}
\newcommand{\diracprop}{\frac{1}{\slf{p} - \slf{k} - m - \go V}}
\newcommand{\dcprop}{\frac{1}{\slf{p} - \slf{k} - m - \go V}}
\newcommand{\Pmunu}{\bra{\bar{\psi}} \gamma^{\mu} 
\diracprop \gamma^{\nu} \ket{\psi}}
\newcommand{\Up}{U^{+}}

\epsfverbosetrue

\usepackage[
   a4paper,      
   dvips,        
   bookmarksopen=true,
   bookmarksnumbered=true,
   pdftitle={Diplomarbeit},
   pdfauthor={Ulrich Jentschura},
   pdfcreator={Ulrich Jentschura},
   pdfstartview={FitB},
   colorlinks,
   linkcolor=blue]{hyperref}   

\begin{document}

%
%

\hrule
\vspace*{0.2cm}
\hrule
\vspace*{0.2cm}
\hrule
\vspace*{0.5cm}
\begin{center}
\sf
\color{red}
\Huge
Theory of the Lamb Shift\\
in Hydrogenlike Systems\\[1ex]
(Theorie der Lamb-Verschiebung\\
in wasserstoffartigen Systemen)\\[1ex]
\end{center}
\vspace*{0cm}
\begin{center}
\LARGE
\color{blue}
Diplomarbeit/Master Thesis \\
an der Fakult\"{a}t f"ur Physik \\
der Ludwig-Maximilians-Universit"at M"unchen \\[11ex]
\color{black}
von/by\\
\end{center}
\vspace*{0.5cm}
\begin{center}
\color{black}
\LARGE
Ulrich D. Jentschura\\[3.5ex]
\color{blue}
1. Auflage/Edition Februar 1996\\
2. Auflage/Edition April 2003\\{}
[with hypertext references and updates]\\[5ex]
\large
\color{green}
E--Mail:\\
jentschura@physik.uni-freiburg.de\\
jentschura@physik.tu-dresden.de\\
ulj@nist.gov\\
\color{black}
\end{center}
\vfill
\vspace*{2cm}
\hrule
\vspace*{0.2cm}
\hrule
\vspace*{0.2cm}
\hrule

\newpage

%
%

\Huge 
Abstract\\[1ex]

\normalsize
Analytic calculations of the Lamb shift represent a considerable
challenge due to the size and the complexity of the expressions that 
occur in intermediate steps. In the current work, we present a 
method for the treatment of the bound-state self-energy in higher orders.
Advantage is taken of computer algebra systems.
The method is applied to the 2P-states, and the (one-loop) self-energy 
is calculated up to the order of $\alpha (Z \alpha)^6$. 
The calculation leads to improved predictions for the 
in atomic hydrogen.

In particular, the focus of the current work is the calculation
of the so-called $A_{60}$-coefficient with improved accuracy.
$A_{60}$-coefficients are calculated
with an accuracy of less than $1 \, {\rm Hz}$ in frequency units,
\begin{displaymath}
A_{60}(2P_{1/2}) = -0.99891(1) \quad \mbox{and} \quad
A_{60}(2P_{3/2}) = -0.50337(1).
\end{displaymath}
An account of the contributions to the Lamb shift of 2P states
is given [status of 1996 when this thesis was completed].
Bound-state quantum electrodynamics combines the complexity of modern
quantum field theory, including renormalization, augmented by the
structure of bound states and the inherent relativistic corrections
with the experimental possibilities of modern high-resolution 
laser spectroscopy.

\vspace*{0.2cm}
\hrule
\vspace*{0.2cm}
\hrule
\vspace*{0.2cm}
\hrule
\vspace*{0.2cm}

This thesis was submitted to the University of Munich on 10 February
1996, and final examinations took place on 8 May 1996. We attempt to 
provide a more complete account of the calculations whose results 
were previously presented in [U. D. Jentschura and K. Pachucki, 
e-print \href{http://arXiv.org/abs/physics/0011008}{physics/0011008}, 
Phys. Rev. A {\bf 54}, 1853 (1996)]. 
The current version is not identical to the master thesis submitted to the 
University of Munich; it contains hyperreferences and updates. 
Part of the results are of relevance for a recent investigation of radiative
corrections induced by local potentials [U. D. Jentschura, e-print
\href{http://arXiv.org/abs/hep-ph/0305066}{hep-ph/0305066}, 
J. Phys. A {\bf 36}, L229 (2003)]. 

The thesis is in German; this English abstract is provided for completeness.
Final results for the Lamb shift of 2P states as reported in 
Ch.~\ref{ergebnisse} represent the status of corrections known by 1996.

\vspace*{0.2cm}
\hrule
\vspace*{0.2cm}
\hrule
\vspace*{0.2cm}
\hrule
\vspace*{0.2cm}

This thesis is the first in the traditional threefold sequence
``diploma(master)--dissertation(PhD)--habilitation'' 
that is being followed in Continental Europe.
Copies of the author's PhD thesis are available as U.~D. Jentschura,
``Quantum Electrodynamic Radiative Corrections in Bound Systems
(Dresdner Forschungen: Theoretische Physik, Band 2)'',
w.e.b. Universit\"{a}tsverlag, Dresden, 1999 (ISBN: 3-933592-65-8), 225 pages
(address of w.e.b. Publishers: Bergstra\ss{}e 78, 01069 Dresden, Germany).
Copies of the habilitation thesis (title:
``Quantum Electrodynamic Bound--State Calculations and 
Large--Order Perturbation Theory'') as well as copies of the 
master thesis in its 1996 version are available via e-mail
or ``snail mail'' (jentschura@physik.uni-freiburg.de or 
U. D. Jentschura, Physikalisches Institut, Universit\"at Freiburg,
Hermann--Herder--Stra\ss{}e 3, 79104 Freiburg im Breisgau, Germany).

\newpage

%
%

\Huge
Zusammenfassung \\[1ex]

\normalsize
Analytische Rechnungen zur Lamb--Verschiebung werden durch Gr"o"se und
Komplexit"at der auftretenden Terme erheblich erschwert. Es wird in dieser
Arbeit eine
Methode vorgestellt, mit welcher der Selbstenergiebeitrag zur
Lamb--Verschiebung wasserstoffartiger Systeme in h"oherer Ordnung analytisch
berechnet werden kann. Durch den Einsatz eines symbolischen Computerprogramms
gelingt es, die Rechenschritte zu vereinheitlichen und zu beschleunigen. 

Die
entwickelte Methode wird auf die $2P$-Zust"ande angewandt und der
(Ein-Schleifen-) Selbstenergiebeitrag bis zur Ordnung 
$\alpha (Z \alpha)^6$ berechnet. 
Die Rechnung zeichnet sich gegen"uber fr"uheren numerischen
Rechnungen durch deutlich geringere Fehlergrenzen aus. 

Insbesondere ist es das Ziel dieser Arbeit, f"ur die $2P$-Zust"ande den 
sogenannten $A_{60}$-Koeffizienten mit verbesserter Genauigkeit zu berechnen. 
F"ur die $2P$-Zust"ande
ist dieser Koeffizient bisher nur durch Extrapolation einer
numerischen Rechnung bekannt [P. J. Mohr, {Phys. Rev. A} {\bf 26}, 2338
(1982)],
\begin{displaymath}
A_{60}(2P_{1/2}) = -0.8(3) \quad \mbox{und} \quad
A_{60}(2P_{3/2}) = -0.5(3).
\end{displaymath}
In dieser Arbeit werden die $A_{60}$-Koeffizienten mit
einer Genauigkeit von $< 1 \, {\rm Hz}$ berechnet, 
\begin{displaymath}
A_{60}(2P_{1/2}) = -0.99891(1) \quad \mbox{und} \quad
A_{60}(2P_{3/2}) = -0.50337(1).
\end{displaymath}
Dadurch kann die Lamb--Verschiebung f"ur die $2P$-Zust"ande mit
verbesserter Genauigkeit angegeben werden (bisher lag 
die theoretische Unsicherheit bei etwa $3 \, {\rm kHz}$).
Es ergeben sich folgende neue theoretische 
Werte f"ur die Lamb--Verschiebung
der $2P$-Zust"ande,
\begin{displaymath}
\delta E_{\rm Lamb}(2P_{1/2}) = -12836.0(3) \, {\rm kHz}
\end{displaymath}
sowie
\begin{displaymath}
\delta E_{\rm Lamb}(2P_{3/2}) = 12517.5(3) \, {\rm kHz},
\end{displaymath}
jeweils mit einer theoretischen Unsicherheit von
$300 \, {\rm Hz}$. 
F"ur die Feinstrukturaufspaltung der $2P$-Zust"ande wird als
theoretischer Wert 
\begin{displaymath}
\delta E_{\rm fs} = E(2P_{3/2}) -  E(2P_{1/2}) = 10969043.2(6) \, {\rm kHz}
\end{displaymath}
erhalten. Ein Vergleich der Voraussagen
der Quantenelektrodynamik mit dem Experiment (h"ochstaufl"osende Spektroskopie)
kann zur "Uberpr"ufung der Quantenfeldtheorien dienen.

\newpage

%
%

\pagenumbering{roman}
\tableofcontents
\newpage
\pagenumbering{arabic}

%
%

\chapter{Einleitung}
\label{einleitung}

{\em ``Quantum Electrodynamics (QED) occupies a unique position in the
hierarchy of theoretical physics. As the basic theory of the 
electromagnetic interaction, it provides the foundation for the
quantum mechanics of atoms and molecules as well as condensed matter
physics, not to mention their numerous applications in chemistry, biology, 
etc. At the same time, it is the first theory that has been applied 
successfully to high energy phenomena.''}\\
T. Kinoshita in \cite{kinokino}\\[3ex]

%
%

\section{Die QED und die Lamb-Verschiebung}
\label{qedlamb}

Die Quantenelektrodynamik beschreibt die Wechselwirkung des 
elektromagnetischen Feldes (des Lichtes) mit geladenen Teilchen. 
Ausgangspunkt f"ur die Entwi"cklung der QED waren unter anderem experimentell 
nachgewiesene geringe Abweichungen von den Voraussagen der Diracschen Theorie des 
Wasserstoffatoms. 

Es sei daher an dieser Stelle an die Diracsche Formel f"ur die 
Energieniveaus des Wasserstoffatoms erinnert. Nach der Diracschen Theorie liegen die 
Niveaus der gebundenen Zust"ande gerade bei
\begin{equation}
E_{n,j}  = m_e f(n,j)
\end{equation}
mit
\begin{equation}
\label{diracsp}
f(n,j) = \bigg( 1 + \frac{(Z \alpha)^2 }{(n-(j+1/2)+ \sqrt{(j+1/2)^2 - 
(Z \alpha)^2)^2}} \bigg) ^ {-1/2}.
\end{equation}
Eine gute M"oglichkeit, die Diracschen Energieniveaus
genauer zu untersuchen, liegt in der Entwi"cklung der Formel
\ref{diracsp} in eine Taylorreihe in der Feinstrukturkonstanten $\alpha$.
In erster N"aherung gilt 
\begin{equation}
E_{n,j} = m_e - (Z \alpha)^2 m_e / (2 n^2) + O(\alpha^4),
\end{equation}
d.h. bis zur Ordnung $\alpha^2$ ist $E_{n,j}$
von $j$ unabh"angig. Wir erkennen hier die Schr"odingersche Formel 
f"ur die gebundenen Energieniveaus ($E_n^S = - \alpha^2 m_e / (2 n^2)$). Zu den 
Schr"odingerschen Niveaus ist im relativistischen Sinne gerade
die Ruhenergie $m_e$ des Elektrons addiert worden.

Zwischen den $2P_{1/2}$ und $2P_{3/2}$-Niveaus liegt nach der Diracschen Theorie
eine Energiedifferenz. In der Ordnung $\alpha^4$ ist diese Differenz das
Feinstrukturintervall
\mathanf
E_{2P3/2} - E_{2P1/2} = {\rm Feinstrukturintervall} = 
  \frac{\alpha^4 m_e}{32} + O(\alpha^6) \quad \mbox{f"ur} \quad Z = 1.
\mathend
In Einheiten der Frequenz ergibt sich daraus eine Differenz von
\mathanf
E_{2P3/2} - E_{2P1/2} \approx 10949 \, {\rm MHz}.
\mathend
Die in der Schr"odingerschen Theorie vorhandene Entartung der
$2P$-Niveaus wird also in der Diracschen Theorie aufgehoben.

Andere Entartungen bleiben jedoch erhalten. So sind die Energieniveaus der
Diracschen Theorie bei vorgegebenem Gesamtdrehimpuls $j$ ($\vec{j} = 
\vec{l} + \vec{s}$) vom Bahndrehimpuls $l$ und Spin $s$ unabh"angig. Daher sind 
zum Beispiel die $2S_{1/2}$ und $2P_{1/2}$ Niveaus in der Dirac-Theorie
energetisch entartet, denn sie tragen beide die Hauptquantenzahl $n=2$ und 
die Gesamtdrehimpuls-Quantenzahl $j=1/2$. Es gilt also in der Dirac-Theorie
\mathanf
E_{2P1/2} - E_{2S1/2} = 0 \quad \mbox{(exakt)}
\mathend
und somit 
\mathanf
E_{2P3/2} - E_{2S1/2} = E_{2P3/2} - E_{2P1/2} = {\rm Feinstrukturintervall}.
\mathend

W. Lamb und R. Retherford stellten jedoch 1947 in einer 
spektroskopischen Messung fest, da"s zwischen dem $2P_{1/2}$ und dem 
$2S_{1/2}$-Niveau im Widerspruch zur Diracschen Theorie ein 
Energieunterschied besteht \cite{lamb}. 
Die Differenz betr"agt nach einer verbesserten Messung von Triebwasser, 
Dayhoff und Lamb \cite{tdl} aus dem Jahr 1953 
\mathanf
E_{2S_{1/2}} - E_{2P_{1/2}} = 1057.8(1) \, {\rm MHz}.
\mathend
Dieser Effekt wird heute als ``klassische Lamb-Verschiebung'' 
bezeichnet. 

Die Laserspektroskopie erm"oglicht heute spektroskopische 
Messungen an wasserstoffartigen Systemen mit einer relativen Genauigkeit
von bis zu $10^{-12}$ \cite{haensch1}. Der angestrebte Vergleich mit den
Voraussagen der Quantenelektrodynamik erfordert Berechnungen zur
Lamb-Verschiebung in h"oherer Ordnung.

Im Falle des Wasserstoffs kommen die Hauptbeitr"age zur 
Abweichung von der Diracschen Theorie durch quantenelektrodynamische 
Effekte zustande. Einerseits sind kleine Abweichungen vom 
Coulomb-Gesetz bei sehr kleinen Abst"anden daf"ur verantwortlich 
(Vakuumpolarisation), andererseits die Wechselwirkung des Elektrons mit dem virtuell 
vorhandenen eigenen Strahlungsfeld (Selbstenergie). Es gelang H. Bethe noch im 
Jahr der ersten Messung (1947), die Rechnungen zu den Hauptbeitr"agen der 
Lamb-Verschiebung (Vakuumpolarisation und Selbstenergie) in erster nichtverschwindender 
Ordnung ($\alpha (Z \alpha)^4$) zu vervollst"andigen \cite{bethe}. 

%
%

\section{Beitr"age zur Lamb-Verschiebung}

Obwohl nicht alle Abweichungen von der Diracschen Theorie auf die Quantenelektrodynamik 
zur"uckzuf"uhren sind, werden dennoch {\em alle} diese Abweichungen unter dem 
Begriff ``Lamb-Verschiebung'' zusammengefa"st. 
Die Quantenelektrodynamik (QED), die Struktur des Kerns (Protons)
und die volle relativistische Behandlung des Wasserstoffs als Zweik"orperproblem 
liefern au"ser der 
Vakuumpolarisation und Selbstenergie noch verschiedene andere Korrekturen 
zur Diracschen Theorie, die wir hier nach der Gr"o"se ihrer Beitr"age
zum Wasserstoffspektrum auflisten wollen:
\begin{itemize}
\item die Selbstenergie (self-energy),
\item die Vakuumpolarisation (vacuum polarization),
\item R"ucksto"skorrekturen (recoil corrections),
\item Strahlungs-R"ucksto"skorrekturen (radiative recoil corrections),
\item Zwei- und Drei-Schleifen-Korrekturen zur Selbstenergie (two- and 
three-loop corrections),
\item Kerngr"o"senkorrektur (finite nuclear size effect).
\end{itemize}
Es sei hier der Klarheit halber erw"ahnt, da"s die Lamb-Verschiebung f"ur jeden 
Zustand des Wasserstoffatoms einzeln ausgewertet wird. Durch die Differenz 
der Verschiebungen der einzelenen Niveaus ergibt sich dann der 
Energieunterschied, der experimentell nachgewiesen wird.

Man spricht bei allen hier erw"ahnten Effekten von (quantenelektrodynamischen) 
Korrekturen zur Diracschen Theorie, weil die Effekte klein sind 
und daher st"orungstheoretisch behandelt werden k"onnen. Dies erleichtert die 
Rechnung bzw. macht die Auswertung der auftretenden Integrale "uberhaupt 
erst m"oglich.

%
%

\section{Gegenstand der vorliegenden Arbeit}
\label{ggstd}

{\em ``Historically, the one-photon radiative corrections (with
all orders of the Coulomb interaction) have been considered in most detail,
and the work required to attain the present accuracy has extended 
over four decades.''}\\
J. Sapirstein in \cite{kinoshita}\\[1ex]

{\parindent0cm
Das Thema der vorliegenden Arbeit ist der
Ein-Schleifen-Selbstenergiebeitrag zur Lamb-Verschiebung.
Dieser Effekt kommt durch den Austausch {\em eines} virtuellen Photons zustande. 
Er stellt den Hauptbeitrag zur Lamb-Verschiebung in Wasserstoff
dar. Es soll eine Methode ($\ep$fw-Methode) vorgestellt werden, mit der dieser
Beitrag in h"oherer Ordnung analytisch berechnet werden kann 
(siehe Kapitel \ref{efwmethod}). Die Methode f"uhrt zu einer 
gewissen Vereinfachung des sogenannten Niedrigenergie-Anteils. 
Der Hauptbeitrag wird von den relativistischen
Korrekturen getrennt, und die Rechnung kann 
in Teilbeitr"age aufgespalten werden. Mit dieser Methode k"onnen
fehlende Koeffizienten h"oherer Ordnung berechnet werden.}

Die Methode wird in dieser Arbeit zur Berechnung der 
(Ein-Schleifen-) Selbstenergie 
der $2P$-Zust"ande in der Ordnung $\alpha (Z \alpha)^6$ angewandt.
Unter Vernachl"assigung der (geringen) Abh"angigkeit von der reduzierten Masse
des Systems, die sp"ater m"uhelos wieder eingesetzt werden kann (siehe
\cite{kinoshita}), ist der Selbstenergiebeitrag zur Lamb-Verschiebung gegeben durch
\mathanf
{\delta E}_{\rm SE} = \frac{\alpha}{\pi} \, m \, \frac{(Z \alpha)^4}{n^3} \, F,
\mathend
wobei sich f"ur den dimensionslosen $F$-Faktor die folgende Entwi"cklung
in Potenzen von $\alpha$ ergibt
\eqnanf
F & = & F(Z) = A_{40} + A_{41} \ln((Z \alpha)^{-2}) +  
   A_{50} (Z \alpha) + \schwups
 & & (Z \alpha)^2 \left[ A_{60} + A_{61} \, \ln((Z \alpha)^{-2}) + 
   A_{62} \, \ln^2((Z \alpha)^{-2}) + O(Z \alpha) \right]. \ok
\eqnend 
Die erste Ziffer in den Indizes der Koeffizienten gibt 
die Potenz von $(Z \alpha)$ an, die zweite Ziffer die Potenz 
des Logarithmus. F"ur $P$-Zust"ande verschwinden einige 
der Koeffizienten, insbesondere gilt 
$A_{41} = A_{50} = A_{62} = 0$ \cite{kinoshita}. 
Einige andere Koeffizienten sind 
f"ur die $2P$-Zust"ande bekannt,
\begin{displaymath}
A_{40}(2P_{1/2}) = -\frac{1}{6} - \frac{4}{3} \ln k_0(2P),
\end{displaymath} 
\begin{displaymath}
A_{61}(2P_{1/2}) = \frac{103}{180}
\end{displaymath} 
sowie
\begin{displaymath}
A_{40}(2P_{3/2}) = \frac{1}{12} - \frac{4}{3} \ln k_0(2P),
\end{displaymath} 
\begin{displaymath}
A_{61}(2P_{3/2}) = \frac{29}{90}
\end{displaymath} 
mit dem Bethe-Logarithmus \cite{kinoshita}
\begin{displaymath}
\ln k_0(2P) = -0.0300167089(1).
\end{displaymath} 
F"ur die $A_{60}$-Koeffizienten der $2P$-Zust"ande existierten bis jetzt keine
genauen Werte. Es besteht jedoch eine M"oglichkeit, die Werte dieser
Koeffizienten mit Hilfe von numerischen Rechnungen abzusch"atzen.   
Numerische Rechnungen existieren f"ur $F(Z)$ lediglich 
f"ur gro"se Kernladungszahlen $Z$, d.h. f"ur schwere wasserstoffartige Ionen.
F"ur kleine $Z$ konvergiert die numerische Rechnung sehr langsam. 
P. Mohr \cite{mohr} konnte in seiner Arbeit trotz gr"o"seren Aufwands
Ergebnisse nur f"ur $Z \geq 5$ angeben. Durch eine Ausgleichsrechnung 
mit einer geeigneten Funktion und Extrapolation in den Bereich f"ur kleine $Z$
l"a"st sich dennoch eine Absch"atzung f"ur die $A_{60}$-Koeffizienten 
angeben. Die zur Ausgleichsrechnung benutzte Funktion lautet:
\mathanf
G_{\rm SE}(Z) = A_{60} + (Z \alpha) A_{70} 
\mathend
Unter Venachl"assigung des f"ur $Z = 1$ kleinen Beitrags 
\mathanf
(Z \alpha) A_{70}
\mathend
ist dann 
\mathanf
G_{\rm SE}(Z = 1) \approx A_{60}.
\mathend
Die auf diese Art und Weise erhaltenen numerischen Resultate f"ur $A_{60}$ 
lauten:
\begin{equation}
A_{60}(2P_{1/2}) \approx G_{\rm SE}(2P_{1/2}) = -0.8(3)
\end{equation}
sowie
\begin{equation}
A_{60}(2P_{3/2}) \approx G_{\rm SE}(2P_{1/2}) = -0.5(3).
\end{equation}
In dieser Arbeit sollen alle Koeffizienten, die in der Ordnung 
$\alpha (Z \alpha)^6$ zur Lamb-Verschiebung beitragen, analytisch 
berechnet werden. Hierbei ist das Wort ``analytisch'' so zu verstehen,
da"s die Rechnung im Hochenergie-Anteil (siehe Abschnitt \ref{hepmethod})
vollkommen analytisch, im Niedrigenergie-Anteil (siehe Abschnitt 
\ref{lepmetho}) nur bis auf verbleibende eindimensionale Integrationen analytisch
erfolgt. Diese (eindimensionalen) 
Integrationen konvergieren sehr schnell und 
k"onnen mit gro"ser Genauigkeit ausgef"uhrt
werden, da die Integranden glatte Funktionen ohne Singularit"aten 
sind (Verfahren: Gau"ssche Quadratur).  

Die Ergebnisse f"uhren auf neue Werte f"ur die Lamb-Verschiebung der
$2P$-Zust"ande und einen neuen Wert f"ur die Feinstrukturaufspaltung.
Wegen der L"ange der Terme werden
die meisten Zwischenschritte in den Rechnungen nicht explizit aufgelistet.
Die Gr"o\3e der Ausdr"ucke "ubersteigt bei den Integranden der Matrixelemente im 
Hochenergie-Anteil und bei den Zwischenschritten im Niedrigenergie-Anteil
oft die Grenze von 1000 Termen. Selbst manche (Teil-) Ergebnisse sind so
lang, da\3 nur deren Struktur (funktionale Abh"angigkeiten) angegeben wird.
Es werden statt einer expliziten Angabe der Zwischenergebnisse
die entscheidenden konzeptuellen Schritte des Verfahrens dargestellt.

%
%

\section{Motivation f\protect"ur die vorliegende Arbeit}
\label{motivation}

Die Quantenelektrodynamik ist bereits experimentell sehr gut best"atigt. 
Experimentelle
"Uberpr"ufungen der Quantenelektrodynamik haben besondere Bedeutung, 
denn die Theorie ist Modelltheorie f"ur das Standardmodell.
Da es nicht viele Wahlm"oglichkeiten bei der Konstruktion renormierbarer 
Quantenfeldtheorien gibt, w"urde eine Abweichung von den 
Vorhersagen der QED zur Entwi"cklung neuer Theorien f"uhren m"ussen. 

Eine weitere Motivation f"ur die aufwendigen Rechnungen ergibt sich
aus der Pr"azision heutiger spektroskopischer Messungen.
Die Genauigkeit der theoretischen Lamb-Verschiebung der $S$-Zust"ande 
wird heute durch die Ladungsverteilung im Atomkern 
(den quadratischen Kernradius) beschr"ankt.
Es besteht also die Hoffnung, durch neue 
Rechnungen und Experimente einige Erkenntnisse "uber Eigenschaften
der Elementarteilchen zu gewinnen.

Der in dieser Arbeit
erhaltene genaue Wert f"ur die Lamb-Verschiebung der $2P$-Zust"ande
kann zum Beispiel als Referenzpunkt zur "Uberpr"ufung der Messung 
der $2S$-Lamb-Verschiebung dienen.

\section{Computer-Algebra Systeme}
\label{compalgsysteme}

Der Einsatz von Computer Algebra Programmen in der Theoretischen Physik 
hat es erm"oglicht, umfangreiche Berechnungen mit vielen tausend Termen in 
den Zwischenschritten in viel k"urzerer Zeit und sehr viel zuverl"assiger 
als bisher durchzuf"uhren. Symbolische Umformungen werden 
vom Rechner vorgenommen. 
Quantenfeldtheorien bieten sich als Anwendungsgebiet f"ur computergest"utzte
symbolische Rechnungen in gewisser Weise an, denn st"orungstheoretische
Rechnungen werden in h"oheren Ordnungen recht umfangreich. 

Auch in dieser Arbeit wird f"ur die Berechnungen ein Computer Algebra System
verwendet, das Programm $Mathematica$ der Firma Wolfram Research. Dieses
System zeichnet sich im Vergleich zu anderen Computer Algebra Systemen durch
eine sehr flexible Programmiersprache aus, die sowohl regelbasierte wie
auch prozedurale und funktionale Elemente enth"alt. 
Die Programmiersprache Mathematica ist auch objektorientiert:
Mit den mathematischen Objekten k"onnen Regeln assoziiert werden. 
Auf diese Weise lassen sich
zum Beispiel Integrationsregeln definieren, die auch Ausdr"ucke mit 
mehreren tausend Termen im Integranden
zuverl"assig und schnell integrieren. Mit diesem Programm k"onnen Matrixelemente 
der Diracschen Wellenfunktion des $2P_{1/2}$ oder des $2P_{3/2}$ 
Zustands automatisiert berechnet werden. Dabei
entstehen Integranden mit mehreren tausend Termen.

Es wurde ein weiteres Algebra-Programm verfa"st, welches Kommutatoren von Operatoren 
automatisch auswertet. Mit Hilfe dieses Programms wird die Foldy-Wouthuysen
Transformation des Dirac-Hamiltonians automatisch ausgewertet (mit 
bekanntem Ergebnis, siehe \cite{bjorkendrell}). 
Das Algebra-Programm wird verwendet, um die Foldy-Wouthuysen Transformation 
des Operators $\alpha^i e^{i \scp{k}{r}} = \gamma^0 \gamma^i e^{i \scp{k}{r}}$ auszuwerten 
(siehe Abschnitt \ref{FWtrafo}). Das Ergebnis dieser Transformation ist neu
und ein zus"atzliches Resultat dieser Arbeit.

%
%

\chapter{Theorie der Selbstenergie}
\label{theorie}

%
%

\section{Konventionen}
\label{vorbem}
 
Es folgt hier eine Liste der am h"aufigsten verwendeten Symbole und 
Konventionen.

\begin{itemize}
\item Es werden nat"urliche Einheiten verwandt, in denen
$\hbar = c = 1$ ist und somit nur eine einzige physikalische
Dimension verbleibt. Als verbleibende Dimension w"ahlen
wir in dieser Arbeit die L"ange. Die Dimension der L"ange wollen wir
mit $\lquer$ bezeichnen.

\item Griechische Indizes durchlaufen die Werte von $0 \dots 3$, 
lateinische Indizes die Werte $1 \dots 3$.

\item Indizes von kontravarianten Vektoren werden als
Superskript plaziert. Zum Beispiel ist $k^0 = \omega$ die $0$-te Komponente des
Vierervektors $k$.

\item Werden griechische oder lateinische Indizes wiederholt, so gilt 
die Summenkonvention, auch wenn beide Indizes oben oder beide
unten stehen (Beispiel $\scp{\alpha}{p} = \alpha^i p^i$). Die Summenkonvention
wird {\em aufgehoben}, wenn der Index mit einem Strich versehen ist 
(zum Beispiel bei $n'$ anstatt $n$).

\item Die Spurbildung einer Matrix erfolgt vermittels
der ${\rm Tr}$-Operation.

\end{itemize}  

\section{Heaviside-Lorentz-Einheiten}
\label{grundzuege}

Um von SI-Einheiten zu Heaviside-Lorentz-Einheiten
zu gelangen, gehen wir aus vom Coulomb-Gesetz
\mathanf
F= k_0 \frac{q_1 q_2}{r^2} = \frac{1}{4 \pi \ep_0} \frac{q_1 q_2}{r^2}
\mathend
und w"ahlen die Einheiten so, da"s $\ep_0 = 1$ wird. Au"serdem w"ahlen wir
die Einheiten der L"ange und Zeit so, da"s gerade $c = 1$ gilt (das hei"st 
insbesondere, da"s L"ange und Zeit die gleiche physikalische Dimension 
besitzen). Wir w"ahlen die Einheit der Energie so, da"s auch $\hbar = 1$ 
ist. In Heaviside-Lorentz-Einheiten ist dann gerade
\mathanf
e^2 = \frac{e^2}{\ep_0 \hbar c} = 
  4 \pi \frac{k_0 e^2}{\hbar c} = 4 \pi \alpha
\mathend
(weil $\ep_0 = \hbar = c = 1$). Die elektrische Ladung ist in diesen 
Einheiten eine dimensionslose Gr"o"se. Das anziehende Coulomb-Potential
zwischen zwei Elementarladungen wird dann
\mathanf
V(\vec{r}) = - \frac{1}{4 \pi \ep_0} \frac{e^2}{r} =  - \frac{1}{4 \pi 
\ep_0} \frac{4 \pi \alpha}{r} = - \frac{\alpha}{r}.
\mathend 
Die erste Maxwell-Gleichung wird zu
\mathanf
{\rm div} \, \vec{E} = \frac{1}{\ep_0} \rho = \rho,
\mathend
so da"s sich die inhomogenen Maxwell-Gleichungen in diesen Einheiten
als
\begin{equation}
\del_{\mu} F^{\mu \nu} = j^{\nu}
\end{equation}
schreiben. 

%
%
\section{Die Selbstenergiefunktion}

\begin{figure}[htb]
\epsfxsize=6cm
\centerline{\epsfbox{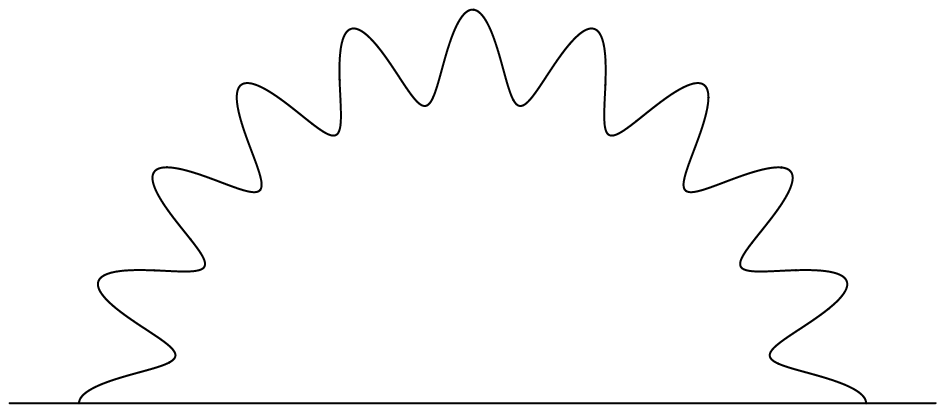}}
\caption{\label{self10} Selbstenergie des freien Elektrons}
\end{figure}

Die Selbstenergie des Elektrons kommt durch die 
Wechselwirkung des Elektrons mit dem eigenen Strahlungsfeld zustande. Der 
quantenelektrodynamische Proze"s der Selbstwechselwirkung l"a"st sich 
durch das Aussenden und das Wiedereinfangen eines virtuellen Photons 
beschreiben. Zun"achst wollen wir uns mit der Selbstenergiefunktion des 
freien Elektrons auseinandersetzen, die durch das Feynman-Diagramm 
\ref{self10} beschrieben ist.
Im Falle eines gebundenen Elektrons m"ussen wir den 
``einfach durchgezogenen ''
freien Feynman-Propagator durch den vollen Propagator des 
Elektrons, welches sich unter dem Einflu"s des Zentralfeldes 
fortbewegt, ersetzen. Dies wird durch den Graphen \ref{self2} 
veranschaulicht, wobei die doppelt durchgezogene Linie den vollen 
Feynman-Propagator f"ur das Elektron (unter dem Einflu"s des 
Coulombfeldes) darstellt:
\begin{figure}[htb]
\epsfxsize=6cm
\centerline{\epsfbox{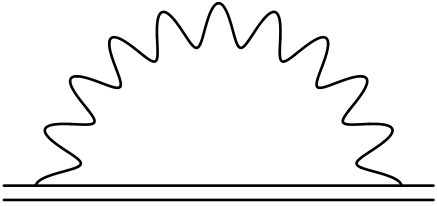}}
\caption{\label{self2} Selbstenergie des gebundenen Elektrons}
\end{figure}
Zun"achst wird die Selbstenergie des freien Elektrons besprochen. Gem"a"s den 
Feynman-Regeln der QED ordnen wir diesem Graphen die folgende Amplitude 
zu. Wir schreiben
\begin{itemize}
\item $-i e \gamma^\mu$ bzw. $-i e \gamma^\nu$ f"ur die zwei Vertizes,
\item $i D_{\mu \nu}(k)$ f"ur den Photonen-Propagator zwischen den 
Vertizes,
\item $i S_F(p)$ f"ur den Feynman-Propagator des Elektrons zwischen den 
Vertizes,
\end{itemize}
5
und wir integrieren "uber den freien unbeobachteten Parameter, den
Vierer-Impuls des Photons $k$. Wir k"onnen die Modifikation
des Feynman-Propagators durch die 
Ein-Schleifen-Korrektur, die wir mit $\Sigma (p)$ bezeichnen
wollen, sofort hinschreiben \cite{greiner7}:
\mathanf
- i \Sigma (p) = \intk \, i D_{\mu \nu}(k) \, ( -i e \gamma^\mu )  \,
  i S_F(p) \, (-i e \gamma^\nu) 
\mathend
oder umgeformt:
\begin{equation}
\label{sigma}
\Sigma (p) = i e^2 \intk \, D_{\mu \nu}(k) \, \gamma^\mu \, S_F(p) \, 
\gamma^\nu.
\end{equation}
Das Problem der ``kleinen'' Modifikation $\Sigma$ zum Feynman-Propagator 
liegt nun darin, da"s $\Sigma (p)$ unendlich gro"s ist. Eine Abz"ahlung
der $k$-Potenzen ergibt 
\mathanf
\begin{array}{rcl}
d^4 k & = & dk \, O(k^3), \\
D_{\mu \nu}(k) & = & O(k^{-2}), \\
S_F & = & O(k^{-1}),
\end{array}
\mathend
da"s das Integral \ref{sigma} linear divergiert. 
Dieses Problem wird durch Renormierung und Regularisierung 
gel"ost.
Dazu wird der durch die Selbstwechselwirkung modifizierte Propagator
des Elektrons 
\mathanf
\label{tildeS} 
  S^I_F = S_F + S_F \, \Sigma (p) \, S_F + S_F \, \Sigma(p) \, S_F \,
    \Sigma(p) \, S_F + \dots
\mathend    
genauer untersucht. Wir k"onnen diese Reihe formal aufsummieren zu:
\begin{equation}
\label{tildeSsum}
  S^I_F(p) = \frac{1}{\slf{p} - m - \Sigma (p) + i \ep}.
\end{equation}
Beachtet man, da"s $S^I_F(p)$ auf Spinor-Wellenfunktionen wirkt, 
die zumindest n"aherungsweise auf der Massenschale liegen, so erscheint 
die folgende Reihenentwicklung um den Punkt $\slf{p} = m$ sinnvoll: 
\mathanf
  \Sigma(p) = A + B (\slf{p} - m) + \Sigma_{\rm ren}(p) (\slf{p} - m)^2.
\mathend
Wir beachten, da"s $\Sigma (p)$ und somit auch $A$, $B$ und 
$\Sigma_{\rm ren}(p)$ von der Ordnung $\alpha$ sind. Wir approximieren  mit 
dieser Vorgabe $\Sigma (p)$ folgenderma"sen (Terme in h"oherer Ordnung in 
$\alpha$ werden weggelassen):
\eqnanf
\label{renormBA}
 S^I_F(p) & = & 
\frac{1}{\slf{p} - m - A - B ( \slf{p} - m ) - (\slf{p} - m)^2 
\Sigma_{\rm ren}(p)} \schwups
& \approx & 
\frac{1}{(\slf{p} - m - A)(1 - B)(1 -  (\slf{p} - m) \Sigma_{\rm ren}(p))} 
\wups
& \approx &
\frac{1 + B}{(\slf{p} - m - A)(1 - (\slf{p} - m ) \Sigma_{\rm ren}(p))} 
\wups
& \approx &
\frac{(1 + B) (1 + (\slf{p} - m ) \Sigma_{\rm ren}(p))}{\slf{p} - m - A} 
\wups
& \approx &
\frac{(1 + B)}{\slf{p} - m - A},
\ok
\eqnend
wobei im letzten Schritt ausgenutzt wurde, da"s wir uns in der 
N"ahe der Massenschale $\slf{p} = m$ befinden. Nun erfolgt die 
Renormierung. Die Konstante $1 + B$ wird in den Faktoren  $- i e 
\gamma^\mu$ der Elektron-Vertizes, die an den Selbstenergiegraphen 
angrenzen, wie folgt absorbiert. Man definiert die Renormierungskonstante 
$Z_2$ als
\begin{equation}
Z_2 = 1 + B
\end{equation}
und die physikalisch observable Ladung des Elektrons als
\mathanf
e'_R = Z_2 \sqrt{Z_3} e.
\mathend
Die $Z_2$-Renormierung ist also eine weitere Ladungsrenormierung
zus"atzlich zur $Z_3$-Ladungsrenormierung (Vakuumpolarisation).

Das Elektron verh"alt sich unter
dem Einflu"s seines eigenen Strahlungsfeldes offenbar so, als h"atte es
die Gesamtmasse $m + A$. Da man aber den Einflu"s dieses Strahlungsfeldes
im Experiment nicht einfach abschalten kann, so mu"s man $A$ 
als Massenrenormierung ansehen. 
Es ist also
\mathanf
A = \delta m.
\mathend
Regularisiert man den Photonenpropagator gem"a"s der 
Vorschrift von Pauli und Villars,
\begin{equation}
\frac{1}{k^2} \to \frac{1}{k^2} - \frac{1}{k^2 - M^2},
\end{equation}
so ergibt sich
\begin{equation}
\label{deltam}
\delta m = \alpha \frac{3 m}{4 \pi} \left[ \ln \left(\frac{M^2}{m^2}\right) + 
  \frac{1}{2} \right].
\end{equation}
Dieser Ausdruck ist einerseits von der Ordnung $\alpha$, also
``klein'', andererseits ist er divergent in $M$, jedoch nur logarithmisch.
Man absorbiert ihn in der renormierten Elektronenmasse. 

Um die Renormierungskonstante $B$ zu berechnen, m"ussen wir
den Photonenpropagator weiter modifizieren gem"a"s
\begin{equation}
\frac{1}{k^2} \to \frac{1}{k^2 - \mu^2} - \frac{1}{k^2 - M^2}\,,
\end{equation}
wobei der urspr"ungliche Propagator im Limes
$\mu \to 0$, $M \to \infty$ erhalten wird. Wir erhalten dann 
\begin{equation}
\label{Z2} 
Z_2 = 1 + B = 1 - \frac{\alpha}{2 \pi} \left[ \f{1}{2} 
  \ln \left( \frac{M^2}{m^2} \right) + 
  \ln \left( \frac{\mu^2}{m^2} \right) + \frac{9}{4} \right].
\end{equation}
Das Ergebnis f"ur $Z_2$ ist sowohl f"ur $\mu \to 0$ als
auch f"ur $M \to \infty$ logarithmisch divergent. Au"serdem ist der Wert
f"ur $Z_2$ nicht eichinvariant und h"angt von der 
fiktiven Elektronmasse $\mu$ ab. Wir werden aber im n"achsten Abschnitt 
sehen, da"s sich die nicht eichinvariante $Z_2$-Renormierung der Selbstenergie 
gerade mit der ebenfalls nicht eichinvarianten $Z_1$-Renormierung der Vertexkorrektur
weghebt und daher die Theorie konsistent bleibt.

Die renormierte Selbstenergiefunktion ergibt sich dann als
\begin{equation}
\Sigma_{\rm ren}(p) = \Sigma(p) - \delta m - (Z_2 - 1) (\slf{p} - m).
\end{equation}

%
%

\section{Die Vertexfunktion}
\label{vertexkorrektur}

\begin{figure}[htb]
\epsfxsize=4.7cm
\centerline{\epsfbox{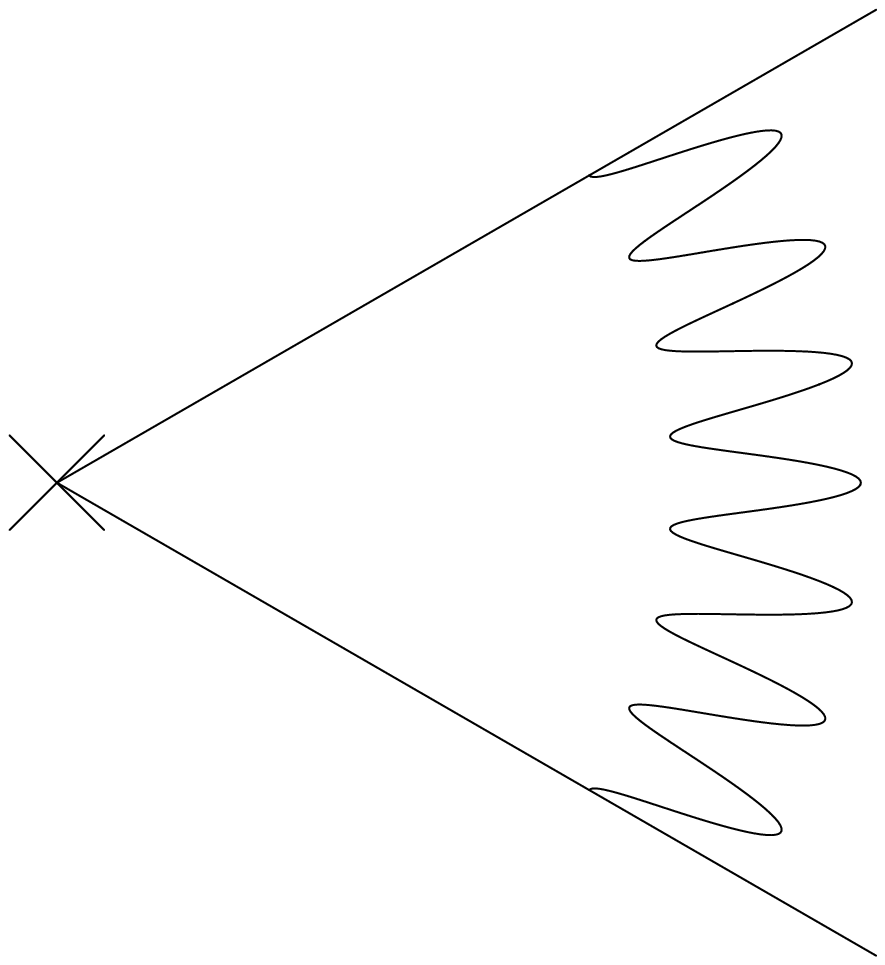}}
\caption{\label{vertexc} Vertexkorrektur}
\end{figure}

Durch die Selbstwechselwirkung eines Teilchens mit seinem eigenen 
Strahlungsfeld ver\"andert 
sich die Amplitude, die einem Wechselwirkungs-Vertex zugeordnet wird
(Vertexkorrektur).
Das entsprechende Feynman-Diagramm ist gegeben durch Abb. \ref{vertexc}.
Wir m"ussen daher
\begin{equation}
- i e \gamma_{\mu} \to - i e \gamma_{\mu} - i e \Gamma_{\mu}(p, p')
\end{equation}
ersetzen. Dabei sind $p$ und $p'$ die Viererimpulse des Elektrons vor bzw. 
nach der Wechselwirkung.  F"ur die Vertexfunktion 
$\Gamma_{\mu}$ kann mit den Feynman-Regeln leicht der folgende Ausdruck 
abgeleitet werden \cite{greiner7}:
\begin{equation}
 \Gamma_{\mu}(p,p') = -i e^2 \intk \frac{1}{k^2} 
  \big( \gamma^{\nu} \frac{1}{\slf{p'} - \slf{k} - m} \gamma_{\mu} 
   \frac{1}{\slf{p} - \slf{k} - m} \gamma_{\nu} \big).
\end{equation}
Auch die Vertexfunktion $\Gamma_{\mu}$ mu"s renormiert werden, denn eine 
Abz"ahlung der $k$-Potenzen im Integranden ergibt eine logarithmische 
Divergenz des Integrals f"ur $\Gamma_{\mu}$. Aus der Entwicklung des 
Propagators
\eqnanf
\frac{1}{\slf{p'} - \slf{k} - m} & = & 
  \frac{1}{\slf{p} - \slf{k} - m + (\slf{p'} - \slf{p})} \schwups
& = & \frac{1}{\slf{p} - \slf{k} - m} - \frac{1}{\slf{p} - \slf{k} - m} 
  (\slf{p'} - \slf{p}) \frac{1}{\slf{p} - \slf{k} - m}
\eqnend
ergibt sich unter Abz\"ahlung der $k$-Potenzen der (divergente)
Hauptbeitrag zur Vertexfunktion gerade f\"ur 
\mathanf
p' = p.
\mathend
Daher kann $\Gamma_{\mu}$ geschrieben werden als
\begin{equation}
\Gamma_{\mu}(p',p) = \Gamma_{\mu}(p,p) + \Gamma^R_{\mu}(p',p)
\end{equation}
mit der renormierten Vertexfunktion $\Gamma^R_{\mu}(p',p)$. Man kann 
nachweisen,
da"s der divergente Anteil der Vertexfunktion $\Gamma_{\mu}$ gerade 
proportional zu $\gamma_{\mu}$ ist:
\begin{equation}
\Gamma_{\mu}(p,p) = \gamma_{\mu} L
\end{equation}
Daher mu"s eine weitere Ladungsrenormierung gem"a"s
\begin{equation}
- i e \gamma_{\mu} \to 
  - i e \gamma_{\mu} - i e \gamma_{\mu} L - i e \Gamma^R_{\mu}(p, p') =
  - i e (1 + L) \Gamma^R_{\mu}(p, p')
\end{equation}
vorgenommen werden, so da"s sich die dreifach renormierte Elektronladung zu
\begin{equation}
e''_R = Z_1^{-1} e'_R
\end{equation}
ergibt mit
\begin{equation}
Z_1^{-1} = 1 + L \quad \mbox{oder} \quad Z_1 = 1 - L 
\end{equation}
(in niedrigster Ordnung in $\alpha$) und
\begin{equation}
e'_R = Z_2 \sqrt{Z_3} e
\end{equation}
($Z_2$-Renormierung aufgrund der Selbstenergie, $Z_3$-Renormierung 
aufgrund der
Vakuumpolarisation). Mit der Ward-Identit"at 
\begin{equation}
\Gamma_{\mu}(p,p) = - \dbyd{\Sigma(p)}{p^{\mu}}
\end{equation}
wird unter Ausnutzung der Entwicklung 
von
$\Sigma(p)$ in Potenzen von $(\slf{p} - m)$ 
\mathanf
\Sigma(p) = A + B (\slf{p} - m) + O((\slf{p} - m)^2)
\mathend
die wichtige Beziehung
\begin{equation}
L \gamma_{\mu} = \Gamma_{\mu}(p,p) = - \dbyd{\Sigma(p)}{p^{\mu}} =
  -B \gamma_{\mu} + O(\slf{p} - m)
\end{equation}
und somit
\begin{equation}
\label{LB}
L = -B
\end{equation}
hergeleitet. Daraus folgt, da"s sich der bei der $Z_2$-Renormierung zur 
Elektronladung
addierte Anteil und der bei der Vertexkorrektur-Renormierung addierte Term
gerade herausheben:
\begin{equation}
e''_R = (1 + L + B) \sqrt{Z_3} e = (1 + (L - L)) e = \sqrt{Z_3} e.
\end{equation}
Die nicht eichinvarianten $Z_1$- und $Z_2$-Renormierungen der Elektronladung
heben sich gegenseitig auf.

%
%

\section{Die Selbstenergie in einem gebundenen Zustand}

Der aufgrund der Selbstenergie modifizierte Feynman-Propagator ist
\begin{equation}
S_F^I = S_F + S_F \Sigma_{\rm ren} S_F +  
  S_F \Sigma_{\rm ren} S_F \Sigma_{\rm ren} S_F + \dots
\end{equation}
(das Superskript $I$ bei $S_F^I$ soll andeuten, da"s hier die 
Interaktion mit dem eigenen Strahlungsfeld ber"ucksichtigt wurde).

Der Feynman-Propagator f"ur die Dirac-Gleichung lautet im Ortsraum,
nach der Zeit fouriertransformiert:
\begin{equation}
\label{SFenergy}
S_F(\vec{r}_1,\vec{r}_2,E) = \swsum{n} \frac{\psi_n(\vec{r_1}) 
\bar{\psi}_n(\vec{r_2})}{E - E_n (1- i \epsilon)}.
\end{equation}
Dabei erstreckt sich die verallgemeinerte Summe sowohl 
"uber den diskreten Teil des Spektrums (normale Summation) wie
auch "uber den kontinuierlichen Teil des Spektrums (Integration).
F"ur diese verallgemeinerten Summen hat J. Schwinger das 
$\schwinger$-Symbol eingef"uhrt.

Die Pole des Propagators liegen gerade bei den Energie-Eigenwerten
$E_n$ der Dirac Gleichung. Ber"ucksichtigt man die leichte 
Modifikation des Propagators durch $\Sigma_{\rm ren}$, so verschieben
sich die Pole geringf"ugig. Die neuen Positionen ${\tilde E}_n = E_n + 
\sigma_n(E_n)$ entsprechen dann den durch die Selbstenergie
verschobenen Energieniveaus. Wir beobachten ("uber $n$ wird nicht summiert!)
\eqnanf
\bra{\bar{\psi}_n} \gamma^0 S_F \gamma^0 \ket{\psi_n} &=& 
\int d^3 r_1 \int d^3 r_2 \swsum{m} \psi_n^{+}(\vec{r}_1)  
\psi_m(\vec{r}_2)  \psi_m^{+}(\vec{r}_2)  
\psi_n(\vec{r}_2) \schwups
& = &  \swsum{m} \delta_{nm} \, \delta_{mn} \frac{1}{E - E_m} \wups
& = &  \frac{1}{E - E_n} \ok 
\eqnend
und 
\eqnanf
\bra{\bar{\psi}_n} \gamma^0 S_F \Sigma_{\rm ren} S_F \gamma^0 \ket{\psi_n} &=& 
\bra{\bar{\psi}_n} \gamma^0 S_F \gamma^0 \gamma^0 
  \Sigma_{\rm ren} S_F \gamma^0 \ket{\psi_n} \schwups
& = & \swsum{m} \swsum{l} \bra{\bar{\psi}_n} \gamma^0 S_F \gamma^0 
\ket{\psi_l}
\bra{\bar{\psi}_l} \Sigma_{\rm ren} \ket{\psi_m}
\bra{\psi_m} S_F \gamma^0 \ket{\psi_n} \wups
& = & \swsum{m} \swsum{l} \frac{1}{E - E_n} \delta_{nl} 
\bra{\bar{\psi}_l} \Sigma_{\rm ren} \ket{\psi_m} \frac{1}{E - E_m} 
\delta_{mn} = \wups
& = & \frac{\bra{\bar{\psi}_n} \Sigma_{\rm ren} \ket{\psi_n}}{(E - E_n)^2}. \ok
\eqnend
Daher ist in erster Ordnung
\begin{equation}
\bra{\psi_n} S_F^I \ket{\psi_n} = \frac{1}{E - E_n} + 
\frac{\bra{\bar{\psi}_n} \Sigma_{\rm ren} \ket{\psi_n}}{(E - E_n)^2} + \dots
= \frac{1}{E - E_n - \bra{\bar{\psi}_n} \Sigma_{\rm ren} \ket{\psi_n}},
\end{equation}
und man kann in erster Ordnung die Energieverschiebung aufgrund der 
Selbstenergie als
\begin{equation}
\label{self}
\delta E_{\rm SE}= \bra{\bar{\psi}_n} \Sigma_{\rm ren} \ket{\psi_n}
\end{equation}
ansetzen.

Aus dieser Gleichung f"ur die Energieverschiebung eines
gebundenen Zustands und der Gleichung \ref{sigma} f"ur die 
Selbstenergiefunktion ergibt sich der Selbstenergiebeitrag zur 
Lamb--Verschiebung im Zustand $\ket{\psi}$ als
\eqnanf
\label{selfE}
\delta E_{\rm SE} & = &  \bra{\bar \psi} i e^2 \intk \, D_{\mu \nu}(k) 
\,\gamma^\mu \, 
S_F^V(p) \, \gamma^\nu - {\rm Ren} \ket{\psi} \schwups
 & = & i e^2 \intk \, D_{\mu \nu}(k) \, \bra{\bar \psi} \gamma^\mu \, 
S_F^V(p) \, \gamma^\nu \ket{\psi} - \bra{\bar \psi} {\rm Ren} \ket{\psi}, 
\ok
\eqnend
wobei  
\begin{equation}
S_F^V = \SFV
\end{equation}
den durch das Coulombpotential zus"atzlich modifizierten 
Feynman-Propagator darstellt. Dieser Propagator wird auch als 
Dirac-Coulomb-Propagator bezeichnet. 

${\rm Ren}$ steht in \ref{selfE} f"ur all diejenigen Ausdr"ucke, die bei der
Renormierung abgezogen werden m"ussen. 

%
%

\section{Renormierung der Selbstenergie}
\label{renormgeb}

Bei der Renormierung der 
Selbstenergiefunktion $\Sigma (p)$ traten sowohl 
die Elektron-Renormierung $Z_2$ als auch die Massenrenormierung $\delta m$
auf. Die Gleichungen \ref{deltam} und \ref{Z2} 
f"ur die Elektron-Renormierung
$Z_2$ und die Massenrenormierung $\delta m$ waren erhalten worden, indem 
zun\"achst das Integral f"ur $\Sigma (p)$ vor der Integration
regularisiert wurde. Das Ergebnis sind dann die logarithmisch
divergenten Massenrenormierugs und Elektronrenormierungskonstanten. Es wurde dabei 
folgende Reihenfolge eingehalten: Regularisierung, Integration, Renormierung. 

In praktischen Rechnungen ist es jedoch
h"aufig g"unstiger, zun\"achst einen divergenten Ausdruck f"ur einen
Proze"s zu berechnen und dann vom Integranden Terme abzuziehen, 
die den Renormierungen entsprechen. Schlie"slich regularisiert
man den Integranden und integriert. Reihenfolge hierbei:
Renormierung, Regularisierung, Integration. Diese Reihenfolge
wollen wir auch in dieser Arbeit einhalten und "uberlegen,
welchen Ausdruck wir in diesem Fall vom Integranden abziehen m"ussen. 

Eine Renormierung und Regularisierung 
der Selbstenergie des Elektrons in einem gebundenen Zustand
\begin{equation}
\delta E_{\rm SE} = \bra{\bar{\psi}} \Sigma(p) \ket{\psi}
\end{equation}
ist nur f"ur hohe Energien des ausgetauschten virtuellen Photons 
(im ultravioletten Limes) notwendig.
Die Selbstenergiefunktion des gebundenen 
Elektrons wird durch das elektrostatische
Zentralkraftfelds modifiziert, denn der freie Propagator f"ur das Elektron 
mu\3 durch
\mathanf
\frac{1}{\slf{p} - \slf{k} - m} \to \dcprop
\mathend
ersetzt werden. F"ur den Hochenergie-Anteil der Selbstenergie 
(``ultravioletter Anteil'') w"ahlen 
wir die Feynman-Eichung des Photonenpropagators. Wir schreiben 
diesen Anteil, den wir mit $E_H$ bezeichnen wollen, 
folgenderma\3en auf:
\begin{equation}
E_H = - i e^2 \int \frac{d^4 k}{(2 \pi)^4} 
  \big(\frac{1}{k^2} - \frac{1}{k^2- M^2} \big) 
  \bra{\bar{\psi}} \gamma^{\mu} \dcprop \gamma_{\mu} \ket{\psi} 
  - \bra{\bar{\psi}} {\rm Ren}\ket{\psi}.
\end{equation}
Wir legen uns hierbei nicht fest, "uber welchen Bereich sich die 
Integration erstreckt. Dieser Punkt  wird im Abschnitt "uber den 
Hochenergie-Anteil in Kapitel \ref{hepmethod} n"aher besprochen. In der 
Formel f"ur $E_H$ steht {\rm ren} f"ur diejenigen Ausdr"ucke, die bei der 
Renormierung der Selbstenergie abgezogen werden m"ussen. 

Wir definieren das Matrixelement
\mathanf
{\tilde P} = \bra{\bar{\psi}} \gamma^{\mu} \dcprop \gamma_{\mu} \ket{\psi}.
\mathend
Au"serdem f"uhren wir f"ur den Nenner im Feynman-Propagator die Abk"urzung 
\begin{equation}
D = \slf{p} - \slf{k} - m
\end{equation}
ein. 

F"ur hohe Energien $\omega$ des virtuellen Photons ist es m"oglich und 
sinnvoll, den 
Dirac-Coulomb-Propagator in Potenzen des Potentials $V$ zu entwickeln. 
Dabei wird ausgenutzt, da"s f"ur gebundene Zust"ande im Coulombpotential
$V = - \frac{\alpha}{r}$ von der Ordnung $\alpha^2$ ist ($r$ als Abstand
vom Kern ist von der Ordnung $a_{Bohr} = 1/(\alpha m)$).

Die Entwicklung des Dirac-Coulomb-Propagators geschieht mit Hilfe der 
Formel
\mathanf
\frac{1}{X - Y} = \frac{1}{X} + \frac{1}{X} Y \frac{1}{X} + 
  \frac{1}{X} Y \frac{1}{X} Y \frac{1}{X} + \dots
\mathend
Wir setzen
$X = \slf{p} - \slf{k} - m = D$ und $Y = \go V$ und erhalten als Resultat 
der Entwicklung 4 Terme, den 0-Vertex, 1-Vertex, 2-Vertex und 3-Vertex-Anteil.
\eqnanf
\frac{1}{\slf{p} - \slf{k} - \go V - m} & = & \frac{1}{D - \go V} \schwups
& = & \frac{1}{D} + \wups
& & \frac{1}{D} \go V \frac{1}{D} + \wups
& & \frac{1}{D} \go V \frac{1}{D} \go V \frac{1}{D} + \wups
& & \frac{1}{D} \go V \frac{1}{D} \go V \frac{1}{D} \go V \frac{1}{D} + 
\wups
& & O(V^4) \ok
\eqnend
H"ohere Terme tragen nicht bei, da $V^3$ bereits von der Ordnung $\alpha^6$
ist und wir nur bis zu dieser Ordnung rechnen wollen.

\begin{figure}[htb]
\epsfxsize=12cm
\centerline{\epsfbox{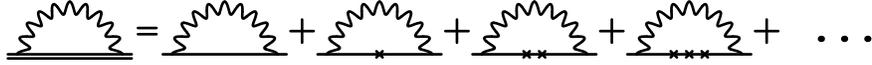}}
\caption{\label{self3x} Entwicklung der Selbstenergie in Termen des 
Coulombfeldes}
\end{figure}

Diese Entwicklung entspricht einer Aufgliederung der Selbstenergie des
gebundenen Elektrons in die Summe aus 4 einzelnen Feynman-Diagrammen 
gem"a"s
der folgenden Abb. \ref{self3x}
Der erste dieser Graphen (der 0-Vertex-Anteil) entspricht gerade der 
Selbstenergie des
freien Elektrons, der zweite dieser Graphen (der 1-Vertex-Anteil) gerade 
der 
Vertexkorrektur zur einfachen Interaktion des Elektrons am Coulombfeld. 
Die weiteren
Graphen aus Abbildung \ref{self3x} tragen keine Divergenzen, wie eine 
Abz"ahlung
der $k$-Potenzen ergibt. 

Daher m"ussen wir bei der Renormierung der Gesamt-Selbstenergie sowohl die 
Renormierungen aufgrund des 0-Vertex-Anteils wie auch aufgrund des 
1-Vertex-Anteils
miteinbeziehen.
 
Zun"achst wird der 0-Vertex-Anteil behandelt, in welchem wir mit
dem Propagator des freien Elektrons rechnen. 
F"ur unsere Rechnungen ben"otigen wir eine 
Integraldarstellung des Massen-Gegenterms.
Als Ansatz w"ahlen wir
\begin{equation}
\label{deltamausdruck}
\delta m = - i e^2 \intk \left[ \frac{1}{k^2} - \frac{1}{k^2 - M^2} \right]
  \frac{2 ( m + \omega )}{k^2 - 2 m \omega}.
\end{equation}
Die G"ultigkeit dieser Darstellung sollte explizit nachgewiesen 
werden.  
Man kann den Ausdruck \ref{deltamausdruck} explizit berechnen und erh"alt 
als 
Ergebnis Gl. \ref{deltam}. Im folgenden soll ein
\"Uberblick \"uber die wichtigsten Rechentricks gegeben werden, die
zur direkten Berechnung des Ausdrucks \ref{deltamausdruck} angewandt
werden. In der Rechnung wird zweckm\"a\3ig zu Hilfsvariablen
$\alpha_1$  und $\alpha_2$ \"ubergegangen, die gem"a\3 
\mathanf
\frac{1}{k^2 + i \ep} = -i \int_0^{\infty} d \alpha_1 \, \exp(i \alpha_1
  (k^2 + i \ep)) 
\mathend
und
\mathanf
\frac{1}{k^2 - 2 m \omega} =  -i \int_0^{\infty} d \alpha_1 \, \exp(i \alpha_1
  (k^2 - 2 m \relscp{k}{t}) ) 
\mathend
eingef\"uhrt werden mit dem zeitartigen Einheitsvektor $t$
\mathanf
t = (1,0,0,0).
\mathend
Der Faktor $\omega$ kann als
\mathanf
\omega = -i \dbyd{}{z^0} \strich_{z=0} \exp(i\relscp{z}{k})
\mathend
geschrieben werden. Mit der verallgemeinerten Gau\3schen
Integralformel
\mathanf
\intk \exp(i(a k^2+ \relscp{b}{k})) = \frac{-i}{(4 \pi)^2 a^2} \,
  \exp(-ib^2/(4a))
\mathend
kann dann die 4-dimensionale 
$k$-Integration ausgef\"uhrt werden. Dabei ist nat\"urlich der
Photonenpropagator gem\"a\3 Pauli-Villars zu regularisieren.
Das Ergebnis der Integration ist
\mathanf
\delta m = \alpha \frac{3 m}{4 \pi} \left[ \ln \left( \frac{M^2}{m^2} \right) + 
\frac{1}{2} \right].
\mathend
Der Abzug des Massen-Gegenterms nach der Integration ist also
\"aquivalent der Subtraktion des Ausdrucks \ref{deltamausdruck} vor der
Integration.

Es ist eine Besonderheit f"ur gebundene Zust"ande, da"s der
mit der Elektronrenormierung $Z_2$ verbundene Term nicht beitr"agt und auch
nicht abgezogen zu werden braucht. Unter Vernachl\"assigung des
Massen-Gegenterms ist der divergente Anteil des 0-Vertex-Diagramms
\begin{equation}
D_0 = \bra{\bar \psi} B (\slf{p} - m) \ket{\psi}
\end{equation}
mit
\mathanf
B = Z_2 - 1.
\mathend
Der divergente Beitrag aufgrund des 1-Vertex-Anteils ist, da zwischen 
den Feynman-Propagatoren gerade der Ausdruck $\go V$ steht, 
gegeben durch
\begin{equation}
D_1 = \bra{\bar \psi} \Gamma^0(p,p) V \ket{\psi}
\end{equation}
mit
\begin{equation}
\Gamma^0(p,p) = L \go
\end{equation}
und somit 
\begin{equation}
D_1 = \bra{\bar \psi} \go L V \ket{\psi}
\end{equation}
mit $L = Z_1^{-1} - 1$. Mit der Ward-Identit"at 
\mathanf
\Gamma_{\mu}(p,p) = - \dbyd{}{p^{\mu}} \Sigma(p)
\mathend
l"a\3t sich nun (siehe Abschnitt \ref{vertexkorrektur}) die 
Beziehung \ref{LB}
\mathanf
L = -B
\mathend
zeigen. Daher gilt:
\eqnanf
D_0 + D_1 &=& 
  \bra{\bar \psi} B (\slf{p} - m) \ket{\psi} + \bra{\bar \psi} \go L 
    V \ket{\psi} = \schwups 
&=& \bra{\bar \psi} (Z_2 - 1) (\slf{p} - m - \gamma^0 V) \ket{\psi} = 0, \ok
\eqnend
weil $\ket{\psi}$ die Dirac-Gleichung mit Feld
\mathanf
(\slf{p} - m - \gamma^0 V) \ket{\psi} = 0
\mathend
erf"ullt. Es ist daher lediglich n"otig, den Massen-Gegenterm $\delta m$ abzuziehen, 
und es ergibt sich in Feynman-Eichung:
\begin{equation}
\label{feynE}
E_H =  - i e^2 \intk \, \frac{1}{k^2} \, \bra{\bar \psi} 
\gamma^\mu \, 
\frac{1}{\slf{p} - \slf{k} - m - \gamma^0 V} \, \gamma_{\mu} \ket{\psi} - 
\bra{\bar \psi} \delta m \ket{\psi}.
\end{equation}
 
%
%

\chapter{Die epsilon-fw-Methode}
\label{efwmethod}

\section{Allgemeines}

Die Rechenmethode f"ur die Selbstenergie ist
eine modifizierte Version der in \cite{krpdiss}
angewandten 
Rechenmethode ($\ep$-Methode). Die Methode st"utzt
sich auf eine Trennung der Rechnung in einen Hoch-
und einen Niedrigenergieanteil (in Bezug auf die Energie des 
virtuellen Photons). Diese Aufteilung war bereits
Gegenstand der urspr"unglichen Rechnungen zur Selbstenergie 
gebundener Zust"ande.
H. Bethe berechnete in \cite{bethe} den 
Niedrigenergie-Anteil f"ur die Selbstenergie in der Ordnung
$\alpha (Z \alpha)^4$. Die hier vorgestellte
Rechnung geht zwei Ordnungen in $Z \alpha$ h"oher. Das bedeutet, da"s in
der Wellenfunktion und im Propagator relativistische Korrekturen zu
der Betheschen Rechnung mitgenommen werden m"ussen. 
  
Die $\ep$-Methode wird in dieser Arbeit
modifiziert. Im Niedrigenergie-Anteil wird eine 
Foldy-Wouthuysen-(FW)-Transformation 
eingef"uhrt. Dadurch erfolgt eine klare Trennung des 
Hauptbeitrages von den relativistischen Korrekturen. 
Au"serdem kann mit dem nichtrelativistischen Propagator f"ur das
gebundene Elektron gerechnet werden. Der nichtrelativistische Propagator 
ist in geschlossener Form sowohl im Ortsraum wie auch im Impulsraum 
bekannt. Die Rechnung l"a"st sich aufgrund der
FW-Transformation in Einzelschritte aufspalten, die nacheinander 
abgearbeitet werden k"onnen. Dadurch wird eine zus"atzliche 
Vereinfachung erzielt (die Ausdr"ucke in den Zwischenschritten 
"ubersteigen nur sehr selten die Grenze von 1000 Termen).
Im Hochenergie-Anteil wird die Rechnung fast vollst"andig 
computergest"utzt ausgef"uhrt. Zun"achst wird in Potenzen des 
Potentials $V$ und des Impulsoperators $\vec{p}$ 
(und damit in $(Z \alpha)$)
entwickelt. Die dabei entstehenden Terme enthalten
Matrixelemente mit der Diracschen Wellenfunktion $\ket{\psi}$.
Die Matrixelemente werden mit symbolischen Programmen 
zur beschleunigten Berechnung von Integralen ausgewertet (wobei hier
die Grenze von 1000 Termen im Integranden oft "uberschritten wird).
Als Name f"ur diese computer-unterst"utze 
Foldy-Wouthuysen-$\ep$-Methode wird ``$\ep$fw-Methode'' vorgeschlagen.

%
%

\section{Einige Definitionen}
\label{defs}

Zun"achst sei an die 4 $\gamma$-Matrizen $\gamma^{\mu}$ ($\mu = 0,1,2,3$) mit ihren 
Kommutatorbeziehungen
\begin{equation}
\anticom{\gamma^{\mu}}{\gamma^{\nu}} = 2 g^{\mu \nu}
\end{equation}
erinnert. Wir verwenden f"ur diese $4 \times 4$-Matrizen die Dirac-Darstellung, 
\begin{equation}
\go = \left( \begin{array}{cc} 1 & 0 \\ 0 & -1 \end{array} \right) 
\quad \mbox{sowie} \quad 
\vec{\gamma} =  \left( \begin{array}{cc} 0 & \vec{\sigma} \\ 
-\vec{\sigma} & 0 \end{array} \right).
\end{equation}
F"ur die $\gamma^5$-Matrix verwenden wir die Konvention
\begin{equation}
\gamma^5 = \gamma^0 \gamma^1 \gamma^2 \gamma^3.
\end{equation}
Wir wollen eine Matrix (einen Operator) als {\em gerade} definieren, 
wenn er von der Form
\begin{equation}
\Es = \left( \begin{array}{cc} A & 0 \\ 0 & B \end{array} \right)
\end{equation}
ist (mit $2 \times 2$-Matrizen $A$ und $B$), d. h. wenn er in einer 
Bispinor-Darstellung Diagonalgestalt hat.
Wir wollen eine Matrix (den Operator) {\em ungerade} nennen, 
falls er von der Form
\begin{equation}
\Os = \left( \begin{array}{cc} 0 & A \\ B & 0 \end{array} \right)
\end{equation}
ist. Es sind zum Beispiel die $\alpha^i$-Matrizen mit 
\begin{equation}
\alpha^i = \gamma^0 \gamma^i = \left( \begin{array}{cc} 0 & \vec{\sigma} \\ 
\vec{\sigma} & 0 \end{array} \right)
\end{equation}
ungerade (die $\alpha$-Matrizen sind nicht mit der Feinstrukturkonstanten $\alpha$ zu verwechseln).
Auch $\scp{\alpha}{p} = \alpha^i p^i$ ist ein ungerader Operator. Demgegen"uber ist zum Beispiel 
die $\gamma^0$-Matrix gerade. 

Nun zur Nomenklatur. Wir wollen den Schr"odinger-Hamiltonian f"ur das Wasserstoffatom 
definieren als
\begin{equation}
H_S = \frac{\psq}{2 m} + V.
\end{equation}
Er wirkt auf dem Raum der komplexwertigen Wellenfunktionen. Der 
Dirac-Hamiltonian
\begin{equation}
H_D = \scp{\alpha}{p} + \beta m + V
\end{equation}
wirkt auf dem Raum der $4$-komponentigen Dirac-Spinoren. 
Nun sei zur Klarheit darauf
hingewiesen, da"s wir 
bez"uglich der Entwicklung in $\alpha$ f"ur gebundene Zust"ande 
\eqnanf
\vec{\alpha} & = & O(\alpha) \schwups
\vec{p} & = & O(\alpha) \schwups
\beta m & = & O(1) \schwups
V & = & O(\alpha^2) \schwups
\modk & = & O(\alpha^2) \glok
\eqnend
ansehen k"onnen. F"ur das Potential $V$ zum Beispiel ergibt sich 
die Begr"undung
folgendernma"sen: $r$ als Abstand
vom Kern ist von der Ordnung $a_{\rm Bohr} = 1/(\alpha m)$, und wir
beziehen uns in dieser Arbeit auf die Massenskala des Elektrons $m$.
Daher f"uhren wir den Parameter $\rho$ mit 
\begin{equation}
\rho = \frac{r}{a_{\rm Bohr}} = \alpha m r
\end{equation} 
ein und erhalten
\mathanf
V = - \frac{\alpha}{r} = - \frac{\alpha^2 m}{\rho}
\mathend
(siehe das Buch Bethe und Salpeter, \cite{bookbethe}).

%
%

\section{Die gamma-Matrizen}

In dieser Arbeit werden 
h"aufig die 16 $\Gamma$-Matrizen verwendet, die eine Basis der 
$4 \times 4$-Matrizen bilden. Es sind dies (siehe zum Beispiel \cite{itzykson},
Seite 54):
\eqnanf
\Gamma^0 & = & I \schwups
\Gamma^{1 \dots 4} & = & \gamma^{\mu} \schwups
\Gamma^{5 \dots 7} & = & i \gamma^0 \gamma^i = i \alpha^i \schwups
\Gamma^{8 \dots 10} & = & \frac{i}{2} \ep^{ijk} \gamma^j \gamma^k = \Sigma^i \schwups
\Gamma^{11 \dots 14} & = & \gamma^5 \gamma^{\mu} \schwups
\Gamma^{15} & = & \gamma^5 \glok
\eqnend
Diese $16$ Matrizen werden als $\Gamma^{\alpha}$-Matrizen bezeichnet, wobei der Index 
$\alpha$ von $0 \dots 15$ l"auft. Die $\Gamma_{\alpha}$ Matrizen ergeben sich 
durch Anwenden der
Metrik $g_{\mu \nu}$ auf die $\gamma$-Matrizen, aus denen die 
$\Gamma$-Matrix konstruiert ist. Zum
Beispiel ist $\Gamma_0 = \Gamma^0$ und $\Gamma_5 = - \Gamma^5$. 
Auf den $4 \times 4$-Matrizen kann 
mit Hilfe der Spur ein Skalarprodukt konstruiert werden,
\begin{equation}
\scalar{A}{B} = \sum_{\alpha} \left( \frac{1}{4} \, {\rm Tr}(\Gamma_{\alpha} A) \right)^{*}
\left( \frac{1}{4} \, {\rm Tr}(\Gamma_{\alpha} B) \right).
\end{equation}
Mit der Beziehung 
\begin{equation}
{\rm Tr}(\Gamma_{\alpha} \Gamma^{\beta}) = 4 \,\delta^{\alpha \beta}
\end{equation}
rechnet man leicht nach, da"s die $\Gamma$-Matrizen eine Orthonormalbasis bez"uglich
des oben definierten Skalarprodukts bilden:
\begin{equation}
\scalar{\Gamma^{\alpha}}{\Gamma^{\beta}} = \delta^{\alpha \beta}.
\end{equation}
Das bedeutet aber, da"s jede $4 \times 4$ Matrix $A$ in den $\Gamma$-Matrizen 
entwickelt werden kann, 
\begin{equation}
\label{expansionGamma}
A = \sum_{\beta} \Gamma^{\beta} \, \scalar{\Gamma^{\beta}}{A} = 
  \sum_{\beta} \Gamma^{\beta} \, \left(\frac{1}{4} 
  {\rm Tr}(\Gamma_{\beta} A) \right).
\end{equation}
Diese Beziehung ist zum Beispiel im Buch von Itzykson und Zuber 
(siehe~\cite{itzykson}) zu finden.
Die Entwicklung in die $\Gamma$-Matrizen kann auf ein Skalarproukt vermittels
der Spurbildung zur"uckgef"uhrt werden.  
F"ur die Berechnung der Spuren ${\rm Tr}(\Gamma_{\beta} M_i)$ wird
ein symbolisches Programm verwandt \cite{HIP}.

%
%

\section{Die Foldy-Wouthuysen-Transformation}
\label{FWtrafo}

Wir wollen an das allgemeine Prinzip erinnern, auf dem 
die Foldy-Wouthuysen-(FW)-Transformation beruht. 
Gegeben sei ein Hamiltonian f"ur Spinoren ,
\begin{equation}
H = \Os + \Es + \beta m,
\end{equation}
mit $\Os$ als ungeradem Operator und $\Es$ als geradem 
Operator (so kann zum Beispiel 
im Dirac-Hamiltonian $\Os = \scp{\alpha}{p}$ und $\Es = V$ 
identifiziert werden). $H$ ist 
hermitesch und somit auch $\Os$ und $\Es$:
\begin{equation}
\Os^{+} = \Os \quad \mbox{und} \quad \Es^{+} = \Es.
\end{equation}
Wir haben aufgrund der expliziten Gestalt der 
Matrix $\go$ die Eigenschaft
\begin{equation}
\anticom{\beta}{\Os} = 0 \quad \mbox{oder} \quad \beta \Os = - \Os \beta.
\end{equation}
Zus"atzlich sei $H$ nicht explizit zeitabh"angig, d. h.
\begin{equation}
\dbyd{H}{t} = 0.
\end{equation}
Da $H$ nicht zeitabh"angig ist, transformiert sich die 
``Schr"odingergleichung''
\begin{equation}
i \dbyd{\ket{\psi}}{t} = H \ket{\psi}
\end{equation}
unter unit"aren Transformationen $U$ wie
\begin{equation}
i \dbyd{U \ket{\psi}}{t} = U H U^{+} (U \ket{\psi}) 
 \quad \mbox{falls} \quad \dbyd{U}{t} = 0.
\end{equation}
Dies k"onnen wir auch so auffassen, als w"urde der 
transformierte Hamiltonian $H' = U H U^{+}$
auf die transformierte Wellenfunktion $\psi' = U \psi$ wirken. 

Es k"onnen folgende Aussagen gezeigt werden.
\begin{enumerate}
\item Der durch
\begin{equation}
U = e^{i S}Ê\quad \mbox{mit} \quad
S = - i \beta \, \frac{\Os}{2 m}
\end{equation}
gegebene Operator ist unit"ar. 
\item Im transformierten Hamiltonian
\begin{equation} 
H' = U \, H \, U^{+}
\end{equation}
verschwinden ungerade Operatoren in niedrigster Ordnung in $\alpha$.
\end{enumerate}
F"ur die Unitarit"at von $U$ reicht es zu zeigen, da"s $S$ 
hermitesch ist. Denn dann ist
\begin{equation}
U^{+} U = e^{-i S^{+}} \, e^{i S} = 1.
\end{equation}
Die Hermitezit"at von $S$ rechnet man leicht nach:
\begin{equation}
S^{+} = i \frac{\Os^{+}}{2 m} \beta^{+} = 
  i \frac{\Os}{2 m} \beta = - i \beta \frac{\Os}{2 m} = S.
\end{equation}
Der transformierte Hamiltonian $H'$ schreibt sich unter Benutzung der 
Campbell-Baker-Haussdorff-Identit"at als
\begin{equation}
H' = e^{i S} \, H \, e^{-i S} = \sum_{n = 0}^{\infty} \frac{i^n}{n!} \, {{^n}[S,H]}
\end{equation}
mit $^n[S,H]$ als $n$-Kommutator von $S$ und $H$ ($^0[S,H] = H$). 
In erster Ordnung ist daher
\begin{equation}
H' = H + i \, [S,H] + \dots,
\end{equation}
wobei wir den Kommutator $[S,H]$ wiederum in niedrigster 
Ordnung in $\alpha$ als
\eqnanf
[S,H] & = & \com{- i \beta \frac{\Os}{2 m}}{\Os + \Es + \beta m} \schwups
& = & \com{- i \beta \frac{\Os}{2 m}}{\beta m} + \dots \wups
& = & i \Os + \dots
\eqnend
schreiben k"onnen. Somit
\begin{equation}
H' = H + i [S,H] = \Os + \Es + \beta m + i (i \Os) + \dots = 
  \Es + \beta m + \dots
\end{equation}
Wie wir sehen, ist im transformierten Hamiltonian der 
ungerade Operator in niedrigster Ordnung 
$\Os$ gerade verschwunden.

Um auch ungerade Operatoren h"oherer Ordnung in $\alpha$ verschwinden zu lassen, m"ussen
zwei und mehr FW-Transformationen hintereinandergeschaltet werden. Der 
dann die Transformation
vermittelnde Operator $U U' U'' \dots$ ist dann selbstverst"andlich 
immer noch unit"ar. 
Dieses Verfahren wollen wir jetzt auf den Dirac-Hamiltonian anwenden.
Wir identifizieren
\begin{equation}
\Os = \scp{\alpha}{p} \quad \mbox{sowie} \quad \Es = V\,.
\end{equation}
Es ist dann
\begin{equation}
S = - i \beta \frac{\Os}{2 m} = - i \beta \frac{\scp{\alpha}{p}}{2 m}.
\end{equation}
In unserem Beispiel reichen zwei FW-Transformationen aus, um auch in Ordnung $\alpha^4$ 
ungerade Operatoren zum Verschwinden zu bringen. 
F"ur die erste der FW-Transformationen erh"alt 
man nach einer l"angeren Rechnung bis zur Ordnung $\alpha^4$ 
(siehe etwa \cite{bjorkendrell}):
\eqnanf
H' & = &
\beta \left( m + \frac{ \Os^2 }{2 m} -  \frac{ \Os^4 }{8 m^3} \right)
+ \Es - \frac{1}{8 m^2} 
\com{ \Os }{\com{ \Os }{ V }} + 
\frac{\beta}{2 m} \com{ \Os }{ \Es } - \frac{ \Os^3 }{3 m^2} \wups
& = &    
\beta \left( m + \frac{\psq}{2 m} -  \frac{(\psq)^2}{8 m^3} \right) 
+ V - \wups
& & - \frac{1}{8 m^2} 
\com{ \scp{\alpha}{p} }{\com{ \scp{\alpha}{p} }{ V }} + 
\frac{\beta}{2 m} \com{ \scp{\alpha}{p} }{V} - 
\frac{(\scp{\alpha}{p})^3}{3 m^2}. \glok   
\eqnend
F"ur die zweite FW-Transformation ist daher
\begin{equation}
S' = - i \beta \frac{\Os'}{2 m} = - i \frac{\beta}{2 m} 
\left( \frac{\beta}{2 m} \com{\Os}{\Es} - \frac{\Os^3}{3 m^2} \right) = 
- i \frac{\beta}{2 m} 
\left( \frac{\beta}{2 m} \com{ \scp{\alpha}{p} }{V} - 
\frac{( \scp{\alpha}{p} )^3}{3 m^2} \right),  
\end{equation}
und nach einer ebenfalls l"angeren Rechnung erh"alt man f"ur den
Foldy-Wouthuysen-transformeierten Hamiltonian $H_{\rm FW}$ 
(siehe zum Beispiel \cite{bjorkendrell})
\begin{equation}
\label{HFW4}
H_{\rm FW} = H' = \go \left( m + \frac{\psq}{2 \, m}  - 
  \frac{(\psq)^2}{8 \, m^3} \right) + V + \frac{\pi \alpha}{2 m^2} 
  \delta(\vec{r})  + \frac{\alpha}{4 m^2 r^3} \vec{\Sigma} \cdot \vec{L}.
\end{equation}
Die Foldy-Wouthuysen-(FW)-Transformation
ist also nichts weiter als die Hintereinanderausf"uhrung 
zweier unit"arer Transformationen $U$ und $U'$,
die durch $\exp(i S)$ und $\exp(i S')$ vermittelt werden 
und die jeweils bis zur Ordnung $\alpha^4$ 
mitgenommen werden. Es ist
\begin{equation}
H_{\rm FW} = U' \, (U \, H_D \, U^{-1}) \, U'^{-1} = (U' \, U) \, H_D \, (U' \, U)^{-1}.
\end{equation}
An der Gleichung \ref{HFW4} ersieht man im "ubrigen, wie 
sich die relativistischen Korrekturn zur Schr"odingerschen Theorie
(Spin-Bahn-Kopplung, Zitterbewegungsterm und Darwin-Term) 
aus der Diracschen Theorie ergeben. 

Was k"onnen wir "uber die transformierte Wellenfunktion $U \ket{\psi}$ aussagen? 
Die Funktion $\ket{\tilde \psi} = U' U \ket{\psi}$ ist bis zur Ordnung $\alpha^4$ 
Eigenzustand des FW-Hamiltonians. Der FW-Hamiltonian ist in dieser Ordnung 
eine gerade Matrix, die obere und untere Komponenten nicht mischt. 
Der Hauptanteil zum Eigenwert ist durch den Beitrag
\begin{displaymath}
\go m = \left( \begin{array}{cc} m & 0 \\ 0 & -m \end{array} \right)
\end{displaymath}
gegeben. $\ket{\tilde \psi}$ ist Eigenzustand zu $H_{\rm FW}$ mit einem
Eigenwert $\approx m$. Daher mu"s $\ket{\tilde \psi}$ bis zu einer dem FW-Hamiltonian
entsprechenden Ordnung frei von unteren Komponenten sein. 

Diese Ordnung entspricht in unserem Fall der f"uhrenden Ordnung der Dirac-Wellenfunktion 
und ihrer ersten relativistischen Korrektur. Wir 
k"onnen daher in den folgenden Rechnungen stets 
die unteren Komponenten 
des Spinors $\ket{\psi}$ vernachl"assigen und 
$\go \to 1$, $\vec{\Sigma} \to \vec{\sigma}$ setzen.

Auf $\ket{\tilde \psi}$ angewandt, ist der FW-Hamiltonian also 
gleichzusetzen mit der oberen Bispinor-Komponente dieses Operators, die 
wir mit $H_{\rm FW}^{\rm up}$ bezeichnen wollen.
\begin{equation}
\label{HFW}
H_{\rm FW}^{\rm up} = m + \left( \frac{\psq}{2 \, m} + V \right) +
  \left( - \frac{(\psq)^2}{8 \, m^3} + \frac{\pi \alpha}{2 m^2} 
  \delta(\vec{r})  + 
  \frac{\alpha}{4 \, m^2 \, r^3} \scp{\sigma}{L} \right) 
\end{equation}
Im folgenden wollen wir nur noch mit der Anwendung des FW-Hamiltonians 
auf die oberen Komponenten eines Spinors besch"aftigen und lassen daher 
das Superskript up wieder weg. Wenn wir uns an die Definition des 
Schr"odinger-Hamiltonians $H_S = \frac{\psq}{2 m} + V$ 
erinnern und als relativistische Korrektur dazu
\begin{equation}
\delta H = - \frac{(\psq)^2}{8 \, m^3} + \frac{\pi \alpha}{2 m^2} 
  \delta(\vec{r})  + 
  \frac{\alpha}{4 \, m^2 \, r^3} \scp{\sigma}{L} 
\end{equation}
definieren, so schreibt sich der FW-Hamiltonian sehr kompakt als
\begin{equation}
\label{HFWcomp}
H_{\rm FW} = m + H_S + \delta H.
\end{equation}
Die Ruhmasse $m$ kann aus dem obigen Hamiltonian wegtransformiert werden.
In nullter Ordnung ist die nichtrelativistische 
Schr"odinger-Wellenfunktion $\ket{\phi}$ Eigenzustand
des FW-Hamiltonians. 
Durch die relativistischen Korrekturen $\delta H$ kommt eine Korrektur 
$\ket{\delta \phi}$ hinzu. Die Normierung
\begin{equation}
\bracket{\phi + \delta \phi}{\phi + \delta \phi} \demand 1 \quad 
  \mbox{in 1. Ordnung}
\end{equation}
f"uhrt (wenn man $\bracket{\phi}{\delta \phi}$ als reell annimmt) auf
\begin{equation}
\bracket{\phi}{\delta \phi} \demand 0.
\end{equation}
Die Korrektur $\delta E$ zum Energieeigenwert ergibt sich aus
\begin{equation}
\delta E = \bra{\phi} \delta H \ket{\phi}.
\end{equation}
$\ket{\phi + \delta \phi}$ soll die transformierte Schr"odingergleichung
(mit relativistischen Korrekturen) approximativ l"osen, d. h.
\begin{equation}
(E_S + \delta E) \, \ket{\phi + \delta \phi} \demand 
(H_S + \delta H) \, \ket{\phi + \delta \phi}
\quad \mbox{in 1. Ordnung}
\end{equation}
Es ergibt sich die folgende Differentialgleichung f"ur $\ket{\delta \phi}$,
\begin{equation}
\label{dgldeltaphi}
(H - \bra{\phi} \delta H \ket{\phi}) \, \ket{\phi} = (E_S - H_S) \, \ket{\delta \phi},
\end{equation}
und nach weiterer Umformung
\begin{equation}
\label{deltaphi}
\ket{\delta \phi} = \frac{1}{(E_S - H_S)'}{\delta H \ket{\phi}}
\end{equation}
mit 
\begin{equation}
G_{\rm red}( E_S - H_S ) = \frac{1}{(E_S - H_S)'}
\end{equation}
als reduzierter Greensfunktion.
   
Das Ergebnis der oben aufgef"uhrten FW-Transformation
des Dirac-Hamiltonians ist bekannt. Es soll nun die FW-Transformation der Ausdr"ucke 
\begin{equation}
y^j = \alpha^j \exp(i \scp{k}{r})
\end{equation}
berechnet werden (diese Ausdr"ucke sollen mit derselben unit"aren
Transformation transformiert werden wie der Hamiltonian). Wir erhalten
\begin{equation}
y_{\rm FW}^j = U' \, (U y^j U^{-1}) \, U'^{-1} = 
  (U' \, U) \, y^j \, (U' \, U)^{-1},
\end{equation}
wobei $U$ und $U'$ diejenigen unit"aren Transformationen 
sind, mit denen auch der FW-Hamiltonian
transformiert wurde. Da $\vec{k}$ eine $\alpha^2$-Potenz 
tr"agt und $\vec{r}$ von der Ordnung 
$\alpha^{-1}$ ist, k"onnen wir $y^j$ in Potenzen von $\alpha$ entwickeln.
\eqnanf
y_0^j & = & \alpha^j, \schwups
y_1^j & = & \alpha^j (i \scp{k}{r}), \schwups
y_2^j & = & \alpha^j (- \frac{1}{2} (\scp{k}{r}))^2. \glok
\eqnend
Wir nehmen jeweils die um $\alpha^2$ h"ohere Korrektur zum 
$y_0^j$-Term noch mit. Nach einer l"angeren Rechnung, die hier nicht vorgef"uhrt werden
soll, erh"alt man als Ergebnis f"ur die FW-Transformation von $y_0^j$:
\eqnanf
y_{0,FW}^j &=& \alpha^j + \go \left( \frac{p^j}{m} - \frac{1}{2 m^3} p^j \psq \right)
  \schwups
& & - \frac{1}{2 m^2} p^j 
\left(\scp{\alpha}{p}\right) - \frac{1}{2 m^2} \alpha^j \com{\scp{\alpha}{p}}{V}
+ \frac{1}{2 m^2} \com{p^j}{V}.
\eqnend
F"ur $y_1^j$ erh"alt man nach der Transformation
\begin{equation}
y_{1,FW}^j = \alpha^j \left(i \scp{k}{r}\right) + \frac{i \go}{2 m} 
  \left( -\alpha^j \com{\scp{\alpha}{p}}{\scp{k}{r}}
- 2 i k^j + 2 (\scp{k}{r}) p^j \right) 
\end{equation}
und schlie"slich f"ur $y_2^j$
\eqnanf
y_{2,FW}^j &=& \alpha^j \left(
- \frac{1}{2} (\scp{k}{r})^2 \right) - \schwups
& & \frac{1}{4 m} 
\left( 2 \go ((\scp{k}{r})^2 p^j +
\com{p^j}{(\scp{k}{r})^2} - 
\go \alpha^j \com{\scp{\alpha}{p}}{(\scp{k}{r})^2)} \right). \ok
\eqnend
F"ur die Berechnung der Lamb--Verschiebung
k"onnen in den letzten drei Ergebnissen die ungeraden Operatoren
weggelassen werden, d. h. die zu den $\alpha$-Matrizen 
proportionalen Terme werden vernachl\"assigt. Man erh"alt nach weiteren Umformungen als 
Beitr"age der geraden Operatoren
\begin{equation}
y_{0,FW}^j = \go \left( \frac{p^j}{m} - \frac{1}{2 m^3} p^j \psq \right) - 
  \frac{1}{2 m^2} \frac{\alpha}{r^3} \left( \vec{r} \times \vec{\Sigma} \right)^j
\end{equation}
sowie
\begin{equation}
y_{1,FW}^j = - \frac{i}{2 m} \go \left( \vec{k} \times \vec{\Sigma} \right)^j + 
  \frac{i}{m} \go \left( \scp{k}{r} \right) p^j
\end{equation}
und
\begin{equation}
y_{2,FW}^j = - \frac{1}{2 m} \go \left(\scp{k}{r}\right)^2 p^j + 
  \frac{1}{2 m} \go \left(\scp{k}{r}\right) 
    \left( \vec{k} \times \vec{\Sigma} \right)^j.
\end{equation}
Da wir uns wie oben nur mit den oberen 
Komponenten der Spinoren besch"aftigen, 
k"onnen wir wie beim FW-Hamiltonian $\go \to 1$ und 
$\vec{\Sigma} \to \vec{\sigma}$ setzen (wir beschr"anken uns auf die
linke obere $2 \times 2$-Submatrix). Wir erhalten dann
\begin{equation}
\label{y0}
y_{0,FW}^j = \left( \frac{p^j}{m} - \frac{1}{2 m^3} p^j \psq \right) - 
\frac{1}{2 m^2} \frac{\alpha}{r^3} \left( \vec{r} \times \vec{\sigma} \right)^j 
\end{equation}
sowie
\begin{equation}
\label{y1}
y_{1,FW}^j = - \frac{i}{2 m} \left( \vec{k} \times \vec{\sigma} \right)^j + 
 \frac{i}{m} \left(\scp{k}{r}\right) p^j 
\end{equation}
und
\begin{equation}
\label{y2}
y_{2,FW}^j = - \frac{1}{2 m} \left(\scp{k}{r}\right)^2 p^j + 
\frac{1}{2 m} \left(\scp{k}{r}\right) 
\big( \vec{k} \times \vec{\sigma} \big)^j. 
\end{equation}
Unter Vernachl"assigung ungerader Operatoren und f"ur die linke
obere $2 \times 2$-Submatrix k"onnen wir daher 
\begin{equation}
\label{yjfw}
y_{\rm FW}^j = 
U' \, (U \alpha^j e^{i \scp{k}{r}} U^{-1}) \, U'^{-1} =
y_{0,FW}^j + y_{1,FW}^j + y_{2,FW}^j
\end{equation}
setzen mit
\begin{eqnarray}
\label{alphaitransformed}
y^j_{\rm FW} & = &
\frac{p^j}{m} \left(1 + i \left( {\vec k} \, \cdot {\vec r} \right) -
\frac{1}{2} \left( {\vec k} \, \cdot \, {\vec r} \right)^2 \right) \\
& & - \frac{1}{2 \, m^3} p^j \psq -
\frac{1}{2 \, m^2} \, \frac{\alpha}{r^3} \, 
  \left( {\vec r} \times \vec{\sigma} \right)^j \nonumber \\
& & + \frac{1}{2 \, m} \left( {\vec k} \, \cdot \, {\vec r} \right)
 \left( {\vec k} \times \vec{\sigma} \right)^j \nonumber 
- \frac{i}{2 \, m}
   \left( {\vec k} \times \vec{\sigma} \right)^j. \nonumber
\end{eqnarray}

%
%

\section{Der Photonen-Propagator in verschiedenen Eichungen}
\label{eichungen}

Der Photonen-Propagator ist definiert als L"osung der Gleichung
\begin{equation}
A_{\mu}(y) = \int d^4 x D_{\mu \nu}(y-x) j^{\nu}(x). 
\end{equation}
Im Impulsraum gilt daher
\begin{equation}
A_{\mu}(k) = D_{\mu \nu}(k) j^{\nu}(k).
\end{equation}
In Lorentz-Eichung lauten die Maxwell-Gleichungen gerade
\begin{equation}
\quabla A^{\mu}(x) = j^{\mu}(x),
\end{equation}
und im Impulsraum folgt daraus in Lorentz-Eichung
\begin{equation}
\quabla A_{\mu}(k) = -k^2 A_{\mu}(k) = j_{\mu}(x) 
  \demand -k^2 D_{\mu \nu}(k) j^{\nu}(k),
\end{equation}
so da"s sich 
\begin{equation}
D_{\mu \nu} = - \frac{g_{\mu \nu}}{k^2} \quad \mbox{in Lorentz- oder 
Feynman-Eichung}
\end{equation}
ergibt. Wir k"onnen zum Propagator beliebige Ausdr"ucke der
Form 
\mathanf
D^{\mu \nu} \to D_{\mu \nu} + \frac{1}{2 k^2} (f_\mu k_\nu + f_\nu k_\mu)
\mathend
addieren, ohne da"s sich an den beobachtbaren Gr"o"sen $F_{\mu \nu} =
\del_{\mu} A_{\nu} - \del_{\nu} A_{\mu}$
etwas "andert. Bei der Rechnung mu"s man nur die Kontinuit"at des
Stromes 
\mathanf
k^{\mu} j_{\mu}(k) = 0
\mathend
ausnutzen und bedenken, da"s im Impulsraum 
\mathanf
\del_{\mu} \to - i k_{\mu}
\mathend
zu ersetzen ist. In der Coulomb-Eichung 
\begin{equation}
\del_i A^i(x) = 0
\end{equation}
lauten die Maxwell-Gleichungen
\begin{equation}
\quabla A_{\mu}(x) - \del_{\mu} \del_0 A_0(x) = j_{\mu}(x)
\end{equation}
oder
\begin{equation}
-k^2 A_{\mu}(k) + k_{\mu} k_0 A_0(k) = j_0(k).
\end{equation}
Es ergibt sich nach einer kurzen Rechnung f"ur die Coulomb-Eichung 
\eqnanf
D_{00}(k) & = & \frac{1}{\vec{k}^2}, \schwups
D_{0i}(k) & = & D_{i0} = 0, \wups
D_{ij}(k) & = & \frac{1}{k^2} (\delta_{ij} - \frac{k_i k_j}{\vec{k}^2}). \ok
\eqnend
Die Transformation zwischen Feynman- und Coulomb-Eichung wird durch
die Funktionen
\begin{equation}
f_0 = \frac{k_0}{\vec{k}^2} \quad \mbox{sowie} \quad 
f_i = - \frac{k_i}{\vec{k}^2}
\end{equation}
vermittelt. 

%
%

\section{Die epsilon-Methode}

Der urspr"ungliche Ausdruck f"ur den Selbstenergiebeitrag sieht 
einfach aus, 
\begin{equation}
\delta E_{SE} = i e^2 \intk D_{\mu \nu} \bra{\bar{\psi}} \gamma^{\mu} 
  \dcprop \gamma^{\nu} \ket{\psi} - \bra{\bar{\psi}} \delta m \ket{\psi} \,.
\end{equation}
Bei einer vollkommen analytischen Rechnung entstehen 
jedoch sehr viele Terme. Dies erschwert die analytische 
Behandlung. Au"serdem ergeben sich sowohl infrarote als auch 
ultraviolette Divergenzen, so da"s eine Teilung des Integrationsbereichs 
sinnvoll erscheint.

Ein weiterer Grund f"ur die Einf"uhrung einer Trennung des 
Integrationsweges ist darin zu suchen, da"s im relativistischen Bereich 
($\omega \approx m$) der Dirac-Coulomb Propagator 
in Potenzen des Coulombfeldes $V$ entwickelt werden kann. Im 
nichtrelativistischen Bereich jedoch ($\omega \ll m$) tragen alle
$V$ Potenzen des Propagators bei.

Es wird daher in Bezug auf die Energie $\omega$ des virtuellen Photons eine 
Teilung des Integrationsweges vorgenommen. Man teilt die Rechnung in zwei Bereiche 
ein, die durch den 
Parameter $\ep \ll m$ voneinander getrennt sind und w"ahlt f"ur die
beiden Bereiche unterschiedliche Eichungen des Photonenpropagators.
Das Ziel ist, f"ur beide Bereiche eine effiziente Berechnungsmethode
zu w"ahlen und dadurch die analytische Integration zu erleichtern.

Im Ausdruck f"ur die Selbstenergie ist der Dirac-Coulomb Propagator 
\mathanf
S_F^V = \dcprop
\mathend
enthalten. Wegen 
\mathanf
\hat{E} \ket{\psi} = i \dbyd{\ket{\psi}}{t} = E_{\psi} \ket{\psi}
\mathend
($\psi$ ist Eigenzustand des Hamiltonians), k"onnen wir im Ausdruck 
f"ur die Selbstenergie den Dirac-Coulomb-Propagator durch
\mathanf
S_F^V \to \frac{1}{\go (E_{\psi} - \omega) - \scp{\gamma}{p} - m - \go V} = 
  S_F^V(E_{\psi} - \omega)
\mathend
ersetzen. Der Propagator $S_F^V(E_{\psi} - \omega)$  
hat, wie man an der Darstellung \ref{SFenergy}
\mathanf
S_F^V(E - \omega) = 
  \swsum{n} \frac{\ket{\psi_n} \bra{\bar{\psi}}}
    {(E_{\psi} - \omega) - E_n (1 - i \eta)}
\mathend
erkennt, Pole bei 
\mathanf
(E_{\psi} - \omega) - E_n (1 - i \eta) = 0 \quad \mbox{oder} \quad 
\omega = E_{\psi} - E_n (1 - i \eta). 
\mathend

Unter Vernachl"assigung des diskreten Spektrums, welches f"ur unsere Rechnung 
keine Rolle spielt, da die erzeugten Pole oberhalb der reellen Achse liegen,
gibt es Energieeigenwerte $E_n$ f"ur $E_n > m$ und $E_n < -m$. F"ur
unseren Fall ist au"serdem $E_{\psi} < m, E_{\psi} \approx m$. Daher gibt es 
Pole als Funktion von $\omega$ gerade bei
\mathanf
\Re(\omega) < 0, \Im(\omega) = i \eta \equiv E_n > 0
\mathend
sowie
\mathanf
\Re(\omega) > 2 m, \Im(\omega) = - i \eta \equiv E_n < 0.
\mathend
Da also im urspr"unglichen Elektronen-Propagator die Pole f"ur die positiven Eigenwerte
$E_n$ leicht unter die reelle Achse gedr"uckt werden und die Pole f"ur negative
Eigenwerte einen leicht positiven Imagin"arteil besitzen, werden die Pole
f"ur $\omega > 2 m$ unter die reelle Achse gedr"uckt, und die Pole f"ur 
$\omega < 0$ bekommen einen leicht positiven Imagin"arteil. 

Es ergeben sich auch Pole des Integranden aus dem Photonenpropagator.
Der Photonen-Propagator ist in Feynman-Eichung 
\mathanf
D_{\mu \nu}(k) = - \frac{g_{\mu \nu}}{k^2 + i \eta} = 
- \frac{g_{\mu \nu}}{\omega^2 - \vec{k}^2 + i \eta}.
\mathend
Aus der Umformung
\mathanf
\frac{1}{\omega^2 - \vec{k}^2 + i \eta} = 
\frac{1}{(\omega - \modk + i \eta)(\omega + \modk - i \eta)}
\mathend
erhalten wir f"ur die Positionen der Singularit"aten:
\mathanf
\omega = \modk - i \eta \quad \mbox{sowie} \quad \omega = - \modk + i \eta. 
\mathend
Diese Pole liegen daher f"ur $\Re(\omega) > 0$ leicht unterhalb
der reellen Achse und f"ur $\Re(\omega) < 0$ leicht dar"uber.

%
%

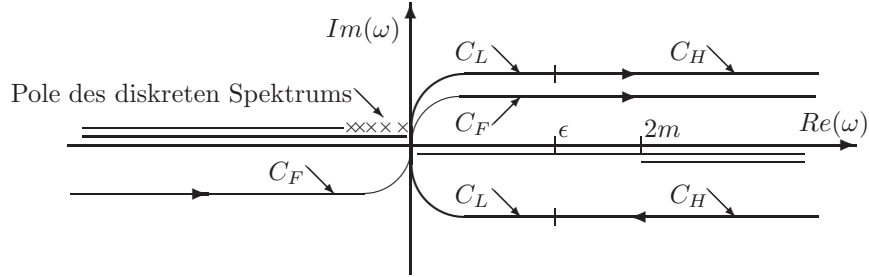
\begin{figure}[htb]
%
%

\begin{center}
\setlength{\unitlength}{0.015in}

\begin{picture}(280,100)(-140,-50)

\thicklines
\put (-120,0){\line(1,0){272}}
\put (-.4,-45){\line(0,1){92}}
\thicklines
\put (151,0){\vector(1,0){4}}
\put (-.4,46) {\vector(0,1){4}}
\thinlines
\put (2,-3){\line(1,0){135}}
\put (-2,3){\line(-1,0){113}}
\put (80,-6){\line(1,0){57}}
\put (-24,6){\line(-1,0){91}}
\put (17,0){\oval(34,34)[tl]}
\put (17,17){\line(1,0){124}}
\put (-17,0){\oval(34,34)[br]}
\put (-17,-17){\line(-1,0){102}}
\thicklines
\put (25,0){\oval(50,50)[l]}
\put (25,25){\line(1,0){117}}
\put (25,-25){\line(1,0){117}}
\put (75,25){\vector(1,0){4}}
\put (80,-25){\vector(-1,0){4}}
\put (75,17){\vector(1,0){4}}
\put (-75,-17){\vector(1,0){4}}
\thinlines
\put (50,-3){\line(0,1){6}}
\put (50,-28){\line(0,1){6}}
\put (50,22){\line(0,1){6}}
\put (80,-3){\line(0,1){6}}
\put (-6,4.7){${\scriptstyle{\times}}$}
\put (-12,4.7){${\scriptstyle{\times}}$}
\put (-17,4.7){${\scriptstyle{\times}}$}
\put (-21,4.7){${\scriptstyle{\times}}$}
\put (-24,4.7){${\scriptstyle{\times}}$}
\put (135,5){$Re(\omega)$}
\put (-30,38){$Im(\omega)$}
\put (15,5){$C_F$}
\put (28,7){\vector(1,1){10}}
\put (-50,-12){$C_F$}
\put (-37,-7){\vector(1,-1){10}}
\put (15,30){$C_L$}
\put (28,35){\vector(1,-1){10}}
\put (15,-20){$C_L$}
\put (28,-15){\vector(1,-1){10}}
\put (90,30){$C_H$}
\put (103,35){\vector(1,-1){10}}
\put (90,-20){$C_H$}
\put (103,-15){\vector(1,-1){10}}
\put (-140,15){$\mbox{Pole des diskreten Spektrums}$}
\put (-20,20){\vector(1,-1){10}}
\put (51,3){$\ep$}
\put (81,3){$2m$}

\end{picture}
\end{center}
\caption{\label{intweg} Integrationswege f"ur die $\omega$-Integration}
\end{figure}

Die $\omega$-Integration erfolgt unter Ber"ucksichtigung der Lage 
der Polstellen entlang des Feynman-Weges $C_F$ (Abb. \ref{intweg}). 
Das Matrixelement ${\tilde P}$ im Integranden geht f"ur gro"se $k$ 
wie $k^{-1}$, der regularisierte Propagator wie $k^{-4}$ und 
$d^4 k$ geht wie $k^3$ (zusammen $k^{-2}$). Daher k"onnen wir
$C_F$ durch einen unendlich gro"sen Halbkreis in der unteren 
komplexen Halbebene schlie"sen (L"ange dieses Halbkreises wie
$k^1$, daher geht das Halbkreis-Integral f"ur gro"se $k$ wie
$k^{-1}$).   

Man verbiegt nun die Integrationskontour $C_F$ zu der modifizierten Kontour
$C_L + C_H$ (dabei werden keine Residuen ausgelassen). Die Integrationskontour 
$C = C_L + C_H$ wird wie in der Abbildung \ref{intweg} angegeben in der vorliegenden 
Arbeit verwandt. Da $C$ infinitesimal "uber bzw. unterhalb der reellen
Achse verl"auft, k"onnen wir sicherlich $C_L$ durch Hinzunahme eines 
Verbindungsweges bei $\Re(\omega) = \ep$ schlie"sen.

Wir definieren als Hochenergie- und Niedrigenergie-Anteile
\mathanf
E_H = i e^2 \int_{C_H} \frac{d^4 k}{(2 \pi)^4} D_{\mu \nu} 
  \bra{\bar{\psi}} \gamma^{\mu} \dcprop \gamma^{\nu} \ket{\psi} 
  - \bra{\bar{\psi}} \delta m \ket{\psi}
\mathend
und
\mathanf
E_L = i e^2 \int_{C_L} \frac{d^4 k}{(2 \pi)^4} D_{\mu \nu} 
  \bra{\bar{\psi}} \gamma^{\mu} \dcprop \gamma^{\nu} \ket{\psi}.
\mathend

Regularisierung und Renormierung sind nur f"ur den Hochenergie-Anteil n"otig.
Die Summe der beiden Anteile ist der gesuchte Selbstenergie-Beitrag
zur Lamb-Shift,
\mathanf
\delta E_{SE} = E_L + E_H\,.
\mathend
$\delta E_{SE}$ h"angt nat"urlich nicht von der Wahl des (kleinen) 
Trennungsparameters $\ep$ ab,
da durch diesen Parameter nur der Integrationsbereich getrennt wird.
Wir haben es also mit zwei Funktionen $E_L$ und $E_H$ zu tun, die
in Abh"angigkeit von der Feinstrukturkonstanten $\alpha$ und dem
Parameter $\ep$ betrachtet werden, und deren Summe nicht von $\ep$
abh"angt:
\mathanf
\delta E_{SE}(\alpha) = E_L(\alpha,\ep) + E_H(\alpha,\ep)
\mathend
Die Entwicklungskoeffizienten von $\delta E_{SE}$ bez"uglich $\alpha$ h"angen 
nicht von $\ep$ ab. Der Entwicklungskoeffizient von $\delta E$ bez"uglich $\alpha$
ist aber gerade die Summe der entsprechenden Entwicklungskoeffizienten 
von $E_L$ und $E_H$. Wir entwickeln daher $ E_L$ und $E_H$ 
zuerst in $\alpha$. Dann wird jeder Entwicklungskoeffizient 
von $E_L$ und $ E_H$ 
bez"uglich $\alpha$ nachtr"aglich in $\ep$ entwickelt. Die Summe dieser
Entwicklung(en) bez"uglich $\ep$ darf aber nicht von $\ep$ abh"angen,
es m"ussen sich alle $\ep$-Koeffizienten (bis auf den von $\ep^0$)
herausheben. Die Koeffizienten von $\ep^0$ tragen daher zur 
Selbstenergie bei, h"ohere und niedrigere (divergente) Koeffizienten heben 
sich gerade heraus. Daher gen"ugt es, die Entwicklungskoeffizienten
von $E_L$ und $E_H$ bez"uglich $\alpha$ nachtr"aglich 
bis zur Ordnung $\ep^0$ zu 
entwickeln. Terme, die f"ur $\ep \to 0$ logarithmisch oder st"arker
divergieren, heben sich gerade heraus.

Dies ist auch ein wichtiger Test f"ur die Rechnung: Denn jeder Rechenfehler bei
$E_L$ und $E_H$
wirkt sich mit hoher Wahrscheinlichkeit auf den Koeffizienten
der $ln(\ep)$ oder $1/\ep$-Divergenz aus. Das Herausheben der Divergenz 
(in der Summe von $E_L$ und $E_H$) ist somit ein 
wichtiger Test der Rechnung.

Es ist f"ur die G"ultigkeit der Methode wichtig, da"s weder $E_L$ noch 
$E_H$ von der Eichung abh"angen. Dazu reicht es aus, die 
Eichinvarianz von $E_L$ zu zeigen. Da $\delta E_{SE} = E_L + E_H$ nicht 
von der Eichung abh"angt, ergibt sich daraus automatisch die Eichinvarianz
von $E_H$. Es ist also zu zeigen, da"s durch Modifikationen am Propagator
der Form (siehe Abschnitt \ref{eichungen})
\mathanf
D_{\mu \nu} \to D_{\mu \nu} + \frac{1}{2 k^2} (f_\mu k_\nu + f_\nu k_\mu)
\mathend
keine Ver"anderung des $\ep^0$ Koeffizienten von $E_L$ eintritt. Unter 
Benutzung der Beziehung
\eqnanf
\diracprop \slf{k} \ket{\psi} & = &
\left( \frac{\slf{k} - \slf{p} + m + \go V}{\slf{p} - \slf{k} - m - \go V} +
   \frac{\slf{p} - m - \go V}{\slf{p} - \slf{k} - m - \go V} \right) \ket{\psi} \wups
& = & (-1 + 0) \, \ket{\psi} = - \ket{\psi} \glok
\eqnend  
k"onnen wir den eichabh"angigen Anteil $E_L^{\rm gauge}$ umformen:
\eqnanf
E_L^{\rm gauge} & = & - i e^2 \intclk \frac{1}{2 k^2} (f_\mu k_\nu + 
f_\nu k_\mu) \times \schwups
 & & \times \Pmunu \wups
& = & - i e^2 \intclk \frac{1}{k^2} \bra{\bar{\psi}} \slf{f} 
\diracprop \slf{k} \ket{\psi} \wups
& = & + i e^2 \intclk \frac{1}{k^2} \bra{\bar{\psi}} \slf{f} 
\ket{\psi}. \ok
\eqnend
Dieser Ausdruck verschwindet f"ur
$\ep \to 0$ und tr"agt somit nicht zur Lamb-Shift bei. Daher 
ist der entscheidende $\ep^0$ Koeffizient in $E_L$ eichinvariant.

Als Eichung w"ahlen wir f"ur den Hochenergie-Anteil den gem"a"s
Pauli-Villars regularisierten Propagator in Feynman-Eichung
(siehe Abschnitt \ref{eichungen})
\mathanf
D_{\mu \nu} = - g_{\mu \nu} \left(\frac{1}{k^2} - \frac{1}{k^2 - M^2}
  \right),
\mathend
f"ur den Niedrigenergie-Anteil jedoch die Coulomb-Eichung (ebenfalls
siehe \ref{eichungen}).

%
%

\section{Der Hochenergie-Anteil}
\label{hepmethod}

Wir schreiben den Hochenergie-Anteil $E_H$ mit der expliziten Form des
regularisierten Photonen-Propagators (in Feynman-Eichung) und mit der 
Massenrenormierung $\delta m$ noch einmal auf,
\eqnanf
\label{EH}
E_H &=& - i e^2 \int_{C_H} \frac{d^4 k}{(2 \pi)^4} 
  \left[ \frac{1}{k^2} - \frac{1}{k^2- M^2} \right] 
  \bra{\bar{\psi}} \gamma^{\mu} \dcprop \gamma_{\mu} \ket{\psi} \wups
 & & - \bra{\bar{\psi}} \delta m \ket{\psi}. \glok
\eqnend
Um diesen Ausdruck bis zur Ordnung $\alpha (Z \alpha)^6$ zu berechnen, ist
es n"otig, das Matrixelement
\begin{equation}
{\tilde P} = \bra{\bar{\psi}} \gamma^{\mu} \dcprop \gamma_{\mu} \ket{\psi} 
\end{equation}
bis Ordnung $\alpha^6$ auszuwerten. Wir wollen f"ur die sp"ateren 
Rechnungen $\bra{\bar{\psi}}$ explizit als $\bra{\psi} \go$ schreiben. 
Damit ist
\begin{equation}
\label{tildeP}
{\tilde P} = \bra{ \psi } \go \, \gamma^{\mu} \, \dcprop \, \gamma_{\mu} \, \ket{\psi}. 
\end{equation}
Wie bereits erw"ahnt, ist die 
Integrationskontour $C_H$ nach 
unten durch $\ep$ begrenzt (Infrarot-Regulator). 
Daher ist es m"oglich und sinnvoll, den 
Dirac-Coulomb-Propagator in Potenzen des Potentials $V$ zu entwickeln. 
F"ur gebundene Zust"ande im Coulombpotential ist
$V = - \frac{\alpha}{r}$ von der Ordnung $\alpha^2$ (siehe Abschnitt 
\ref{defs}). Wir m"ussen daher das Matrixelement $\tilde P$ bis zur
Ordnung $V^3$ entwickeln.

Um diese Entwicklung besonders "ubersichtlich ausf"uhren zu k"onnen,
vereinbaren wir f"ur den Nenner im Feynman-Propagator die Abk"urzung 
(siehe Abschnitt \ref{vertexkorrektur})
\begin{equation}
D = \slf{p} - \slf{k} - m.
\end{equation}
Die Entwicklung des Dirac-Coulomb-Propagators geschieht mit Hilfe 
der Formel
\mathanf
\frac{1}{X - Y} = \frac{1}{X} + \frac{1}{X} Y \frac{1}{X} + 
  \frac{1}{X} Y \frac{1}{X} Y \frac{1}{X} + \dots
\mathend
Wir setzen $X = \slf{p} - \slf{k} - m = D$ und $Y = \go V$ und erhalten 
als Resultat der
Entwicklung 4 Terme, den 0-Vertex, 1-Vertex, 2-Vertex und 3-Vertex-Anteil.
\eqnanf
\frac{1}{\slf{p} - \slf{k} - \go V - m} & = & \frac{1}{D - \go V} \schwups
& = & \frac{1}{D} + \wups 
& & \frac{1}{D} \go V \frac{1}{D} + \wups
& & \frac{1}{D} \go V \frac{1}{D} \go V \frac{1}{D} + \wups
& & \frac{1}{D} \go V \frac{1}{D} \go V \frac{1}{D} \go V \frac{1}{D} + 
O(V^4) \ok
\eqnend
H"ohere Terme tragen nicht bei, da $V^3$ bereits von der Ordnung $\alpha^6$
ist und wir nur bis zu dieser Ordnung rechnen wollen.
Wir definieren dann als Matrizen
\eqnanf
M_0 &=& \go \gamma^{\mu} \frac{1}{D} \gamma_{\mu} \schwups
M_1 &=& \go \gamma^{\mu} \frac{1}{D} \go V \frac{1}{D} \gamma_{\mu} \schwups 
M_2 &=& \go \gamma^{\mu} \frac{1}{D} \go V \frac{1}{D} \go V
  \frac{1}{D} \gamma_{\mu} \schwups
M_3 &=& \go \gamma^{\mu} \frac{1}{D} \go V
   \frac{1}{D} \go V \frac{1}{D} \go V 
   \frac{1}{D} \gamma_{\mu}. \glok
\eqnend
Wir k"onnen somit das Matrixelement ${\tilde P}$ entwickeln als
\begin{equation}
\label{Psum}
{\tilde P} = {\tilde P}_0 + {\tilde P}_1 + {\tilde P}_2 + {\tilde P}_3
\end{equation}
mit     
\eqnanf
{\tilde P}_0 &=& 
\bra{\psi} \go \gamma^{\mu} \frac{1}{D} \gamma_{\mu} \ket{\psi} =  
  \bra{\psi} M_0 \ket{\psi}  \schwups
{\tilde P}_1 &=& \bra{ \psi} \go \gamma^{\mu} \frac{1}{D} \go V 
\frac{1}{D} 
  \gamma_{\mu} \ket{\psi} = \bra{\psi} M_1 \ket{\psi}  \schwups
{\tilde P}_2 &=& \bra{ \psi} \go  \gamma^{\mu} \frac{1}{D} \go V 
  \frac{1}{D} \go V \frac{1}{D}  \gamma_{\mu} \ket{\psi} = 
  \bra{\psi} M_2 \ket{\psi} \schwups
{\tilde P}_3 &=& \bra{ \psi} \go \gamma^{\mu} \frac{1}{D} \go V
   \frac{1}{D} \go V \frac{1}{D} \go V 
  \frac{1}{D}  \gamma_{\mu}  \ket{\psi} = \bra{\psi} M_3 \ket{\psi}. 
\glok
\eqnend 
Diese Matrixelemente (0-Vertex, 1-Vertex, 2-Vertex und 3-Vertex-Anteile) 
sind auszuwerten. Diese Auswertung erfolgt, etwas vereinfacht beschrieben, 
in den folgenden Schritten:

\begin{enumerate}

%
%

\item Entwicklung der $M_i$-Matrizen in die 16 $\Gamma$-Matrizen.
Gem"a"s der Gl. \ref{expansionGamma} gilt:
\begin{equation}
M_i = \sum_{\beta} a_{\beta} \Gamma^{\beta}
\end{equation}
mit
\mathanf 
a_{\beta} = 
  \frac{1}{4} \, {\rm Tr}(\Gamma_{\beta} M_i).
\mathend  
Die Entwicklung in die $\Gamma$-Matrizen 
macht die Rechnung "ubersichtlich und systematisiert in gewisser Weise 
das Vorgehen, was sich besonders beim Einsatz des Computers positiv 
bemerkbar macht.
%
%
\item Entwicklung der $a_{\beta}$-Koeffizienten in
Potenzen der Feinstrukturkonstanten $\alpha$. Die Entwicklungskoeffizienten
$a_{\beta}$ sind rationale Ausdr"ucke in den Impulsen des Elektrons, im Viererimpuls
des virtuellen Photons und im Potential $V$ und k"onnen daher in Potenzen
von $\alpha$ entwickelt werden (da jeder der Operatoren eine bestimmte
Potenz von $\alpha$ tr"agt). Es gilt:
\begin{equation}
a_{\beta} = \sum_{j=0}^{6} \alpha^j \, b_j(a_{\beta}).
\end{equation}
Es wird also jeder $a_{\beta}$-Koeffizient
bis $\alpha^6$ entwickelt. 

\item Mittelung "uber die r"aumlichen Anteile des Photonenimpulses.
"Uber die r"aumlichen Komponenten in den $b_j(a_{\beta})$-Koeffizienten
von $k$ kann gemittelt werden, da sich die Integration "uber den
gesamten $R^3$ erstreckt. 
\begin{equation}
b_j(a_{\beta}) = f(\vec{k}, {\rm andere \,\, Arg.}) \to
{\bar f}(\modk, {\rm andere \,\, Arg.}) = {\bar b}_j(a_{\beta}).
\end{equation}
Dabei entsteht der Ausdruck ${\bar b}(a_{\beta})$, welcher nur noch
von $\modk$, $\omega$ und nat"urlich von $\vec{p}$ und $V$ abh"angt. 
%
%
\item Ersetzung der Matrixelemente. Es werden nun (f"ur jede 
$\Gamma^{\beta}$-Matrix einzeln) die ${\bar b}_j(a_{\beta})$-Koeffizienten
durch die entsprechenden Matrixelemente mit $\ket{\psi}$ ersetzt, gem"a"s
\begin{equation}
{\bar b}_j(a_{\beta}) \to 
  \bra{\psi} \Gamma^{\beta'} {\bar b}_j(a_{\beta'}) \ket{\psi},
\end{equation}   
(keine Summation "uber $\beta$, daher $\beta'$, siehe Konventionen). 

Es entstehen aus den vorangegangenen
Schritten Beitr"age wie zum Beispiel (f"ur die Matrix $M_1$, 
Entwicklungskoeffizient von $\go$):
\mathanf
{\bar b}_4(a_{\go}) =  \left(\vec{p} \,\, V \vec{p} \right) f(\omega,\modk) +
  {\rm weitere Terme},
\mathend
wobei $f(\omega,\modk)$ eine rationale Funktion von $\omega$ und $\modk$
ist. Zwecks Berechnung von ${\tilde P}$ wird dann gem\"a\ss{}
\mathanf
\left( \vec{p} \, V \, \vec{p} \right) \, f(\omega,\modk) \to 
\bra{\psi} \left( \vec{p} \, V \, \vec{p} \right) \ket{\psi} \, 
f(\omega,\modk)
\mathend
ersetzt. 
Die Auswertung der Matrixelemente wird von in $Mathematica$ geschriebenen
symbolischen Programmen erledigt (vom Autor entwickelt). Bis zur Ordnung 
$\alpha^6$ gilt in unserem Fall zum Beispiel 
\mathanf
\bra{\psi} \left( \vec{p} \, V \, \vec{p} \right) \ket{\psi} = 
  -\frac{5}{48} \, \alpha^4 \, m^3  
    - \frac{283}{1152} \, \alpha^6 \, m^3\,.
\mathend
Die Auswertung der Matrixelemente ist auch von Hand mit Hilfe 
von Kommutatorbeziehungen unter intensiver Ausnutzung der Algebra der 
Dirac-Gleichung mit Coulombfeld m"oglich. Dieser Weg  
wurde zur Verifikation der Ergebnisse ebenfalls beschritten.
%
%
\item Zusammenfassung des Ergebnisses. F"ur alle $\Gamma^{\alpha}$'s, 
deren Koeffizienten nicht Null sind, wird das oben beschriebene 
Verfahren angewandt. Das Ergebnis
\mathanf
{\tilde P}_i = \sum_{\beta = 0}^{15} \sum_{j=0}^{6} \alpha^j 
  \bra{\psi} \Gamma^{\beta} {\bar b}_j(a_{\beta}) \ket{\psi}
\mathend
wird aufaddiert, vereinfacht und in Potenzen
der Feinstrukturkonstante $\alpha$ geordnet.

\end{enumerate}

Den letzten Schritt bei der Auswertung des Hochenergie-Anteils
bilden die Integrationen bez"uglich $\modk$ und $\omega$.
Die $k$-Integration kann mit Residuenintegration ausgef"uhrt werden.
 
F"ur die $\omega$-Integration werden ultraviolett divergente Ausdr"ucke 
regularisiert, es werden Feynman-Parameter eingef"uhrt und die Integration
kovariant ausgef"uhrt.

F"ur die ultraviolett konvergenten Ausdr"ucke 
wird zun"achst die Integrationsvariable reskaliert und auf eine 
dimensionslose Gr"o"se zur"uckgef"uhrt. Bei der
$\omega$-Integration beschr"ankt man sich auf den oberen
Ast des $C_H$-Weges (das Gesamtergebnis ist gerade
das Integral "uber den oberen Ast, plus das komplex Konjugierte dieses
Integrals). 

%
%

\section{Der Niedrigenergie-Anteil}
\label{lepmetho}

Bei der Behandlung des Niedrigenergie-Anteils wird
die $\omega$-Integration als erstes ausgef"uhrt. Wir gehen 
aus vom allgemeinen Ausdruck f"ur den Niedrigenergieanteil 
zur Selbstenergie
\mathanf
 E_L = i e^2 \int_{C_L} \frac{d^4 k}{(2 \pi)^4} D_{\mu \nu} 
  \bra{\bar{\psi}} \gamma^{\mu} \dcprop \gamma^{\nu} \ket{\psi}.
\mathend
Zun"achst formen wir das Matrixelement mit dem
Propagator in einer f"ur den Niedrigenergieanteil
g"unstigen Weise um. Unter Ausnutzung der Beziehung
\begin{equation}
\label{einsdurch}
A \frac{1}{B} C = \frac{1}{C^{-1} B A^{-1}},
\end{equation}
und der Identit"at
\mathanf
e^{i \vec{k} \cdot \vec{r}} p^i e^{- i \vec{k} \cdot \vec{r}} = p^i + 
[i \vec{k} \cdot \vec{r}, p^i] = p^i - k^i,
\mathend
die man mit der Campbell-Baker-Hausdorff-Formel oder durch explizites 
Differenzieren nachweist, ergibt sich mit der Definition $\alpha^0 = \go 
\go = I$ (diese Definition erfolgt hier rein aus Zweckm"a"sigkeit und ist 
nicht ``w"ortlich'', im Sinne eines Lorentz-Index, zu verstehen),
\eqnanf
\label{gammatoalpha}
{\tilde P}_{\mu \nu} & = & 
\bra{\bar{\psi}} \gamma^{\mu} \dcprop \gamma^{\nu} \ket{\psi} \schwups
& = & \bra{\bar{\psi}} \gamma^{\mu} 
\frac{1}{e^{i \scp{k}{r}}(\slf{p} - \go \omega - \go V - m) e^{-i 
\scp{k}{r}}} 
\gamma^{\nu} \ket{\psi} \wups
& = & \bra{\psi^{+}} \go \gamma^{\mu} e^{i \scp{k}{r}} 
\frac{1}{\go ((E_{\psi} - \omega) - \scp{\gamma}{p} - \go V - m)}  
\gamma^{\nu} e^{-i \scp{k}{r}} \ket{\psi} \wups 
& = & \bra{\psi^{+}} \alpha^{\mu} e^{i \scp{k}{r}} 
\frac{1}{(E_{\psi} - \omega) - \scp{\alpha}{p} - V - \go m}  
\alpha^{\nu} e^{-i \scp{k}{r}} \ket{\psi} \wups
& = & \bra{\psi^{+}} \alpha^{\mu} e^{i \scp{k}{r}} 
\frac{1}{(E_{\psi} - \omega) - H_D}  
\alpha^{\nu} e^{-i \scp{k}{r}} \ket{\psi}. \ok
\eqnend
Wir verwenden f"ur den Niedrigenergieanteil die Coulomb-Eichung. 
In dieser Eichung ist wie bereits in Abschnitt \ref{eichungen} 
erw"ahnt

\eqnanf
D_{00} & = & \frac{1}{\vec{k}^2}, \wups
D_{0i} & = & D_{i0} = 0, \wups
D_{ij} & = & \frac{1}{k^2} 
  \left( \delta_{ij} - \frac{k_i k_j}{\vec{k}^2} \right) 
= \frac{1}{k^2} \delta^T_{ij}.  \ok
\eqnend
Zun"achst wollen wir denjenigen Anteil behandeln, der vom $D_{00}$-Element
herr"uhrt. Es ist 
\eqnanf
\delta E_{L,00} &=& i e^2 \int_{C_L} \frac{d^4 k}{(2 \pi)^4} 
  \, D_{00}(k) \, \bra{\bar{\psi}} \go \, \dcprop \, \go \ket{\psi} \schwups
& = & i e^2 \int_{C_L} \frac{d^4 k}{(2 \pi)^4} 
  \, \frac{1}{\vec{k}^2} \, \bra{\psi^{+}} e^{i \scp{k}{r}} \, 
\frac{1}{(E_{\psi} - \omega) - H_D}  
   \, e^{-i \scp{k}{r}} \ket{\psi}. \ok
\eqnend
Mit 
\begin{equation}
\bra{\vec{r}_1} \, \frac{1}{(E - \omega) - H_D} \,
\ket{\vec{r}_2} =  
\swsum{n} \frac{\psi_n(\vec{r_1}) \,
\psi_n^{+}(\vec{r_2})}{(E - \omega) - E_n (1 - i \eta)}
\end{equation}
k"onnen wir den von $D_{00}$ hervorgerufenen Teilbeitrag zur 
Selbstenergie wie folgt hinschreiben:
\mathanf
\delta E_{L,00} = 
i e^2 \int \frac{d^3 k}{(2 \pi)^3} \int_{C_L} \frac{d \omega}{2 \pi} 
  \, \frac{1}{\vec{k}^2} \, \bra{\psi} \, \frac{\ket{\psi_n}
\bra{\psi_n}}{(E_{\psi} - \omega) - E_n (1 - i \eta)} \, \ket{\psi}.
\mathend
Da aufgrund der speziellen Wahl des $C$-Integrationsweges nur diejenigen Pole
des  Dirac-Coulomb-Propagators mit $E_n < 0$ entscheidend sind (der Pol mit 
$E_n = E_{\psi} > 0 $ tr"agt somit nicht bei), ist der Integrand innerhalb
des $C_L$-Weges analytisch. Damit ergeben sich aber bei der 
$\omega$-Integration auf $C_L$ (mit $\omega < \ep$) keine Residuen, die
zur Lamb-Shift beitragen k"onnten. Geht man mit $\ep \to 0$, so verschwindet
der von  der von $\delta E_{L,00}$ erzeugte Beitrag. 

Wir schreiben daher $\delta E_L$ folgenderma"sen unter Vernachl"assigung
von $D_{00}$ auf:
\mathanf
E_L = - i e^2 \int \frac{d^3 k}{(2 \pi)^3} 
\int_{C_L} \, \frac{d \omega}{2 \pi} \, \frac{1}{\omega^2 - \vec{k}^2} \,
\delta^{T, ij} \,
{\tilde P}^{ij}(\omega).
\mathend
Dabei ergibt sich ${\tilde P}^{ij}(\omega)$ unter Ausnutzung von 
\ref{gammatoalpha} wie folgt:
\begin{equation}
{\tilde P}^{ij}(\omega) = \bra{\psi} \alpha^i \, e^{i \scp{k}{r}} \,
\frac{1}{H_D - (E_{\psi} - \omega)} \,
\alpha^j \, e^{-i \scp{k}{r}} \, \ket{\psi}.
\end{equation}
${\tilde P}^{ij}(\omega)$ stellt eine Funktion dar, die innerhalb der
$C_L$-Kontour analytisch ist. Der $E - \omega$-Ausdruck 
im Nenner des Elektronpropagators steuert kein
Residuum bei, da $\omega < \ep$ nach oben beschr"ankt ist (mit
$\omega \ll m \approx E$). Daher k"onnen wir die $\omega$-Integration 
mit dem Residuensatz ausf"uhren. Es ist
\mathanf
\frac{1}{\omega^2 - \vec{k}^2 + i \eta} = 
\frac{1}{(\omega - \modk + i \eta)(\omega + \modk - i \eta)}.
\mathend
Der Pol bei $\omega = \modk - i \eta$ tr"agt zum Integral bei, das Residuum
ist
\eqnanf
\delta E_L & = & -i e^2 \int_{\modk < \ep} \frac{d^3 k}{(2 \pi)^3}
\, \frac{1}{2 \pi} \, (- 2 \pi i) \, \frac{1}{2 \modk} \, \delta^{T, ij} \,
{\tilde P}^{ij}(\modk) \schwups
& = & -e^2 \int_{\modk < \ep} \frac{d^3 k}{(2 \pi)^3}
\, \frac{1}{2 \modk} \, \delta^{T, ij} \, 
{\tilde P}^{ij}(\modk) \wups
& = & -e^2 \int_{\modk < \ep} \frac{d^3 k}{(2 \pi)^3 2 \modk} 
\, \delta^{T,ij} \,
\bra{\psi^{+}} \, \alpha^i e^{i \scp{k}{r}} \,
\frac{1}{H_D - (E_{\psi} - \omega)} \,
\alpha^j e^{-i \scp{k}{r}} \, \ket{\psi} \ok
\eqnend
mit
\begin{equation}
\omega \equiv \modk.
\end{equation}
F"ur den Niedrigenergie-Anteil wird
daher $\omega$ mit $\modk$ identifiziert.

Jetzt wird im relevanten Matrixelement ${\tilde P}^{ij}$ eine unit"are 
Foldy-Wouthuysen-Transformation (eigentlich zwei 
hintereinandergeschaltete unit"are Transformationen $U$ und $U'$, siehe Abschnitt 
\ref{FWtrafo}) eingef"ugt. Die Produkt-Transformation 
$U_{\rm FW} = U \, U'$ wird hier 
der Einfachheit halber wieder mit $U$ bezeichnet. Es gilt
\eqnanf
{\tilde P}^{ij} & = & \bra{\psi^{+}} \, \alpha^i e^{i \scp{k}{r}} \,
\frac{1}{H_D - (E - \omega)}  
\, \alpha^j e^{-i \scp{k}{r}} \, \ket{\psi} \schwups
& = & \bra{\psi^{+}} \Up \, U \, \alpha^i \, e^{i \scp{k}{r}} 
\, \Up \, U \, \frac{1}{H_D - (E - \omega)} \, \Up \, U \,   
\alpha^j \, e^{-i \scp{k}{r}} \, \Up \, U \ket{\psi} \wups
& = &  \bra{U \psi^{+}}  \left( U \alpha^i e^{i \scp{k}{r}} 
\Up \right) \frac{1}{U \left(H_D - (E - \omega)\right) \Up} \left( U   
\alpha^j e^{-i \scp{k}{r}} \Up \right) \ket{U \psi} \wups
& = &  \bra{\tilde \psi^{+}}  \left(U \alpha^i e^{i \scp{k}{r}} 
\Up\right) \, \frac{1}{U \left(H_D - (E - \omega)\right) \Up} ) 
\, \left( U \alpha^j e^{-i \scp{k}{r}} \Up \right) \ket{\tilde \psi}\,,\ok
\eqnend
mit $\ket{\tilde \psi} = U \ket{\psi}$ (bei dieser Rechnung wurde wieder 
\ref{einsdurch} benutzt).

Die Foldy-Wouthuysen-Transformation $U$ hat zum Ziel, die ungeraden Anteile
des Hamiltonians bis zur Ordnung $\alpha^4$, d.h. bis zur ersten
relativistischen Korrektur zum Schr"odinger-Hamiltonian, zum Verschwinden 
zu bringen. Wir wollen noch einmal das Ergebnis der FW-Transformation aus 
dem Abschnitt \ref{FWtrafo} (Gl. \ref{HFW}) in Erinnerung rufen:
\begin{equation}
H_{FW} = U \, H_D \, \Up =
  U \left(\scp{\alpha}{p} + \beta m + V\right) U^{-1} = m + H_S + \delta H
\end{equation}
$H_{FW}$ soll den Foldy-Wouthuysen-(FW)-Hamiltonian bezeichnen. Hierbei
ist:
\mathanf
H_S = \frac{\psq}{2 \, m} + V
\mathend
und 
\mathanf
\delta H = - \frac{(\psq)^2}{8 \, m^3} + \frac{\pi \alpha}{2 m^2} 
  \delta(\vec{r}) + \frac{\alpha}{4 m^2 r^3} \vec{\sigma} \cdot \vec{L} \,.
\mathend
Der Nenner in ${\tilde P}^{ij}$ kann 
folgenderma"sen umgeformt werden ($E_S = $ Schr"odinger-Energie, $\delta 
E =$ $\alpha^4$-Korrektur zur Schr"odinger-Energie):
\eqnanf
U \, \left(H_D - (E - \omega)\right) \, \Up & = &
H_{FW} - (E - \omega) = \schwups
& = & (m + H_S + \delta H) - (m + E_S + \delta E - \omega) -\wups
& = & (H_S - (E_S - \omega)) + \delta H - \delta E. \ok
\eqnend
Weil
\begin{equation}
\frac{1}{H_S - (E_S - \omega)} = G_{\rm nr}(E_S - \omega)
\end{equation}
gerade dem nichtrelativistischen Propagator $G_{nr}$ entspricht, sind
wir in der Lage, mit dem (viel einfacheren) nichtrelativistischen 
Propagator zu rechnen und alle relativistischen Korrekturen dazu 
st"orungstheoretisch zu behandeln. 

Dazu ist es n"otig, auch die erste relativistische Korrektur
zur Schr"odinger-Eigenfunktion $\phi$ mitzunehmen.
Selbstverst"andlich ist die  urspr"ungliche Eigenfunktion $\ket{\phi}$ 
des Schr"odinger-Hamiltonians 
nicht mehr Eigenfunktion des FW-Hamiltonians. Mit der ersten 
relativistischen 
Korrektur ergibt sich aus Gl. \ref{deltaphi}
\begin{equation}
\ket{\tilde \psi} = \ket{\phi} + \ket{\delta \phi} = 
\ket{\phi} + \frac{1}{(E_S - H_S)'} \, {\delta H \ket{\phi}}.
\end{equation}
Das Ergebnis der 
FW-Transformation der Ausdr"ucke $y^j = \alpha^j \exp(i \scp{k}{r})$
ist (siehe Gl. \ref{alphaitransformed}):
\begin{eqnarray}
y^j_{FW} & = &
\frac{p^j}{m} \left(1 + i \left( {\vec k} \, \cdot {\vec r} \right) -
\frac{1}{2} \left( {\vec k} \, \cdot \, {\vec r} \right)^2 \right) 
  \nonumber \\
& & - \frac{1}{2 \, m^3} p^j \psq -
\frac{1}{2 \, m^2} \, \frac{\alpha}{r^3} \, 
  \left( {\vec r} \times \vec{\sigma} \right)^j \nonumber \\
& & + \frac{1}{2 \, m} \left( {\vec k} \, \cdot \, {\vec r} \right)
 \left( {\vec k} \times \vec{\sigma} \right)^j \nonumber 
- \frac{i}{2 \, m}
   \left( {\vec k} \times \vec{\sigma} \right)^j. \nonumber
\end{eqnarray}
Wir k"onnen daher
\begin{equation}
U \, \alpha^j e^{i \scp{k}{r}} \, \Up = \frac{p^j}{m} + \delta y^j
\end{equation}
schreiben mit $p^j/m$ als Hauptbeitrag und $\delta y^j$ als den 
relativistischen Korrekturen dazu. Der Ausdruck ${\tilde P}^{ij}$ 
setzt sich somit unter Vernachl"assigung h"oherer Ordnungen
folgenderma"sen zusammen,
\eqnanf
{\tilde P}^{ij} & = &
\bra{\tilde \psi^{+}}  \left[U \alpha^i e^{i \scp{k}{r}} 
\Up\right] \, \frac{1}{U (H_D - (E - \omega)) \Up} \, \left[U   
\alpha^j e^{-i \scp{k}{r}} \Up\right] \ket{\tilde \psi} \schwups
& = &
\bra{\phi + \delta \phi} \left[\frac{p^i}{m} + \delta y^i\right] \,
\frac{1}{(H_S - (E_S - \omega)) + \delta H - \delta E}
\, \left[\frac{p^j}{m} + \delta y^j\right] \ket{\phi + \delta \phi} \wups
& = &
\bra{\phi} \frac{p^i}{m} \, 
\frac{1}{H_S - (E_S - \omega)} \, 
\frac{p^j}{m} \ket{\phi} \wups
& &
+ 2 \times \bra{\phi} \delta y^i \, 
\frac{1}{H_S - (E_S - \omega)} \, 
\frac{p^j}{m} \ket{\phi} \wups
& &
+ \bra{\phi} \frac{p^i}{m} \,
\frac{1}{H_S - (E_S - \omega)} \, \delta E \,
\frac{1}{H_S - (E_S - \omega)} \,
\frac{p^j}{m} \ket{\phi} \wups
& &
- \bra{\phi} \frac{p^i}{m} \,
\frac{1}{H_S - (E_S - \omega)} \, \delta H \, \frac{1}{H_S - (E_S - \omega)} \, 
\frac{p^j}{m} \ket{\phi} \wups
& &
+ 2 \times \bra{\phi} \frac{p^i}{m} \,
\frac{1}{H_S - (E_S - \omega)} \,
\frac{p^j}{m} \ket{\delta \phi} + O(\alpha^4) \ok
\eqnend
F"ur die praktische Auswertung des 
Niedrigenergieanteils ist die Einf"uhrung einer dimensionslosen
Gr"o"se $P$ zweckm"a"sig, die sich aus ${\tilde P}$ gem"a"s
\begin{equation}
P = \frac{m}{2} \delta^{T,ij} \, {\tilde P}^{ij}
\end{equation}
ergibt. ${\tilde P}^{ij}$ hat aufgrund des Propagators die Dimension
$\lquer$, $m$ dagegen die Dimension $\lquer^{-1}$ (in nat"urlichen 
Einheiten). Es ergeben sich somit bis zur Ordnung $\alpha^2$ folgende 
Beitr"age zu
${\tilde P}^{ij}$. Zun"achst lautet der Hauptbeitrag
\eqnanf
\label{Pnd}
P_{nd} &=& \frac{m}{2} \delta^{T,ij}
\bra{\phi} \frac{p^i}{m} \, 
\frac{1}{H_S - (E_S - \omega)} \,  
\frac{p^j}{m} \ket{\phi} \schwups
&=& \frac{1}{3 m} \bra{\phi} p^i \, 
\frac{1}{H_S - (E_S - \omega)}  
\, p^i \ket{\phi}\,. \ok 
\eqnend
Die Indizes $nd$ stehen hierbei f"ur den 
``nichtrelativistischen Dipol''. $P_{nd}$ ist der einzige Beitrag zu $P$
in der Ordnung $\alpha^0 = 1$ und damit der einzige Beitrag 
zum Niedrigenergieanteil der Selbstenergie in der Ordnung
$\alpha (Z \alpha)^4$. Alle anderen Beitr"age
tragen in der Ordnung $\alpha^2$ zu $P$ und damit in
der Ordnung $\alpha (Z \alpha)^6$ zur Lamb-Shift bei. 

Diese weiteren Beitr"age wollen wir jetzt der Reihe nach auflisten.
Zuerst wird der Beitrag $P_{\delta y}$, der durch die FW-Transformation 
der $\alpha^i$-Matrizen zustande kommt, aufgef"uhrt,
\begin{equation}
\label{Pdeltay}
P_{\delta y} = 
2 \times \frac{m}{2} \delta^{T,ij} \bra{\phi} \delta y^i \,
\frac{1}{H_S - (E_S - \omega)}
\, \frac{p^j}{m} \ket{\phi}.
\end{equation}
Weitere Beitr"age ergeben sich aufgrund der relativistischen Korrektur
zum Energieeigenwert,
\eqnanf
\label{PdeltaE}
P_{\delta E} &=&
\frac{m}{2} \delta^{T,ij} \bra{\phi} \frac{p^i}{m} \,
\frac{1}{H_S - (E_S - \omega)} \, \delta E \, \frac{1}{H_S - (E_S - \omega)}
\, \frac{p^j}{m} \ket{\phi} \schwups
&=& \frac{1}{3 m} \bra{\phi} p^i \,
\frac{1}{H_S - (E_S - \omega)} \, \delta E \, 
\frac{1}{H_S - (E_S - \omega)} \,
p^i \ket{\phi}, \ok
\eqnend
aufgrund der relativistischen Korrekturen zum 
Schr"odinger-Hamiltonian,
\eqnanf
\label{PdeltaH}
P_{\delta H} &=&
- \frac{m}{2} \delta^{T,ij} \bra{\phi} \frac{p^i}{m} \,
\frac{1}{H_S - (E_S - \omega)} \, \delta H \, 
\frac{1}{H_S - (E_S - \omega)}
\, \frac{p^j}{m} \ket{\phi} \schwups
&=& -\frac{1}{3 m} \bra{\phi} p^i \,
\frac{1}{H_S - (E_S - \omega)} \, \delta H \, 
\frac{1}{H_S - (E_S - \omega)} \,
p^i \ket{\phi}, \ok 
\eqnend
und aufgrund der relativistischen Korrektur zur 
Energie-Eigenfunktion, 
\eqnanf
\label{Pdeltaphi}
P_{\delta \phi} &=&
2 \times \frac{m}{2} \delta^{T,ij} \bra{\phi} \frac{p^i}{m} \,
\frac{1}{H_S - (E_S - \omega)} \,
\frac{p^j}{m} \ket{\delta \phi} \schwups
&=& \frac{2}{3 m} \bra{\phi} p^i \,
\frac{1}{H_S - (E_S - \omega)}
\, p^i \ket{\delta \phi}. \ok  
\eqnend
Den letzten Punkt bei der Behandlung des Niedrigenergieanteils bildet
die $d^3 k$-Integration. F"ur den Niedrigenergie-Anteil ist $\modk \equiv 
\omega$, daher gilt
\eqnanf
\label{EausP}
E_L &=& - e^2 \int_{\modk < \ep} \frac{d^3 k}{(2 \pi)^3}
\frac{1}{2 \modk} \, \delta^{T, ij} \, 
{\tilde P}^{ij}(\modk) \schwups
& = & - e^2 \int_{\modk < \ep} \frac{d^3 k}{(2 \pi)^3}
\, \frac{1}{m \modk} \, P(\modk) \wups
&=& - (4 \pi \alpha) \int_{\modk < \ep} \frac{dk \, k^2 \, 4 \pi}{(2 \pi)^3}
\, \frac{1}{m \modk} \, P(\modk) \wups
&=& - \frac{2 \alpha}{\pi m} \int_0^\ep d\omega \, \omega \, P(\omega) \,.
\eqnend
Wir f"uhren nun eine Variablentransformation gem"a"s
\begin{equation}
\label{deft}
t \equiv \frac{\sqrt{-2 m E_{\phi}}}{\sqrt{-2 m (E_{\phi} - \omega)}}
= \frac{(\alpha m)/2}{\sqrt{-2 m (E_{\phi} - \omega)}}
\end{equation}
zur dimensionslosen Gr"o"se $t$ durch. Wir verwenden die
Abk"urzung
\begin{equation}
\lambda = \frac{\alpha m}{2}. 
\end{equation}
Damit folgt 
\begin{equation}
\int_0^\ep d\omega \, \omega  P(t(\omega)) = 
  \frac{\lambda^4}{2 m^2} \int_{t_\ep}^1 dt \, \frac{1-t^2}{t^5} P(t)
\end{equation}
mit 
\begin{equation}
\label{teps}
t_\ep = \frac{\lambda}{\sqrt{\lambda^2 + 2 m \ep}} = 
\frac{\alpha}{\sqrt{\alpha^2 + 8 \ep / m}}.
\end{equation}
$F$ ergibt sich dann zu
\begin{equation}
\label{FausP}
F = - \frac{1}{2} \int_{t_\ep}^1 dt \, \frac{1-t^2}{t^5} \, P(t).
\end{equation}
Wie wir in Abschnitt \ref{lep} sehen werden, ist der "Ubergang 
zur dimensionslosen Variablen $t$ auch dadurch gut motiviert, da"s die 
Propagatoren in der gebr"auchlichen Darstellung die Energie gerade mit 
einer zu $t$ direkt proportionalen Variablen parametrisieren. $t$ ist also 
in gewissem Sinne eine ``nat"urliche'' Parametrisierung von $E$.

%
%
\chapter{Die Selbstenergie des 2P-Zustands (j=1/2) zur sechsten Ordnung in Z\_alpha}

%
%

\section{Der Hochenergie-Anteil}
\label{hep}

\subsection{Allgemeines zur Rechenmethode}
\label{allghep}

Bereits im Abschnitt \ref{hepmethod} war gezeigt worden, da"s man den 
Hochenergie-Anteil in Potenzen des bindenden 
Coulomb-Potentials aufspalten kann. 
Bis zur Ordnung $\alpha^6$ tragen zum Matrix-Element
\mathanf
{\tilde P} = \bra{\bar{\psi}} \gamma^{\mu} \dcprop \gamma_{\mu} \ket{\psi} 
\mathend
gerade der 0-,1-,2- und 3-Vertex-Anteil bei. 
Wir wollen uns noch einmal die Form der Matrix-Elemente am Beispiel des 
2-Vertex-Anteils in Erinnerung rufen.
\eqnanf
{\tilde P}_2 &=& \bra{\psi} \go \gamma^{\mu} \frac{1}{D} \go V
   \frac{1}{D} \go V 
  \frac{1}{D}  \gamma_{\mu}  \ket{\psi} \schwups
  &=& \bra{\bar{\psi}} M_2 \ket{\psi} \ok 
\eqnend
mit der Matrix
\begin{equation}
M_2 = \go \gamma^{\mu} \frac{1}{D} \go V
    \frac{1}{D} \go V 
   \frac{1}{D} \gamma_{\mu}. 
\end{equation}
Hier soll auf 
die rechentechnischen Details der in Kapitel \ref{hepmethod} beschriebenen
Methode zur Auswertung eines Vertex-Anteils
n"aher eingegangen werden.
\begin{enumerate}  
\item Entwicklung der $M_i$-Matrizen in die 16 $\Gamma$-Matrizen
gem"a"s der Gl. \ref{expansionGamma}:
\begin{equation}
M_i = \sum_{\beta} a_{\beta} \Gamma^{\beta}.
\end{equation}
Welche Matrizen hier beitragen, h"angt vom Matrix-Element ab. F"ur den 
3-Vertex-Anteil bekommen wir Beitr"age nur von der $\id$-Matrix und von 
der $\go$-Matrix, f"ur den 2-Vertex tragen die $\id$-Matrix, die $\go$-Matrix,
die $\alpha^i$-Matrizen und auch die $\Sigma$-Matrizen bei.

\item Entwicklung der $a_{\beta}$-Koeffizienten in
Potenzen der Feinstrukturkonstanten $\alpha$. Diese Entwicklung beruht im 
wesentlichen auf folgender Beobachtung: Die Matrizen sind aus freien 
Feynman-Propagatoren
\mathanf
S_F = \frac{1}{\slf{p} - \slf{k} - m}
\mathend
mit dazwischen liegenden Potentialtermen aufgebaut. Den freien 
Feynman-Propagator kann man auch schreiben als
\mathanf
S_F = \frac{\slf{p} - \slf{k} + m}{(p-k)^2 - m^2}.
\mathend
Mit der Bindungsenergie $E_b$ des Elektrons 
\mathanf
E_b = E_S + \delta E + \dots
\mathend
($E_S =$ Schr"odinger-Energie und $\delta E = $ erste relativistische 
Korrektur) l"a"st sich der Nenner umformen zu
\eqnanf
(p-k)^2 - m^2 &=& (m + E_b - \omega)^2 - (\vec{p} - \vec{k})^2 - m^2 \wups
&=& m^2 + 2 m E_b + E_b^2 + 2 m \omega + 2 E_b \omega + \omega^2 \wups
& &  - \vec{p}^2 + 2 \scp{k}{p} - \vec{k}^2 - m^2 \wups
&=& (\omega^2 - \vec{k}^2 - 2 m \omega) + (2 m E_b + E_b^2 +
   2 E_b \omega  - \psq + 2 \scp{k}{p}). \ok
\eqnend
Wir erkennen an diesem Ausdruck den $\alpha^0$-Hauptteil 
\mathanf
\omega^2 - \vec{k}^2 - 2 m \omega
\mathend
sowie h"ohere Terme
\mathanf
2 m E_b + E_b^2 + 2 E_b \omega  - \psq + 2 \scp{k}{p}.
\mathend
In diesen h"oheren Termen wird nun bis zur notwendigen Potenz in $\alpha$ 
entwickelt. Dadurch ``rutschen'' die Impulsoperatoren des Elektrons
$\vec{p}$ und Ausdr"ucke mit den Photonen-Impulsen 
$\scp{k}{p}$ in den Z"ahler, und es entstehen neue Matrix-Elemente, die 
auszuwerten sind. Bei der Entwicklung mu"s man etwas Vorsicht walten
lassen, denn die verschiedenen Impuls-Operatoren, die von den einzelnen 
Feynman-Propagatoren kommen, sind zu unterscheiden, da die von ihnen 
erzeugten Matrix-Elemente unterschiedliche Werte haben.
   
\item Mittelung "uber die r"aumlichen Anteile des Photonenimpulses.

Wie wir im vorigen Abschnitt gesehen haben, kommen bei der Entwicklung 
des Nenners des Elektronenpropagators Terme der Gestalt $\scp{k}{p}$
vor. Da wir jedoch sp"ater "uber den gesamten $R^3$ integrieren wollen, 
k"onnen wir an dieser Stelle bereits die Mittelung "uber die r"aumlichen 
Winkel vornehmen und dadurch die Ausdr"ucke stark vereinfachen.

Die Mittelung "uber die Winkel erfolgt durch die Ersetzungs-Vorschrift:
\begin{equation}
f(\vec{k}) \to \frac{1}{4 \pi} \int d \Omega f(\vec{k}).
\end{equation}
F"ur diese Mittelung werden unter anderem die folgenden Formeln ben"otigt:
\eqnanf
\frac{1}{4 \pi} \int d \Omega \, \left( \scp{k}{p} \right)\, 
\left(\scp{k}{q}\right) &=& 1/3 \, \vec{k}^2 \,\, \left(\scp{p}{q}\right) \schwups
\frac{1}{4 \pi} \int d \Omega \, (\scp{k}{p})^3 \, \left(\scp{k}{q}\right) &=&
 1/5 \, (\vec{k}^2)^2 \, \psq \, \left( \scp{p}{q} \right) \schwups
\frac{1}{4 \pi} \int d \Omega \, (\scp{k}{p})^2 \, (\scp{k}{q})^2 &=&
  1/15 \, (\vec{k}^2)^2 \, \left(2 (\scp{p}{q})^2 + \psq \vec{q}^2\right) \schwups
\frac{1}{4 \pi} \int d \Omega \, (\scp{k}{p})^5 \, (\scp{k}{q}) &=&
  1/7 \, (\vec{k}^2)^3 \, (\scp{p}{p})^2 \, \left( \scp{p}{q} \right) \schwups
\frac{1}{4 \pi} \int d \Omega \, (\scp{k}{p})^4 \,  (\scp{k}{q})^2 &=&
  1/35 \, \left( \vec{k}^2 \right)^3 \, (\scp{p}{p}) \, 
\left(4 (\scp{p}{q})^2 + \psq {\vec{q} \,}^2 \right) 
  \schwups
\frac{1}{4 \pi} \int d \Omega \, (\scp{k}{p})^3 \, (\scp{k}{q})^3 &=&
1/35 \, (\vec{k}^2)^3 \, 
\left(2 (\scp{p}{q})^2 + 3 \psq \vec{q}^2\right). \glok
\eqnend
Diese Formeln gelten f"ur zwei beliebige Vektoren $\vec{p}$ und $\vec{q}$. 

\item Ersetzung der Matrixelemente. In Abschnitt \ref{hepmethod} war die 
Ersetzung der Matrix-Elemente am Beispiel 
\mathanf
\left( \vec{p} \, V \, \vec{p} \right) \, f(\omega,\modk) \to 
\bra{\psi} \left( \vec{p} \, V \, \vec{p} \right) \ket{\psi} \, 
  f(\omega,\modk).
\mathend
erl"autert worden. Bis zur Ordnung 
$\alpha^6$ ist das gesuchte Matrix-Element gegeben durch 
\mathanf
\bra{\psi} \left( \vec{p} \, V \, \vec{p} \right) \ket{\psi} = 
  -\frac{5}{48} \alpha^4 m^3 - \frac{283}{1152} \alpha^6 m^3.
\mathend
Zur computergest"utzten Berechnung dieses Matrix-Elements wird die
Wellenfunktion des $2P_{1/2}$-Zustandes explizit eingegeben und 
bis zur Ordnung $\alpha^6$ entwickelt. Die  
Impuls- und andere Operatoren werden im Ortsraum explizit auf die
Wellenfunktionen angewandt. Das Resultat wird
in Potenzen von $\alpha$ geordnet. Dabei entstehen Ausdr"ucke mit 
mehreren tausend Termen. Alle diese Terme haben allerdings eine 
Standard-Struktur, so da"s die Integration (bis auf einige Ausnahmen) 
mit wenigen Integrationsregeln durchgef"uhrt werden kann. Dadurch wird 
die Rechnung um Gr"o"senordnungen beschleunigt. W"urde man die in 
$Mathematica$ eingebauten Integrations-Routinen benutzen, so w"urden 
die Rechnungen auch auf den schnellsten Maschinen bei weitem zu lange dauern.  

Die Auswertung der Matrixelemente ist auch von Hand mit Hilfe 
von Kommutatorbeziehungen unter intensiver Ausnutzung der Algebra der 
Dirac-Gleichung mit Coulombfeld m"oglich. Bei Matrix-Elementen, die nur 
in der $\alpha^6$-Potenz beitragen, ist "ubrigens die Verwendung der 
nichtrelativistischen Approximation an die Dirac-Wellenfunktion 
$\ket{\psi}$, das hei"st der Schr"odinger-Wellenfunktion $\ket{\phi}$,
m"oglich. Hier wollen wir nur einige sehr h"aufig verwendete Beziehungen 
unter den Operatoren vorstellen, die zur Auswertung der Matrix-Elemente 
von Hand n"utzlich sind. Es sind dies die f"ur $\ket{\psi}$ geltende 
Dirac-Gleichung
\mathanf
\scp{\alpha}{p} = E - \go m - V,
\mathend
die f"ur $\ket{\phi}$ geltende Schr"odinger-Gleichung
\mathanf
\psq = 2 m (E - V)
\mathend
sowie die Kommutatorbeziehung
\mathanf
A B A = 1/2 ([A,[B,A]] + A^2 B + B A^2).
\mathend
Diese Kommutatorbeziehung kann man zum Beispiel anwenden, um das 
Matrix-Element $\bra{\psi} \left( \vec{p} V \vec{p} \right) \ket{\psi}$
folgenderma"sen umzuformen:
\mathanf
\bra{\psi} \left( \vec{p} V \vec{p} \right) \ket{\psi} =
  \bra{\psi} \left( - \Delta(V) + 2 \psq V \right) \ket{\psi}. 
\mathend
Dabei ergibt $\Delta(V)$ einen der $\delta$-Funktion proportionalen Term. 
In erster Ordnung tr"agt dieser Term nicht bei, da die 
$"P$-Wellenfunktion am Ursprung verschwindet. Die erste relativistische 
Korrektur indes verschwindet nicht, so da"s sich in Ordnung $\alpha^6$ 
ein weiterer Beitrag ergibt. 
\item Zusammenfassung des Ergebnisses: Nach der Ersetzung aller 
Matrix-Elemente wird das Ergebnis in Potenzen von $\alpha$ zusammengefa"st 
und geordnet. Alle Vertex-Anteile tragen zur Ordnung $\alpha^6$ 
bei, da wir einen gebundenen Zustand behandeln.
\end{enumerate}  

Zuletzt eine Bemerkung zu den angegebenen Teilergebnissen:
In allen Zwischenergebnissen f"ur die Matrix-Elemente und auch f"ur die 
Resultate zu den einzelnen Vertex-Termen ersetzen wir der 
Einfachheit halber den Faktor $Z \alpha$ durch $\alpha$, um ihn dann im 
Endergebnis voll auszuschreiben.

%
%

\subsection{Der 3-Vertex Anteil}

\begin{figure}[ht]
\epsfxsize=6cm
\centerline{\epsfbox{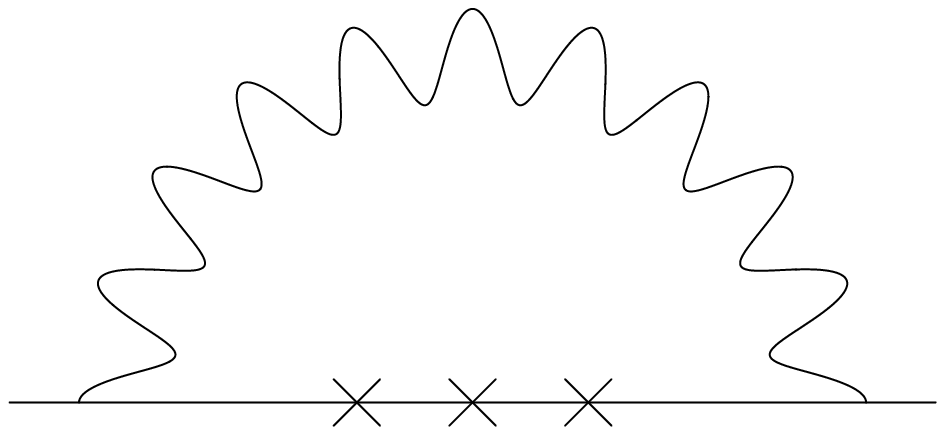}}
\caption{\label{vertex3} 3-Vertex-Anteil}
\end{figure}

Der 3-Vertex-Anteil (3 Vertizes des Coulomb-Feldes zwischen den
Photonen-Vertizes) wird durch das Feynman-Diagramm \ref{vertex3} wiedergegeben.
Da dieser Beitrag am einfachsten auszuwerten ist, wollen wir uns ihm 
zuerst zuwenden.
Es ist f"ur den 3-Vertex-Anteil das Matrixelement 
\eqnanf
{\tilde P}_3 &=& \bra{\psi} \go \gamma^{\mu} \frac{1}{D} \go V
   \frac{1}{D} \go V \frac{1}{D} \go V 
  \frac{1}{D}  \gamma_{\mu}  \ket{\psi} \schwups
  &=& \bra{\bar{\psi}} M_3 \ket{\psi} \ok 
\eqnend
mit der Matrix
\begin{equation}
M_3 = \go \gamma^{\mu} \frac{1}{D} \go V
   \frac{1}{D} \go V \frac{1}{D} \go V 
   \frac{1}{D} \gamma_{\mu} 
\end{equation}
auszuwerten. Zun"achst zur Entwicklung von $M_3$ in die 16 
$\Gamma$-Matrizen. Bis zur Ordnung $\alpha^6$ sind nur die 
Entwicklungs-Koeffizienten von $\id$ und $\go$ von Null verschieden.
Der Entwicklungskoeffizient von $\id$ lautet
\eqnanf
a_{\id} &=& 
  V^3 \bigg( 2\,\big( -(\vec{k}^2)^2 - 8\,\vec{k}^2\,{m^2} - 8\,{m^4} + 
       12\,\vec{k}^2\,m\,\omega  + 16\,{m^3}\,\omega  - 
       \schwups
& & - 6\,\vec{k}^2\,{{\omega }^2} -  - 12\,{m^2}\,{{\omega }^2} + 
    4\,m\,{{\omega }^3} - 
  {{\omega }^4} \big)  \bigg) \bigg/ 
   {{\left( \vec{k}^2 + 2\,m\,\omega  - {{\omega }^2} \right) }^4}. \ok
\eqnend
F"ur den Entwicklungskoeffizienten von $\go$ ergibt sich:
\mathanf
a_{\go} = 
  V^3 {{16\,\left( \vec{k}^2\,{m^2} + 2\,{m^4} - \vec{k}^2\,m\,\omega  - 
       4\,{m^3}\,\omega  + 3\,{m^2}\,{{\omega }^2} - m\,{{\omega }^3} \right) }
       \over
     {{{\left( \vec{k}^2 + 2\,m\,\omega  - {{\omega }^2} \right) }^4}}}.  
\mathend
Sowohl $a_{\id}$ wie auch $a_{\go}$ enthalten aufgrund des $V^3$-Terms
ausschlie"slich Anteile in $\alpha^6$. Da keine Entwicklung des
Nenners des Elektronenpropagators 
in den Koeffizienten n"otig ist, enthalten die Koeffizienten 
keine Skalarprodukte
von $\vec{k}$ mit anderen Vektoren, und wir brauchen keine Mittelung 
"uber die Winkel vorzunehmen. Die Koeffizienten liegen bereits in der Form
\mathanf
V^3 f(\omega,\modk)
\mathend
vor. Wir k"onnen also die Ersetzung bez"uglich
der Matrixelemente gem"a"s 
\mathanf
a_{\id} = V^3 f(\omega,\modk) \to \bra{\psi} V^3 \ket{\psi} f(\omega,\modk)
\mathend
und
\mathanf
a_{\go} = V^3 f(\omega,\modk) \to \bra{\psi} \go V^3 \ket{\psi} f(\omega,\modk)
\mathend
vornehmen. Die Matrixelemente lauten explizit (bis zur Ordnung $\alpha^6$):
\mathanf
\bra{\psi} V^3 \ket{\psi} = \bra{\psi} \go V^3 \ket{\psi} =
  - \frac{1}{24} \, \alpha^6 \, m^3.
\mathend
Die besondere Form der $P$-Zustand-Wellenfunktion bedingt,
da"s die obigen Matrixelemente in $\alpha^6$ konvergieren und somit
die Auswertung durch eine Entwicklung des Ausdrucks \ref{tildeP}
in Potenzen des Coulomb-Feldes m"oglich ist.  
Damit ist das Endergebnis f"ur den 3-Vertex-Anteil
\eqnanf
{\tilde P}_3 &=& 
  {{\alpha}^6}\,\bigg( (\vec{k}^2)^2\,{m^3} - 8\,{m^7} - 
       4\,\vec{k}^2\,{m^4}\,\omega  + 16\,{m^6}\,\omega  + 
       6\,\vec{k}^2\,{m^3}\,{{\omega }^2} - 
       \schwups
& & 
      - 12\,{m^5}\,{{\omega }^2} +  4\,{m^4}\,{{\omega }^3} + 
        {m^3}\,{{\omega }^4} \bigg) \bigg/
   \bigg( 12\,{{\left( \vec{k}^2 + 2\,m\,\omega  - {{\omega }^2} \right) }^4}
     \bigg) \ok.
\eqnend
Somit ist der einfachste der Vertex-Anteile berechnet.
Es gibt f"ur den 3-Vertex-Anteil nur einen $\alpha^6$-Koeffizienten, 
da hier bereits 3 Coulomb-Vertizes enthalten sind, die einen Faktor
$(Z \alpha)^6$ beitragen. 

%
%

\subsection{Der 2-Vertex Anteil}

\begin{figure}[htb]
\epsfxsize=6cm
\centerline{\epsfbox{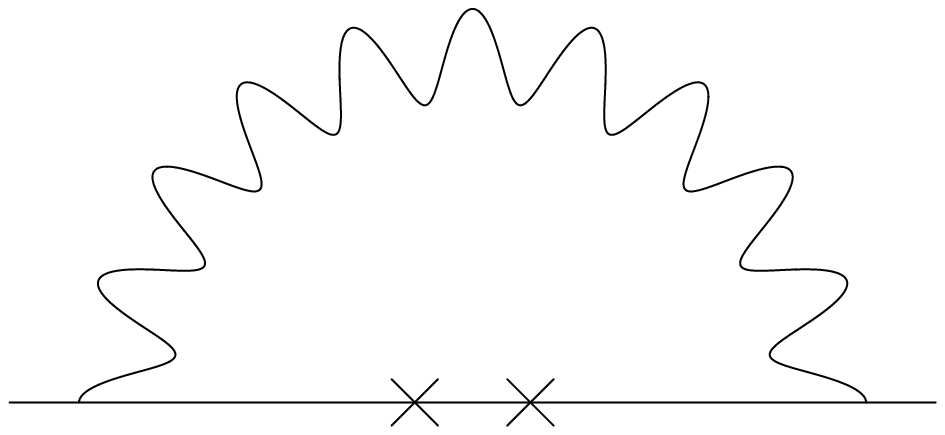}}
\caption{\label{vertex2} 2-Vertex-Anteil}
\end{figure}

Der 2-Vertex-Anteil ist durch das Feynman-Diagramm \ref{vertex2} gegeben.
Zwischen den Vertizes der virtuellen Photonen wirkt das Coulomb-Feld 
zweimal auf das Elektron. Es ist somit das Matrixelement 
\eqnanf
{\tilde P}_2 &=& \bra{\psi} \go \gamma^{\mu} \frac{1}{D} \go V
   \frac{1}{D} \go V 
  \frac{1}{D}  \gamma_{\mu}  \ket{\psi} \schwups
  &=& \bra{\bar{\psi}} M_2 \ket{\psi} \ok 
\eqnend
mit der Matrix
\begin{equation}
M_2 = \go \gamma^{\mu} \frac{1}{D} \go V
    \frac{1}{D} \go V 
   \frac{1}{D} \gamma_{\mu} 
\end{equation}
auszuwerten. Bei der Entwicklung von $M_2$
in die $\Gamma$-Matrizen erhalten wir weit gr"o"sere
Terme als beim 3-Vertex-Anteil. Daf"ur sind zwei Gr"unde verantwortlich. 
Zum einen erhalten wir nichtverschwindende Koeffizienten sowohl f"ur die
$\id$-Matrix, die $\go$-Matrix, die $\alpha^i$-Matrizen wie auch f"ur die 
$\Sigma$-Matrizen (alle anderen Projektionen verschwinden in der 
Ordnung $(Z \alpha)^6$).
Zum anderen mu"s der Nenner im Elektronenpropagator in 
$\alpha$ entwickelt werden, was zu gr"o"seren Termen f"uhrt.

Aufgrund der L"ange der auftretenden Terme verzichten wir auf eine 
Darstellung der Rechnung und gehen gleich zu den Matrix-Elementen "uber.
F"ur die Ersetzung durch Matrix-Elemente in den Koeffizienten 
werden folgende Ausdr"ucke ben"otigt:
\eqnanf
%
%
\bra{\psi} V^2 \ket{\psi} &=& 
 {{{{\alpha}^4}\,{m^2}}\over {12}} + {{47\,{{\alpha}^6}\,{m^2}}\over {288}},
 \schwups
%
%
\bra{\psi} V^2 \psq \ket{\psi} &=& 
  {{{{\alpha}^6}\,{m^4}}\over {16}},
   \schwups
%
%
\bra{\psi} V \vec{p} \,\, V \vec{p} \ket{\psi} &=&
  {{{{\alpha}^6}\,{m^4}}\over {12}} ,
  \schwups
%
%
\bra{\psi} V \psq V \ket{\psi} &=& 
 {{5\,{{\alpha}^6}\,{m^4}}\over {48}},
   \schwups 
%
%
\bra{\psi} \vec{p} \,\, V^2 \vec{p} \ket{\psi} &=&
  {{5\,{{\alpha}^6}\,{m^4}}\over {48}}. 
  \glok
\eqnend     
Zur Auswertung der Matrix-Elemente werden im Abschnitt \ref{allghep} 
einige Erkl"arungen gegeben. F"ur den $\go$-Beitrag ben"otigen wir das folgende
Matrix-Element
\begin{equation}
%
%
\bra{\psi} \go V^2 \ket{\psi} = 
 {{{{\alpha}^4}\,{m^2}}\over {12}} + {{{{\alpha}^6}\,{m^2}}\over {36}}.
\end{equation}
Alle anderen Matrix-Elemente f"ur den $\go$-Koeffizienten 
stimmen bis zur Ordnung $\alpha^6$ mit den 
entsprechenden Matrix-Elementen f"ur den Koeffizienten von $\id$ "uberein.
F"ur den Koeffizienten der $\alpha$-Matrizen ben"otigen wir folgende
Matrixelemente,
\eqnanf
%
%
\bra{\psi} V^2 \left( \scp{\alpha}{p} \right) \ket{\psi} &=&
  {{{{\alpha}^6}\,{m^3}}\over 6}, \schwups
%
%
\bra{\psi} V \left( \scp{\alpha}{p} \right) V \ket{\psi} &=&
  {{{{\alpha}^6}\,{m^3}}\over 6}. \glok
\eqnend
Schlie"slich ben"otigen wir f"ur den Koeffizienten der $\Sigma$-Matrizen 
noch folgende Matrix-Elemente:  
\eqnanf
%
%
\bra{\psi} V \, \vec{\Sigma} \cdot \left( \vec{p} \, V \right) \times
  \vec{p} \ket{\psi} &=&
{{-i}\over {12}}\,{{\alpha}^6}\,{m^4}, \schwups
%
%
\bra{\psi} \vec{\Sigma} \cdot \left( \vec{p} \, V^2 \right) \times
  \vec{p} \ket{\psi} &=& 
{{-i}\over 6}\,{{\alpha}^6}\,{m^4}. \glok
\eqnend
Das Endresultat f"ur den 2-Vertex-Anteil lautet:
\eqnanf
& & {\tilde P}_2 = \alpha^4\,
     \bigg( \vec{k}^2\,{m^3} - 4\,{m^5} - 3\,\vec{k}^2\,{m^2}\,\omega  + 6\,{m^4}\,
       \omega - \wups 
& &         - 3\,{m^3}\,{{\omega }^2} - {m^2}\,{{\omega }^3} \bigg)  \bigg/
           \bigg( 6\, \left( \vec{k}^2 + 2\,m\,\omega  - {\omega }^2 \right)^3 
    \bigg)  + \wups
& &  \alpha^6\,  
  \bigg( 132\,(\vec{k}^2)^3\,{m^3} + 492\,(\vec{k}^2)^2\,{m^5} 
    - 432\,\vec{k}^2\,{m^7} - 141\,(\vec{k}^2)^3\,{m^2}\,\omega  - \wups 
& &      - 340\,(\vec{k}^2)^2\,{m^4}\,\omega  + 968\,\vec{k}^2\,{m^6}\,\omega  + 
          672\,{m^8}\,\omega  - 558\,(\vec{k}^2)^2\,{m^3}\,{{\omega }^2} - \wups 
& &       - 264\,\vec{k}^2\,{m^5}\,{{\omega }^2} - 1904\,{m^7}\,{{\omega }^2} +  
             235\,(\vec{k}^2)^2\,{m^2}\,{{\omega }^3} - \wups 
& &       - 644\,\vec{k}^2\,{m^4}\,{{\omega }^3} + 2120\,{m^6}\,{{\omega }^3}  
       + 280\,\vec{k}^2\,{m^3}\,{{\omega }^4} - 1052\,{m^5}\,{{\omega }^4} - \wups 
& &       - 47\,\vec{k}^2\,{m^2}\,{{\omega }^5} + 88\,{m^4}\,{{\omega }^5} + 
          146\,{m^3}\,{{\omega }^6} - 47\,{m^2}\,{{\omega }^7} 
          \bigg) \bigg/  \wups
& &       \bigg( 144\,{{\left( \vec{k}^2 + 2\,m\,\omega  - 
            {{\omega }^2} \right) }^5} \bigg).
\eqnend
 
%
%

\subsection{Der 1-Vertex Anteil}

\begin{figure}[htb]
\epsfxsize=6cm
\centerline{\epsfbox{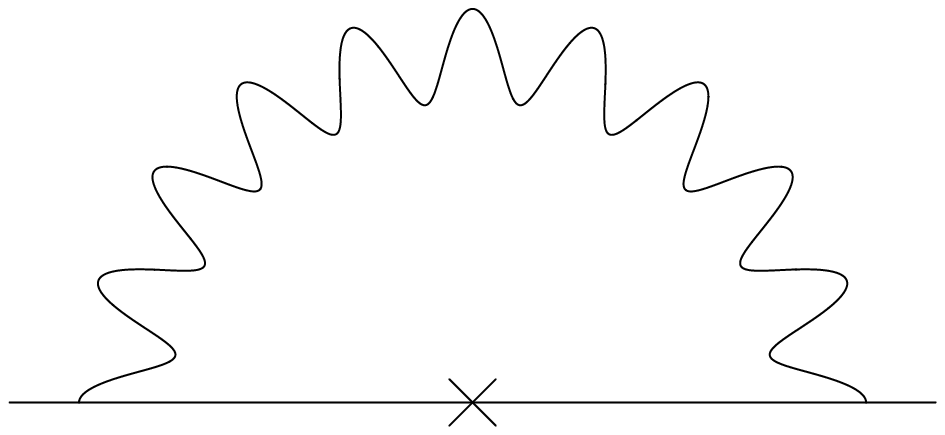}}
\caption{\label{vertex1} 1-Vertex-Anteil}
\end{figure}

Dem 1-Vertex-Anteil (1 Vertex des Coulomb-Feldes zwischen den
Photonen-Vertizes) entspricht dem Feynman-Diagramm \ref{vertex1}. 
Zwischen den Photonen-Vertizes wirkt das bindende Coulomb-Feld
einmal auf das Elektron. Es ist somit das Matrixelement 
\eqnanf
{\tilde P}_1 &=& \bra{\psi} \go \gamma^{\mu} \frac{1}{D} \go V 
  \frac{1}{D}  \gamma_{\mu}  \ket{\psi} \schwups
  &=& \bra{\bar{\psi}} M_1 \ket{\psi} \ok 
\eqnend
auszuwerten, wobei f"ur die Matrix $M_1$ gilt:
\begin{equation}
M_1 = \go \gamma^{\mu} \frac{1}{D} \go V 
   \frac{1}{D} \gamma_{\mu}. 
\end{equation}
Wie beim 2-Vertex-Anteil verzichten wir wegen 
der L"ange der Terme auf eine Auff"uhrung der 
Entwicklungskoeffizienten.
In der Ordnung $\alpha^6$ erhalten wir Beitr"age f"ur die
$\id$-Matrix, die $\go$-Matrix, die $\alpha^i$-Matrizen wie auch f"ur die 
$\Sigma$-Matrizen (alle anderen Projektionen verschwinden in der 
Ordnung $\alpha^6$).
Anbei eine Liste der f"ur diesen 1-Vertex Anteil berechneten Matrixelemente:
\eqnanf
%
%
\bra{\psi} V \ket{\psi} &=& 
{{-{{\alpha}^2}\,m }\over 4} - {{5\,{{\alpha}^4}\,m}\over {32}} - 
  {{63\,{{\alpha}^6}\,m}\over {512}}, \schwups
%
%
\bra{\psi} V \psq \ket{\psi} &=&
 {{-5\,{{\alpha}^4}\,{m^3}}\over {48}} - {{337\,{{\alpha}^6}\,{m^3}}\over 
 {1152}}, \schwups
%
%
\bra{\psi} \vec{p} \,\, V \vec{p} \ket{\psi} &=&
  {{-5\,{{\alpha}^4}\,{m^3}}\over {48}} - {{283\,{{\alpha}^6}\,{m^3}}\over {1152}},
  \schwups
%
%
\bra{\psi} V \left( \psq \right)^2  \ket{\psi} &=& 
  {{-35\,{{\alpha}^6}\,{m^5}}\over {192}}, \schwups
%
%
\bra{\psi} \psq V \psq \ket{\psi} &=&
  {{-19\,{{\alpha}^6}\,{m^5}}\over {192}}, \schwups
%
%
\bra{\psi} \left( \vec{p} \,\, V \vec{p} \right) \psq \ket{\psi} &=&
  {{-9\,{{\alpha}^6}\,{m^5}}\over {64}}, \schwups
%
%
\bra{\psi} \left( \vec{p} \left( \vec{p} \,\, V \vec{p} \right) 
  \vec{p} \right) \ket{\psi} &=& -5\,{{\alpha}^6}\,{m^5}\over {64}.  \glok
\eqnend  
F"ur den $\go$-Koeffizienten ben"otigen wir zus"atzlich weitere 
Matrix-Elemente,
\eqnanf
%
%
\bra{\psi} \go V \ket{\psi} &=& 
  {{-\left( {{\alpha}^2}\,m \right) }\over 4} - {{{{\alpha}^4}\,m}\over {16}} - 
  {{{{\alpha}^6}\,m}\over {32}}, \schwups
%
%
\bra{\psi} \go V \psq \ket{\psi} &=&
  {{-5\,{{\alpha}^4}\,{m^3}}\over {48}} - {{43\,{{\alpha}^6}\,{m^3}}\over {288}}, \schwups
%
%
\bra{\psi} \go \left( \vec{p} \,\, V \vec{p} \right) \ket{\psi} &=&
  {{-5\,{{\alpha}^4}\,{m^3}}\over {48}} - {{113\,{{\alpha}^6}\,{m^3}}\over {576}}, \schwups
%
%
\bra{\psi} \go V \left( \psq \right)^2  \ket{\psi} &=& 
  {{-35\,{{\alpha}^6}\,{m^5}}\over {192}}, \schwups
%
%
\bra{\psi} \go \psq V \psq \ket{\psi} &=&
   {{-19\,{{\alpha}^6}\,{m^5}}\over {192}}, \schwups
%
%
\bra{\psi} \go \left( \vec{p} \,\, V \vec{p} \right) \psq \ket{\psi} &=&
   {{-9\,{{\alpha}^6}\,{m^5}}\over {64}}, \schwups
%
%
\bra{\psi} \go \left( \vec{p} \,\, 
  \left( \vec{p} \,\, V \vec{p} \right) \vec{p} \right) \ket{\psi} &=&
  {{-5\,{{\alpha}^6}\,{m^5}}\over {64}}, \glok
\eqnend  
F"ur den Koeffizienten der $\alpha^i$-Matrizen erhalten wir die folgenden 
Matrixelemente:
\eqnanf
%
%
\bra{\psi} V \left( \scp{\alpha}{p} \right) \ket{\psi} &=&
  {{-7\,{{\alpha}^4}\,{m^2}}\over {48}} - {{65\,{{\alpha}^6}\,{m^2}}\over {288}},  \schwups
%
%
\bra{\psi} V \left( \scp{\alpha}{p} \right) V \ket{\psi} &=&
  {{-37\,{{\alpha}^6}\,{m^4}}\over {192}}, \schwups
%
%
\bra{\psi} V \psq \left( \scp{\alpha}{p} \right) \ket{\psi} &=&
  {{-15\,{{\alpha}^6}\,{m^4}}\over {64}}, \schwups
%
%
\bra{\psi} \left( \scp{\alpha}{p} \right) 
  \left( \vec{p} \,\, V \vec{p} \right) \ket{\psi} &=&
  {{-23\,{{\alpha}^6}\,{m^4}}\over {192}}. \glok  
\eqnend
Schlie"slich ben"otigen wir f"ur den Koeffizienten der $\Sigma$-Matrizen 
noch folgende Matrixelemente:  
\eqnanf
%
%
\bra{\psi} \vec{\Sigma} \cdot \left( \vec{p} \,\, V \right) \times
  \vec{p} \ket{\psi} &=& 
   {i\over {12}}\,{{\alpha}^4}\,{m^3} + {{121\,i}\over {576}}\,{{\alpha}^6}\,{m^3}, \schwups
%
%
\bra{\psi} \left(\vec{\Sigma} \cdot \left( \vec{p} \,\, V \right) \times
  \vec{p}\right) \psq \ket{\psi} &=& 
  {{7\,i}\over {48}}\,{{\alpha}^6}\,{m^5}, \wups
%
%
\bra{\psi} \vec{p} \left( \vec{\Sigma} \cdot \left( \vec{p} \, V^2 \right) 
    \times
  \vec{p} \right) \vec{p} \, \ket{\psi} &=& 
  {i\over {48}}\,{{\alpha}^6}\,{m^5}. \glok  
\eqnend
Das Ergebnis f"ur den 1-Vertex-Anteil lautet:
\eqnanf
& & {\tilde P}_1 =  
{{{{\alpha}^2}\,\left( \vec{k}^2\,m - 2\,{m^3} + 2\,{m^2}\,\omega  + 
        m\,{{\omega }^2} \right) }\over 
    {2\,{{\left( \vec{k}^2 + 2\,m\,\omega  - {{\omega }^2} \right) }^2}}} 
    + \schwups 
& & +
  {{\alpha}^4}\,\bigg( 45\,(\vec{k}^2)^3\,m + 46\,(\vec{k}^2)^2\,{m^3} - 
        48\,\vec{k}^2\,{m^5} + 188\,(\vec{k}^2)^2\,{m^2}\,\omega  -  
        \wups
& &        
     - 256\,\vec{k}^2\,{m^4}\,\omega  + 384\,{m^6}\,\omega  - 
        45\,(\vec{k}^2)^2\,m\,{{\omega }^2} + \wups
& &        
     + 576\,\vec{k}^2\,{m^3}\,{{\omega }^2} - 888\,{m^5}\,{{\omega }^2} - 
        56\,\vec{k}^2\,{m^2}\,{{\omega }^3} + 720\,{m^4}\,{{\omega }^3} 
        -  \wups
& & - 45\,\vec{k}^2\,m\,{{\omega }^4} - 102\,{m^3}\,{{\omega }^4} - 
        132\,{m^2}\,{{\omega }^5} + 45\,m\,{{\omega }^6} \bigg) \bigg/     
    \bigg( 144\,{{\left( \vec{k}^2 + 2\,m\,\omega  - {{\omega }^2} 
    \right) }^4} \bigg) + \wups 
& &    
  {{\alpha}^6}\,\bigg( 8505\,(\vec{k}^2)^5\,m + 19790\,(\vec{k}^2)^4\,{m^3} - 
        43296\,(\vec{k}^2)^3\,{m^5} - 16128\,(\vec{k}^2)^2\,{m^7} + \wups
& & + 61420\,(\vec{k}^2)^4\,{m^2}\,\omega  - 
        22104\,(\vec{k}^2)^3\,{m^4}\,\omega  - 
        477312\,(\vec{k}^2)^2\,{m^6}\,\omega  + \wups
& & + 430080\,\vec{k}^2\,{m^8}\,\omega  - 
        25515\,(\vec{k}^2)^4\,m\,{{\omega }^2} + 
        279240\,(\vec{k}^2)^3\,{m^3}\,{{\omega }^2} - 
        34224\,(\vec{k}^2)^2\,{m^5}\,{{\omega }^2} - \wups
& & - 474240\,\vec{k}^2\,{m^7}\,{{\omega }^2} - 368640\,{m^9}\,{{\omega }^2} - 
        135000\,(\vec{k}^2)^3\,{m^2}\,{{\omega }^3} + 
        1011864\,(\vec{k}^2)^2\,{m^4}\,{{\omega }^3} - \wups
& & - 1040480\,\vec{k}^2\,{m^6}\,{{\omega }^3} + 1413120\,{m^8}\,{{\omega }^3} + 
        17010\,(\vec{k}^2)^3\,m\,{{\omega }^4} - 
        561180\,(\vec{k}^2)^2\,{m^3}\,{{\omega }^4} + \wups
& & + 2020400\,\vec{k}^2\,{m^5}\,{{\omega }^4} - 2366880\,{m^7}\,{{\omega }^4} + 
        36480\,(\vec{k}^2)^2\,{m^2}\,{{\omega }^5} - 
        1269960\,\vec{k}^2\,{m^4}\,{{\omega }^5} + \wups
& & + 2243040\,{m^6}\,{{\omega }^5} + 
        17010\,(\vec{k}^2)^2\,m\,{{\omega }^6} + 
        205480\,\vec{k}^2\,{m^3}\,{{\omega }^6} - 1212480\,{m^5}\,{{\omega }^6} + \wups
& & + 86360\,\vec{k}^2\,{m^2}\,{{\omega }^7} + 280200\,{m^4}\,{{\omega }^7} - 
        25515\,\vec{k}^2\,m\,{{\omega }^8} + 56670\,{m^3}\,{{\omega }^8} - \wups
& & - 49260\,{m^2}\,{{\omega }^9} + 8505\,m\,{{\omega }^{10}} \bigg) \bigg/
%
  \bigg( 34560\,{{\left( \vec{k}^2 + 2\,m\,\omega  - {{\omega }^2} 
       \right)}^6} \bigg). \glok       
\eqnend 

%
%

\subsection{Der 0-Vertex Anteil}

\begin{figure}[ht]
\epsfxsize=6cm
\centerline{\epsfbox{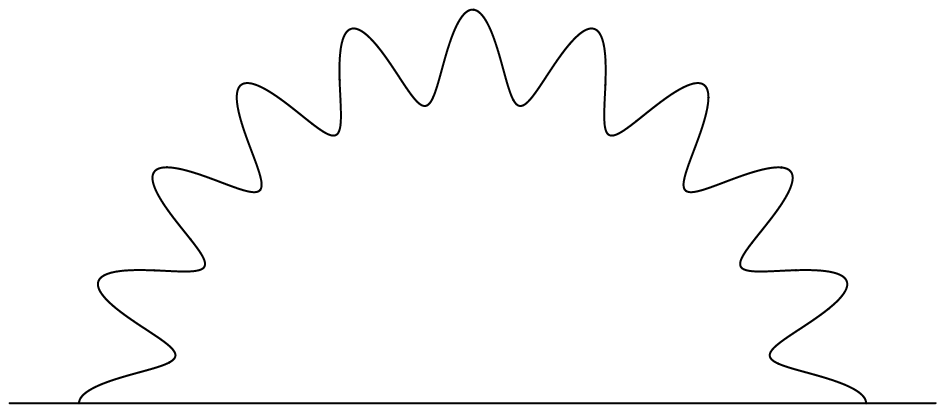}}
\caption{\label{vertex0} 0-Vertex-Anteil}
\end{figure}

Im 0-Vertex-Anteil rechnen wir mit der Selbstenergiefunktion des freien
Elektrons, das hei"st, wir werten das Diagramm \ref{vertex0} aus (die \"au\ss{}eren 
Spinoren sind wie in jeder Berechnung der gebundenen Selbstenergie die 
Dirac--Coulomb--Wellenfunktionen, nicht die freien Spinoren).
Es ist somit das Matrixelement 
\eqnanf
{\tilde P}_0 &=& \bra{\psi} \go \gamma^{\mu} \frac{1}{D}  \gamma_{\mu}  \ket{\psi} \schwups
  &=& \bra{\bar{\psi}} M_3 \ket{\psi}, \glok 
\eqnend
auszuwerten, wobei f"ur die Matrix $M_0$ gilt:
\begin{equation}
M_0 = \go \gamma^{\mu} \frac{1}{D} \gamma_{\mu}. \glok
\end{equation}
Wie "ublich beschr"anken wir uns auf die Angabe der 
relevanten Matrix-Elemente. Wir erhalten  
Beitr"age f"ur die $\id$-Matrix, die $\go$-Matrix und die $\alpha^i$-Matrizen 
(alle anderen Projektionen verschwinden in der Ordnung $\alpha^6$).
F"ur den Beitrag der $\id$-Matrix ben"otigen wir die folgenden 
Matrix-Elemente
\eqnanf
%
%
\langle \psi | \psi \rangle &=& 1 \quad \mbox{(Normierung)}, \schwups
%
%
\bra{\psi} \psq \ket{\psi} &=& 
{{{{\alpha}^2}\,{m^2}}\over 4} + {{13\,{{\alpha}^4}\,{m^2}}\over {48}} + 
  {{23\,{{\alpha}^6}\,{m^2}}\over {72}}, \schwups
%
%
\bra{\psi} \left( \psq \right)^2 \ket{\psi} &=&  
  {{7\,{{\alpha}^4}\,{m^4}}\over {48}} + {{167\,{{\alpha}^6}\,{m^4}}\over {288}},
    \schwups
%
%
\bra{\psi} \left( \psq \right)^3 \ket{\psi} &=&   
  {{21\,{{\alpha}^6}\,{m^6}}\over {64}}. \glok
\eqnend
F"ur die $\go$-Matrix werden die nachfolgenden Elemente gebraucht:
\eqnanf
%
%
\bra{\psi} \go \ket{\psi} &=&  
  1 - {{{{\alpha}^2}}\over 8} - {{5\,{{\alpha}^4}}\over {128}} - 
    {{21\,{{\alpha}^6}}\over {1024}}, \schwups
%
%
\bra{\psi} \go \psq \ket{\psi} &=& 
 {{{{\alpha}^2}\,{m^2}}\over 4} + {{19\,{{\alpha}^4}\,{m^2}}\over {96}} + 
  {{943\,{{\alpha}^6}\,{m^2}}\over {4608}}, \schwups
%
%
\bra{\psi} \go \left( \psq \right)^2 \ket{\psi} &=& 
  {{7\,{{\alpha}^4}\,{m^4}}\over {48}} +
   {{479\,{{\alpha}^6}\,{m^4}}\over {1152}}.  \glok
\eqnend  
F"ur die Matrix-Elemente mit den $\alpha$-Matrizen gilt:
\eqnanf
%
%
\bra{\psi} \left(\scp{\alpha}{p}\right) \, \ket{\psi} &=& 
  {{{{\alpha}^2}\,m}\over 4} + {{5\,{{\alpha}^4}\,m}\over {32}} + 
  {{63\,{{\alpha}^6}\,m}\over {512}}, \schwups
%
%
\bra{\psi} \left( \scp{\alpha}{p} \right) \, \psq \ket{\psi} &=&  
  {{7\,{{\alpha}^4}\,{m^3}}\over {48}} + 
    {{419\,{{\alpha}^6}\,{m^3}}\over {1152}}, \schwups
%
%
\bra{\psi} \left( \scp{\alpha}{p} \right) \, \left( \psq \right)^2 \ket{\psi} &=&  
  {{21\,{{\alpha}^6}\,{m^5}}\over {64}}. \glok 
\eqnend
Auf eine Angabe des Endergebnisses f"ur den 0-Vertex-Anteil wird 
hier aufgrund der L"ange des Ausdrucks verzichtet. Der Ausdruck w"urde mehr als eine
Seite f"ullen, ist jedoch von seiner Struktur her dem 
Ergebnis f"ur den 1-Vertex-Anteil sehr "ahnlich. Er ist von der Gestalt:
\eqnanf
{\tilde P}_0 &=&  
  {{-2\,\left( m + \omega  \right) }\over 
    {\vec{k}^2 + 2\,m\,\omega  - {{\omega }^2}}} +  \schwups
& & + \alpha^2 ( \dots ) + \alpha^4 ( \dots ) + \alpha^6 ( \dots ) \ok
\eqnend
Den $\alpha^0$-Koeffizient kennen wir bereits. Er entspricht gerade dem
Integranden der parametrischen Darstellung des Massen-Gegenterms, die wir
in Abschnitt \ref{deltamausdruck} kennengelernt hatten. 
Somit hebt sich dieser (bei Integration "uber $\modk$ und $\omega$ 
divergente) Ausdruck gerade heraus, und das Integral konvergiert.

%
%

\subsection{Der Massen-Gegenterm}

Wie leicht zu erkennen, w"urde der erste Term im Matrix-Element ${\tilde 
P}_0$ auch bei bei Integration bzgl. des 4-Impulses des virtuellen Photons
auch nach Regularisierung des Photonenpropagators noch divergieren.

Der Massen-Gegenterm (die Massen-Renormierung) gleicht, wie bereits im 
Kapitel "uber die Theorie des Selbstenergie-Beitrages, diese Divergenz 
gearde aus. Wir erhalten mit der parametrischen Darstellung des 
Massen-Gegenterms \ref{deltamausdruck}
\mathanf
\delta m = - i e^2 \intk \left[\frac{1}{k^2} - \frac{1}{k^2 - M^2}\right]
  \frac{2 ( m + \omega )}{k^2 - 2 \, m \, \omega}
\mathend
den Integranden des Massen-Gegenterms als
\mathanf
\frac{2 ( m + \omega )}{k^2 - 2 \, m \, \omega}
\mathend
mit $k^2 = \omega^2 - \vec{k}^2$. Wenn man dann noch das Matrix-Element
\mathanf
\langle{\bar \psi} | \psi \rangle = \bra{\psi} \go \ket{\psi} =  
  1 - {{{{\alpha}^2}}\over 8} - {{5\,{{\alpha}^4}}\over {128}} - 
    {{21\,{{\alpha}^6}}\over {1024}} 
\mathend
einsetzt, so ergibt sich der Integrand des Massen-Gegenterms, 
$\bra{\bar \psi} \delta m \ket{\psi}$, als
Matrix-Element mit $\ket{\psi}$ geschrieben, zu
\eqnanf
\bra{\psi} \go \delta m \ket{\psi} &=& {{-2\,\left( m + \omega  \right) }\over 
    {\vec{k}^2 + 2\,m\,\omega  - {{\omega }^2}}} + 
  {{{{\alpha}^2}\,\left( m + \omega  \right) }\over 
    {4\,\left( \vec{k}^2 + 2\,m\,\omega  - {{\omega }^2} \right) }} +  
    \wups
& & + {{5\,{{\alpha}^4}\,\left( m + \omega  \right) }\over 
    {64\,\left( \vec{k}^2 + 2\,m\,\omega  - {{\omega }^2} \right) }} + 
  {{21\,{{\alpha}^6}\,\left( m + \omega  \right) }\over 
    {512\,\left( \vec{k}^2 + 2\,m\,\omega  - {{\omega }^2} \right) }}. \glok
\eqnend

%
%

\subsection{Die k- und omega-Integration}

Den letzten Schritt bei der Berechnung des Hochenergie-Anteils bilden die
$\modk$ und die $\omega$-Integrationen.
Die Ergebnisse f"ur den 0-,1-,2- und 3-Vertex-Anteil werden aufaddiert 
(siehe Gl. \ref{Psum}) 
\mathanf
{\tilde P} = {\tilde P}_0 + {\tilde P}_1 + {\tilde P}_2 + {\tilde P}_3
\mathend
und ergeben das Matrixelement ${\tilde P}$ bis zur Ordnung $\alpha^6$. Dieses 
Matrix-Element wird durch Abzug des Massen-Gegenterms renormiert und 
bez"uglich $\modk$ und $\omega$ integriert. 

Den ersten Schritt bildet die $\modk$-Integration. Wir definieren die
Abk"urzung
\begin{equation}
X \equiv \sqrt{2 \, m \, \omega - \omega^2},
\end{equation}
um die Ausdr"ucke in der Rechnung in ihrer L"ange zu kontrollieren. 
Nach gewissen Umformungen des Integranden ist die 
einzige Formel, die zur Integration bzgl. $\modk$ ben"otigt wird  
\eqnanf
\label{intformel}
& & \frac{1}{2 \pi i} \int_{-\infty}^{\infty} \frac{1}{k^2 - \omega^2}
  \frac{1}{(X^2 + k^2)^n} = \schwups
& & = \frac{1}{2 \modw} \frac{1}{\omega^2 + X^2} +
  \frac{(-1)^n}{(n-1)!} \frac{\del^{n-1}}{\del(X^2)^{n-1}} 
  \frac{i}{2 X} \frac{1}{\omega^2 + X^2},
\eqnend 
wobei die Betragsfunktion f"ur komplexes Argument $\omega$
wie folgt erkl"art ist,
\begin{equation}
\modw = {\rm sgn}(\Im(\omega)) \,\, \omega,
\end{equation}
d. h.
\eqnanf
\modw = + \omega \quad \mbox{falls} \quad \Im(\omega) > 0, \schwups  
\modw = - \omega \quad \mbox{falls} \quad \Im(\omega) < 0. \ok  
\eqnend
Die Formel \ref{intformel} kann man durch Vorziehen der Differentiation 
nach $X^2$ vor das Integral und mit Anwendung des Residuensatzes leicht 
nachrechnen.

F"ur die $\omega$-Integration werden ultraviolett divergente Ausdr"ucke 
regularisiert, es werden Feynman-Parameter eingef"uhrt und die Integration
kovariant ausgef"uhrt.
F"ur die ultraviolett konvergenten Ausdr"ucke 
wird zun"achst die Integrationsvariable reskaliert und auf eine 
dimensionslose Gr"o"se zur"uckgef"uhrt. Es wird definiert
\begin{equation}
u = \frac{X + i \, \omega}{X - i \, \omega}
\end{equation}
(mit $X \equiv \sqrt{2 \, m \, \omega - \omega^2}$).
Die $\omega$-Integration wird durch folgende "Uberlegung erleichtert:
Der Integrand hat eine Verzweigung auf der positiven reellen Achse, d.h.
der Realteil des Integranden leicht unterhalb der reellen Achse ist gerade 
das Negative des Realteils oberhalb dieser Achse, wie man an der Formel 
\ref{intformel} erkennt, und der Imagin"arteil beh"alt gerade sein
Vorzeichen bei. Da auf dem Integrationsweg $C_H$ unterhalb
der rellen Achse in negativer Richtung integriert wird, 
ergibt sich daraus, da"s der Wert 
des Integrals auf dem unteren Ast das konjugiert-Komplexe des Integrals
auf dem oberen Ast ist. 
Die Integration nach $u$ und Entwicklung im Infrarot-Regulator $\ep$
erfolgt mit Formeln, die zum Beispiel in \cite{krpdiss} zu finden sind.

\subsection{Ergebnis des Hochenergie-Anteils}

Nach der Auf"uhrung der $\modk$ und $\omega$-Integration und Ersetzung 
von $\alpha$ durch $Z \alpha$ lautet das Ergebnis f"ur den $F$-Faktor des 
Hochenergie-Anteils 
\begin{equation}
\label{Fhep}
F_H =   
  -{1\over 6} + 
  \left[ \frac{4177}{21600} - \frac{103}{180} \ln(2) \right] (Z \alpha)^2 -  
  {{2\,{(Z \alpha)^2}}\over {9\,\epsilon }} - 
  \frac{103}{180} \,(Z \alpha)^2 \, \ln (\epsilon)
\end{equation}
Die Divergenzen in Termen des Abschneideparameters $\epsilon$ werden sich 
durch entsprechende Terme im Niedrigenergie-Anteil gerade herausheben.

%
%

\section{Der Niedrigenergie-Anteil}
\label{lep}           
   
\subsection{Allgemeines zur Rechenmethode}
\label{allglep}

In dem gesamten Abschnitt "uber den Niedrigenergie-Anteil wird es darum 
gehen, ein allgemeines Matrix-Elemente der Form
\begin{equation}
\label{Pallg}
P_{\rm allg.}(\omega) =  \bra{\phi} {\cal O}_1 \, p^i \, 
\frac{1}{H_S - (E_S - \omega)} \, p^j \, {\cal O}_2 \ket{\phi}
\end{equation}
zu berechnen. Dabei ist $E_S = E_{\phi}$ die Schr"odinger-Energie des 
Zustands $\ket{\phi}$, und wir f"uhren eine Variablen-Transformtion zum 
Parameter $t$ mit 
\mathanf
t = \frac{\lambda}{X}
\mathend
durch, wobei
\begin{equation}
\lambda = \frac{\alpha m}{2} \quad \mbox{sowie} \quad X = 
\sqrt{-2 m (E_{\phi} - \omega)}
\end{equation}
(siehe Abschnitt \ref{lepmetho}, insbesonders Gl. \ref{deft}). 
${\cal O}_1$ und ${\cal O}_2$ sind Operatoren, die den Drehimpuls nicht 
"andern, also beispielsweise skalare Funktionen von $r$. 

Haben wir $P(\omega)$ bzw. 
$P(t)$ erst einmal berechnet, so ergibt sich der gesuchte Beitrag zu F 
durch
\mathanf
F = - \frac{1}{2} \int_{t_\ep}^1 dt \, \frac{1-t^2}{t^5} \, P(t)
\mathend
(siehe Gl. \ref{FausP}). Wie ein solches Matrix-Element, welches den 
Propagator beinhaltet, im einzelnen berechnet wird, wollen wir hier 
skizzieren. 

Wir beschr"anken uns in dieser allgemeinen Diskussion auch auf den Fall 
der Darstellung des Propagators im Ortsraum und mit Laguerre-Polynomen. 
Wie wir im Abschnitt \ref{lepmathe} noch sehen werden, wird f"ur manche 
Matrix-Elemente eine andere Darstellung des Photonenpropagators gew"ahlt.
Bis auf ein einziges Matrix-Element wird jedoch im Ortsraum gerehnet, und 
wir haben als Ansatz f"ur den Photonenpropagator
\begin{equation}
G(\vec{r}_1, \vec{r}_2, E) = \sum_{lm} g_l(r_1,r_2,\nu) 
 \, Y_{lm}(\theta_1, \phi_1) \, Y^{*}_{lm}(\theta_2, \phi_2)
\end{equation}  
mit noch zu bestimmenden Funktionen $g_l(r_1,r_2,\nu)$. Die Ausdr"ucke 
f"ur $g_l(r_1,r_2,\nu)$ beinhalten noch eine Summation "uber 
Laguerre-Polynome $k$-ter Stufe von $k=0$ bis $\infty$ (siehe Abschnitt 
\ref{lepmathe}). Die Berechnung von $P_{\rm allg.}$ erfolgt nun in folgenden 
Schritten:

\begin{enumerate}

\item Ausf"uhren der Winkelintegration. Da die $P$-Wellenfunktionen den 
Drehimpuls $l=1$ haben und durch die Wirkung von $p^i$ in Komponenten 
f"ur $l=0$ und $l=2$ aufgespalten werden, ist die Winkelintegration 
bez"uglich der Kugelfunktionen relativ einfach durchzuf"uhren. 
Einzelheiten dazu finden sich im Abschnitt \ref{lepmathe}, in dem auf die
mathematischen Grundlagen der Auswertung von $P$-Matrix-Elementen 
eingegangen wird.

\item Ausf"uhren der Integration des Radialteils. Die verbleibende 
Integration bez"uglich $r_1$ und $r_2$ wird mit Formeln erledigt, die 
ebenfalls in Abschnitt \ref{lepmathe} diskutiert werden.

\item Summation "uber $k$. Diese Summation wird mit Formeln erledigt, die 
zum Teil durch computergest"tzte Rechnungen hergeleitet worden sind. Alle 
so nicht in der Literatur verzeichneten und eigens hergeleiteten 
Summenformeln wurden intensiv an Beispielen gepr"uft.

\item Zusammenhangsformeln zur Vereinfachung der hypergeometrischen 
Funktionen. Durch die Anwendung von Gau"sschen Zusammenhangsformeln der 
hypergeometrischen Funktionen werden diese vereinfacht, und das Ergebnis 
wird in eine ``lesbare'' Form gebracht.

Durch Integration nach $t$ gem"a"s Formel \ref{FausP} ergibt sich dann 
der gesuchte $F$-Faktor. Diese Integration erfolgt in folgenden Schritten:
\begin{enumerate}
\item Die hypergeometrischen Funktionen werden um die Stelle $\omega = 0$ 
($t=1$) entwickelt, um einen bei $t=1$ konvergenten Ausdruck zu erhalten. 
Der entstehende Ausdruck (der ``divergente Anteil'') enth"alt alle 
Divergenzen f"ur $t \to 0$ ($\omega \to \infty$). Der verbleibende 
(``nicht-divergente'') Ausdruck
ist bei $t=1$ und $t=0$ konvergent und liefert einen weiteren Beitrag zum
$F$-Faktor. Da dieser Ausdruck hypergeometrische Funktionen enth"alt, 
wird er numerisch integriert (es handelt sich um eine einfache 
eindimensionale Integration). 
\item Integration des divergenten Anteils mit $Mathematica$-eigenen 
Routinen von $t=t_\ep$ bis $t=1$ (mit $t_\ep$ gegeben durch \ref{teps}). 
\item Symbolische Entwicklung des Resulats der Integration des 
divergenten Anteils in $\alpha$ bis $\alpha^2$, danach in $\ep$ bis
$\ep^0$. 
\item Berechnung des nicht-divergenten numerischen Anteils  
als einfache eindimensionale numerische Integration (Gau"ssche Quadratur).
\item Addition des divergenten und des 
nicht-divergenten Anteils ergibt $F$.
\end{enumerate}
\end{enumerate}
Da die letztgenannten Schritte zur Berechnung von $F$ mathematisch nicht 
besonders anspruchsvoll sind, werden sie in dieser Arbeit nicht n"aher erl"autert.
Wie bereits im Hochenergie-Anteil ersetzen wir in den Zwischenergebnissen der 
Einfachheit halber den Faktor $Z \alpha$ durch $\alpha$, um ihn dann im 
Endergebnis voll auszuschreiben.

%
%

\subsection{Mathematische Vorbereitungen}
\label{lepmathe}

\subsubsection{Verschiedene Darstellungen des Propagators}

F"ur unsere Rechnungen brauchen wir h"aufig den Propagator
\mathanf
\label{GofE}
G(E) = \frac{1}{H_S - E}
\mathend
der Schr"odinger-Gleichung. Im Ortsraum lautet die Gl. \ref{GofE} 
gerade
\begin{equation}
(H_S - E) \, G(\vec{r}_1, \vec{r}_2, E) = \delta(\vec{r}_1 - \vec{r}_2). 
\end{equation}
Man kann diese Gleichung l"osen mit dem Ansatz (siehe \cite{swainsondrake})
\begin{equation}
G(\vec{r}_1, \vec{r}_2, E) = \sum_{lm} g_l(r_1,r_2,\nu) 
  \, Y_{lm}(\theta_1, \phi_1) \, Y^{*}_{lm}(\theta_2, \phi_2)
\end{equation}  
mit $\nu$ als ``Energie-Parameter'' gem"a"s 
\mathanf
\nu^2\,  a^2 = - \frac{1}{2 \,m \,E},
\mathend
wobei $a = a_{Bohr} = 1/(\alpha m)$ der Bohrsche Radius ist.
Setzt man nun f"ur unsere Arbeit
\mathanf
E = E_{\phi} - \omega,
\mathend
so folgt nach einer kurzen Rechnung
\begin{equation}
\nu = 2 \frac{\alpha m / 2}{\sqrt{2 m (E_{\phi} - \omega)}} = 2 t,
\end{equation}
wobei der $t$-Parameter gerade der in Abschnitt \ref{allglep} 
eingef"uhrte Integrationsparameter ist

F"ur die Funktion $g_l(r_1,r_2,E)$ (den Radialteil des 
Schr"odinger-Coulomb-Propagators) werden in dieser Arbeit zwei 
Darstellungen benutzt. Zun"achst wollen wir uns der Darstellung
\eqnanf
\label{scpgLaguerre}
g_l(r_1, r_2, \nu) &=& 2 m \, \left( \frac{2}{a \nu} \right)^{2 l + 1}
\left( r_1 r_2 \right)^l \, \exp\left( - \frac{r_1 + r_2}{a \nu} \right) \times 
\schwups
& & \times \sum_{k=0}^{\infty} \frac{k!}{(2 l + 1 + k)! (l + 1 + k - \nu)}
\, L_k^{2 l + 1}\left(\frac{2 r_1}{a \nu} \right) 
\, L_k^{2 l + 1}\left(\frac{2 r_2}{a \nu} \right) \ok
\eqnend
mit $L_a^b$ als dem assoziierten Laguerre-Polynom der Stufe $(a,b)$
zuwenden.

Den Radialteil des Schr"odinger-Coulomb-Propagators kann man auch
vermittels Whittaker-Funktionen darstellen.
\begin{equation}
\label{scpgWhittaker}
g_l(r_1, r_2, \nu) = m \, a \, \nu \, \frac{\Gamma(l + 1 - \nu)}{(2 l + 1)!}
  \frac{1}{r_1 r_2} \, M_{\nu,l+1/2}\left(\frac{2 r_{\leq}}{a \nu}\right)
  \, W_{\nu,l+1/2}\left(\frac{2 r_{\geq}}{a \nu}\right)
\end{equation}
F"ur zwei der Matrix-Elemente ben"otigen wir noch die Schwingersche
Darstellung des Propagators im Impulsraum
\eqnanf
G(\vec{p},\vec{p}\,',E_{\phi} - \omega) & = & 4 \pi m X^3 \left( 
\frac{i e^{i\pi\tau}}{2 \sin \pi \tau}
\right) \int_{1}^{0^+} d\rho
\rho^{-\tau} \frac{\del}{\del \rho} \schwups
& & \times \frac{1 - \rho^2}{\rho} \frac{1}{\Bigl[X^2 (\vec{p}-\vec{p}\,')^2 + 
(p^2 + X^2)(p'^2 + X^2) \frac{(1-\rho)^2}{4\rho} \Bigr]^2} \ok
\eqnend
mit $X=\sqrt{-2m (E_{\phi} - \omega)}$ und $\tau = m \alpha / X = 2 t$.

\subsubsection{Winkelintegration}

Da die drei $P$-Wellenfunktion die Drehimpuls-Quantenzahl $l=1$ haben, 
transformieren sie sich unter Drehungen wie ein Vektor (siehe zum 
Beispiel \cite{landau}, Seite 425 ff.). Man kann daher $P_x$-, $P_y$ und 
$P_z$ Orbitale identifizieren mit
\begin{equation}
\phi^l (\vec{r}) = r^l {\tilde \phi}(r)
\end{equation}
wobei $r^1 = x$, $r^2 = y$ und $r^3 = z$ (siehe Konventionen). Es ist 
\begin{equation}
{\tilde \phi}(r) =  
{{{{\alpha}^{{5\over 2}}}\,{m^{{5\over 2}}}}\over 
   {4\,\,{\sqrt{2\,\pi }}}} {e^{-{{\alpha\,m\,r}\over 2}}}
\end{equation}
eine Funktion, die nur von $r$ abh"angt. Die zu berechnenden 
Matrix-Elemente ergeben sich dann aus der Mittelung "uber die drei 
verschiedenen $P$-Orbitale. 

Nun folgt eine generelle Eigenschaft von Funktionen des Typs $p^i r^j f(r)$.
Aus dem Additionsgesetz f"ur Drehimpulse bzw. ganz explizit aus der Formel
\begin{equation}
p^i r^j f(r) = -i \frac{1}{3} \delta^{ij} 
  \left[\left(3 f(r) + r f'(r)\right) \right] - i 
  \left[ \hat{r}^i \hat{r}^j - \frac{1}{3} \delta^{ij} \right] 
  \left( r f'(r) \right)
\end{equation}
erkennt man, da"s die Funktionen $p^i r^j f(r)$ aufspalten in einen 
Anteil mit $l=0$ und einen Anteil mit $l=2$. Der $l=0$-Anteil ist gerade  
\mathanf
-i\frac{1}{3} \delta^{ij} \left[(3 f(r) + r f'(r)) \right],
\mathend
und der $l=2$-Anteil ergibt sich aus
\mathanf
-i\left[\hat{r}^i \hat{r}^j - \frac{1}{3} \delta^{ij} \right] 
\left(r f'(r) \right).
\mathend
Au"serdem haben wir folgende generelle Eigenschaft der Kugelfunktionen. 
F"ur zwei Funktionen $h_1(\theta,\phi)$ und $h_2(\theta,\phi)$ mit
\begin{equation}
h_i(r,\theta,\phi) = \sum_{l,m} c_{l,m}(r) \, Y_{l,m}(\theta,\phi)
\end{equation}
mit der $l = l'$-Komponente
\begin{equation}
h_{i,l=l'}(\theta,\phi) = \sum_m c_{l',m}(r) \, Y_{l',m}(\theta,\phi)
\end{equation} 
(keine Summation "uber $l'$) gilt 
\eqnanf
\int d \Omega_1 \int d \Omega_2 \, h_1^{*}(r_1, \theta_1, \phi_1) \,
  \sum_m Y_{l',m}(\theta_1,\phi_1) \, Y_{l',m}^{*}(\theta_2,\phi_2) 
     \, h_2^{*}(r, \theta_2, \phi_2) = & &  \schwups
\int d \Omega  \, h^{*}_{1,l=l'}(r_1, \theta , \phi ) \,
  h_{2,l=l'}(r_2, \theta , \phi ). & & \ok
\eqnend
Daher gilt mit 
\begin{equation}
f_1(r) = {\cal O}_1 {\tilde \phi}(r) 
\end{equation}
und 
\begin{equation}
f_2(r) = {\cal O}_2 {\tilde \phi}(r) 
\end{equation}
die folgende Umformung:
\eqnanf
P_{\rm allg.}(\omega) &=&  \bra{\phi} \, {\cal O}_1 \, p^i \,  
\frac{1}{H_S - (E_S - \omega)} \, p^i \, {\cal O}_2 \ket{\phi} \schwups
& = & \frac{1}{3} \bra{\phi^l} \, {\cal O}_1 \, p^i \,  
\frac{1}{H_S - (E_S - \omega)} \, p^i \, {\cal O}_2 \, \ket{\phi^l } \wups
& = & \frac{1}{3} \int d r_1 \, d r_2 \, r_1^2 \, r_2^2 \, d \Omega_1 \, d \Omega_2 \,
  \left( (p^i \, r^l \, f_1)(r_1,\theta_1,\phi_1) \right)^* \,
    \sum_{lm} g_l(r_1,r_2,\nu) \times \wups
& & \times Y_{lm}(\theta_1, \phi_1) \, Y^{*}_{lm}(\theta_2, \phi_2) \,
      \left( (p^i \,  r^l \, f_2)(r_2,\theta_2,\phi_2) \right) \wups
& = & \frac{1}{3} \int d r_1 \, d r_2 \, r_1^2 \, r_2^2 \, d \Omega \bigg(
  \left( (p^i r^l f_1)(r_1,\theta,\phi) \right)_{l=0}^* \,
     g_0(r_1,r_2,\nu) \,
        \left( (p^i r^l f_2)(r_2,\theta,\phi) \right)_{l=0} \wups
& & + \left( (p^i r^l f_1)(r_1,\theta,\phi) \right)_{l=2}^* \,
     g_2(r_1,r_2,\nu) \,
        \left( (p^i r^l f_2)(r_2,\theta,\phi) \right)_{l=2} \bigg) \wups         
& = &  \int d r_1 \, d r_2 \, r_1^2 \, r_2^2 \,
  \bigg( \frac{4 \pi}{9}  \,
  (3 f_1(r_1) + r_1 f_1'(r_1)) \,
    g_0(r_1,r_2,\nu) \,
      (3 f_2(r_2) + r_2 f_2'(r_2)) +  \wups
& &  \frac{8 \pi}{9} \,
       (r_1 f_1'(r_1)) \, g_2(r_1,r_2,\nu) \, (r_2 f_2'(r_2)) \bigg). \ok
\eqnend   
Die Winkelintegration ist somit ausgef"uhrt, und es verbleibt die 
Integration bez"uglich $r_1$ und $r_2$.

%
%

\subsubsection{$r_1$ und $r_2$-Integration}

Wir wollen uns hier nur mit den Matrix-Elementen besch"aftigen, die mit 
der Darstellung des Schr"odigner-Coulomb Propagators durch 
Laguerre-Polynome gerechnet sind. Nach der Winkelintegration 
sind folgende Formeln f"ur die Berechnung der verbleibenden Integrale 
bez"uglich $r_1$ und $r_2$ hilfreich. Zun"achst ist
\begin{equation}
\int_0^\infty dr \, r^\alpha \ln(r) \exp(-\lambda r) =  
   {\lambda^{-1 - \alpha }}\,\Gamma(\alpha + 1)\,
   \left( \Psi(\alpha + 1) - \ln (\lambda )  \right)
\end{equation}
mit der logarithmischen Ableitung der $\Gamma$-Funktion
\begin{equation}
\Psi(x) = \frac{\Gamma'(x)}{\Gamma(x)}.
\end{equation}
Eine weitere wichtige Formel ist
\eqnanf
& & \int_0^{\infty} dr \, \exp(-\lambda r) \,  r^{\gamma} \, L_n^{\mu}(r) = 
\schwups
& & \frac{\lambda^{-1 - \gamma }\,
      {_2}F_1(-n, \gamma + 1, \mu + 1,{1\over \lambda})\,
      \Gamma(\gamma + 1)\,\Gamma(\mu  + n + 1)}
    {n!\,\Gamma(\mu + 1)}. \ok
\eqnend    
F"ur die Ausf"uhrung der $r_1$ und $r_2$-Integrationen bei Verwendung der 
Whittaker-Funktionen existieren ebenfalls Formeln \cite{buchholz}.

%
%

\subsubsection{Summation "uber $k$}

Im Schr"odinger-Coulomb-Propagator tritt bei Darstellung im Ortsraum mit 
Laguerre-Polynomen (siehe Gl. \ref{scpgLaguerre}) die Summe
"uber den Summationsindex $k$ auf. Es treten dabei zum Beispiel
folgende Summen auf:
\eqnanf 
\label{standardsums}
\sum_{k=0}^{\infty} \xi^k &=& 1 / (1 - \xi ), \schwups  
\sum_{k=0}^{\infty} k^n \xi^k &=& \Phi(\xi^k,-n,0), \schwups
\sum_{k=0}^{\infty} \frac{\xi^k}{a + k} &=& 
  \frac{1}{a} \, {_2}F_1(1, a, a + 1, \xi). \glok
\eqnend  
Die Lerchsche $\Phi$-Funktion ist definiert als
\begin{equation}
\Phi(z, \xi, a) = \sum_{k=0}^{\infty} 
  \frac{z^k}{\left( a + k \right)^{\xi}}.
\end{equation}
Die hypergeometrische Funktion ${_2}F_1$ ist erkl"art als
\begin{equation}
{_2}F_1(a,b,c,z) = \sum_{j=0}^{\infty} \frac{(a)_j (b)_j}{(c)_j}
  \frac{z^k}{k!}
\end{equation}
mit dem Pochhammer-Symbol
\mathanf
(a)_j = \frac{\Gamma(a + j)}{\Gamma(a)}.
\mathend
Neben diesen recht einfachen Summationen treten in den Berechnungen auch
kompliziertere Summen suf. Wir haben zum Beispiel die folgende Formel
\begin{equation}
\label{daschauher}
\sum_{k=0}^{\infty}
\frac{1}{k!} \,\Gamma(k + \lambda ) \, \xi^k \, {_2}F_1(-k,b,c,z) =
\frac{1}{\left( 1 - \xi \right)^{\lambda}} \,\Gamma(\lambda ) \, 
  {{_2}F_1(\lambda ,b,c,-{{\xi\,z}\over {1 - \xi}})}
\end{equation}      
Diese und alle anderen Formeln, die mit der Theorie der 
hypergeometrischen Funktionen zusammenh"angen, finden sich in 
\cite{bateman} und \cite{buchholz}. Mit $\lambda = 1$ folgt aus der
Formel \ref{daschauher}
\begin{equation}
\label{sum1}
\sum_{k=0}^{\infty}
{\xi^k}\,{_2}F_1(-k,b,c,z) =
  {{{_2}F_1(1,b,c,-{{\xi\,z}\over {1 - \xi}})}\over {1 - \xi}}.
\end{equation}
Mit der Beziehung
\begin{equation}
\label{nhoch}
k^{n+1} \xi^k = \xi \frac{\del}{\del \xi} \left( k^n \xi^k \right)
\end{equation}    
lassen sich aus der Formel \ref{sum1} auch Formeln f"ur Summen der Form
\begin{equation}
\label{gestaltn}
\sum_{k=0}^{\infty} k^n {\xi^k}\,{_2}F_1(-k,b,c,z) 
\end{equation}
aufstellen. 
Bei der Anwendung von Gl. \ref{nhoch} auf Gl. \ref{sum1} treten auf der 
rechten Seite Ableitungen nach dem Argument der hypergeometrischen Funktion
auf. Diese Ableitungen lassen sich mit der Formel
\begin{equation}
{_2}F_1(1 + a,b,c,z) = {_2}F_1(a,b,c,z) + 
   {{z\, \del_4 \left( {_2}F_1 \right) (a,b,c,z)}\over {{a^2}}}
\end{equation}
beseitigen. Durch computergest"utzte Rechnungen wurden so Formeln f"ur die
Summen der Gestalt (\ref{gestaltn}) f"ur $n = 1 \dots 4$ hergeleitet, da 
diese Summen in den sp"ateren Rechnungen vorkommen. Es erfolgt hier nur eine Angabe des
Ergebnisses f"ur $n=4$:
\eqnanf
\label{k4}
\sum_{k=0}^{\infty} k^4 {\xi^k}\,{_2}F_1(-k,b,c,z) &=&
{{{_2}F_1(1,b,c,-{{\xi\,z}\over {1 - \xi}})}\over {1 - \xi}} - 
  {{15\,{_2}F_1(2,b,c,-{{\xi\,z}\over {1 - \xi}})}\over 
    {{{\left( -1 + \xi \right) }^2}}} \wups
& &  + {{50\,{_2}F_1(3,b,c,-{{\xi\,z}\over {1 - \xi}})}\over 
         {{{\left( 1 - \xi \right) }^3}}} - 
       {{60\,{_2}F_1(4,b,c,-{{\xi\,z}\over {1 - \xi}})}\over 
          {{{\left( -1 + \xi \right) }^4}}} \wups
& &   + {{24\,{_2}F_1(5,b,c,-{{\xi\,z}\over {1 - \xi}})}\over 
          {{{\left( 1 - \xi \right) }^5}}} \glok
\eqnend
mit der Eulerschen Konstanten $\gamma = 0.577216\dots$.
Mit Hilfe dieser Formeln ist es m"oglich, Summen wie
\mathanf
\sum_{k=0}^{\infty} k^n {\xi^k}\,\del_2 \left( {_2}F_1 \right)(-k,b,c,z) 
\mathend
auszuwerten. Es wird dabei zun"achst die Ableitung nach dem zweiten 
Parameter der hypergeometrischen Funktion durch den Ausdruck f"ur den
entsprechenden Differenzen-Quotienten ersetzt,
\mathanf
\del_2 \left( {_2}F_1 \right)(-k,b,c,z) \to 
  \frac{{_2}F_1 (-k,b + \ep ,c,z) - {_2}F_1 (-k,b,c,z)}{\ep},
\mathend
mit einem $\epsilon$, welches wir sp"ater wieder gegen Null schicken. 
Nach der obigen Ersetzung lassen sich die hergeleiteten Summations-Formeln
anwenden, und im Ergebnis k"onnen, wie im n"achsten Abschnitt besprochen 
wird, die Parameter der hypergeometrischen Funktion durch 
Zusammenhangsformeln geeignet ver"andert werden. Zuletzt werden Regeln 
f"ur die explizite Form von hypergeometrischen Funktionen, die 
bereits im Programm $Mathematica$ implementiert sind, verwandt, um die 
Diffenrentiation ("Ubergang zum Differentialquotienten) ausf"uhren zu 
k"onnen. 

F"ur die folgenden Summen konnte keine geschlossenen 
Ausdr"ucke in Termen hypergeometrische Funktionen gefunden werden:
\mathanf
\sum_{k=0}^{\infty} \frac{\xi^k}{a+k}\,\del_2 
  \left( {_2}F_1 \right)(-k,b,c,z).
\mathend
Diese Summen werden numerisch behandelt.
Um Singularit"aten bei $t=1$ und $t = 0$ auszuschlie"sen, summieren wir
nicht von $k=0$ aufw"arts und multiplizieren mit $t^2$. Wir definieren die 
``spezielle Funktion''
\begin{equation}
{\tilde F}_{abc} (t) = t^2 \,
  \sum_{k=2}^{\infty} \frac{\left( \frac{t-1}{t+1} \right)^k}{a-2t+k}\,
  \del_2 \left( {_2}F_1 \right)(-k,b,c,\frac{2}{1+t}).
\end{equation}
In unserer Rechnung kommt diese Funktion in der Gestalt
\begin{equation}
{\tilde F}_{122} (t) =
  t^2 \, \sum_{k=0}^{\infty} \frac{\left( \frac{t-1}{t+1} \right)^k}{1-2t+k}\,
  \del_2 \left( {_2}F_1 \right)(-k,2,2,\frac{2}{1+t})
\end{equation}
und   
\begin{equation}
{\tilde F}_{366} (t) =
  t^2 \, \sum_{k=0}^{\infty} \frac{\left( \frac{t-1}{t+1} \right)^k}{3-2t+k}\,
  \del_2 \left( {_2}F_1 \right)(-k,6,6,\frac{2}{1+t})
\end{equation}
vor. Man kann zeigen, da"s ${\tilde F}_{122}(t)$ und 
${\tilde F}_{366} (t)$ f"ur $t \to 0$ gegen Null und f"ur 
$t \to 1$ schneller als $(1-t)^2$ gegen Null gehen.
Da Integrale mit diesen Funktionen numerisch behandelt werden
und daf"ur eine Gausssche Quadratur verwandt wird, die bei glatten 
Funktionen ohne Singularit"aten schnell konvergiert, ist dieses Verhalten 
sehr hilfreich. 

Weitere Summationen, die bei der Berechnung der $P$-Matrixelemente 
auftreten, involvieren die logarithmische Ableitung der $\Gamma$-Funktion,
\begin{equation}
\Psi(x) = \frac{\Gamma'(x)}{\Gamma(x)}
\end{equation}
Durch zum Teil computergest"utze Rechnungen lassen sich auch f"ur diese 
Summationen Formeln angeben. Zwei Beispiele seien hier erw"ahnt:
\begin{equation}
\sum_{k=0}^{\infty} \xi^m \Psi(2 + k) = 
  {{\gamma\,\xi  + \ln (1 - \xi )}\over 
     {\left( -1 + \xi  \right) \,\xi }}
\end{equation}   
wiederum mit $\gamma$ als der Eulerschen Konstanten und
\eqnanf
\sum_{k=0}^{\infty} k^4 \xi^m \Psi(2 + k) &=&  
{{1 - 6\,\xi  - 13\,{{\xi }^2} - 32\,{{\xi }^3}}\over 
    {{{\left( -1 + \xi  \right) }^5}}} + \schwups
& & + {{\gamma\,\xi \,\left( 1 + \xi  \right) \,
      \left( 1 + 10\,\xi  + {{\xi }^2} \right) }\over 
    {{{\left( -1 + \xi  \right) }^5}}} + \wups
& & + {{\left( 1 - 5\,\xi  + 11\,{{\xi }^2} + {{\xi }^3} + 
        16\,{{\xi }^4} \right) \,\ln (1 - \xi )}\over 
    {{{\left( -1 + \xi  \right) }^5}\,\xi }}. \ok
\eqnend    
Es gibt jedoch auch f"ur diejenigen Summen, 
welche die $\Psi$-Funktion beinhalten,
bestimmte F"alle, f"ur die keine L"osung in geschlossener 
Form gefunden werden konnte. Es sind Funktionen des Typs:
\begin{equation}
{\tilde \Psi}_{ab}(t) = 
  t^2 \sum_{k=2}^{\infty} \frac{\left( \frac{1-t}{1+t} \right)^{2 k}}{b - 
  2 t + k} \Psi(a + k).
\end{equation}
Diese Funktionen kommen in unserer Rechnung in der Gestalt 
\begin{equation}
{\tilde \Psi}_{21}(t) = 
  t^2 \sum_{k=2}^{\infty} \frac{\left( \frac{1-t}{1+t} \right)^{2 k}}{1 - 
  2 t + k} \Psi(2 + k)
\end{equation}
und
\begin{equation}
{\tilde \Psi}_{63}(t) = 
  t^2 \sum_{k=2}^{\infty} \frac{\left( \frac{1-t}{1+t} \right)^{2 k}}{3 - 
  2 t + k} \Psi(6 + k)
\end{equation}
vor. Auch diese Summen werden numerisch behandelt. 
Zum Schlu"s definieren wir noch die folgenden ``spezielle Funktionen'', die 
mit der Lerchschen $\Phi$-Funktion zusammenh"angen,
\begin{equation}
{\tilde \Phi}_{ab}(t) =  
  t^2 \sum_{k=2}^{\infty} 
    \frac{\left( \frac{1-t}{1+t} \right)^{2 k}}{\left( b - 
       2 t + k \right)^a},
\end{equation}
die in dieser Arbeit als 
\begin{equation}
{\tilde \Phi}_{21}(t) =  
  t^2 \sum_{k=2}^{\infty} 
     \frac{\left( \frac{1-t}{1+t} \right)^{2 k}}{\left( 1 - 
       2 t + k \right)^2}
\end{equation}
und
\begin{equation}
{\tilde \Phi}_{23}(t) =  
  t^2 \sum_{k=2}^{\infty} 
    \frac{\left( \frac{1-t}{1+t} \right)^{2 k}}{\left( 3 - 
       2 t + k \right)^2}
\end{equation}
vorkommen. Auch diese Funktionen werden numerisch behandelt. 

\subsubsection{Zusammenhangsformeln}

Die hypergeometrischen Funktionen im Ergebnis \ref{k4} haben hohe Werte f"ur den 
ersten Parameter. Daher erscheint es ratsam, diesen Parmeter durch
die bekannten Gaussschen Zusammenhangsformeln zu verkleinern
(siehe zum Beispiel \cite{bateman}). Auch durch 
die Anwendung der letzten Formel aus den Standard-Summationen 
\ref{standardsums} treten hypergeometrische Funktionen mit unangenehmen 
Werten f"ur die Parameter auf (in diesem Fall sind die $b$- und 
$c$-Parameter zu hoch). Um die Parameter geeignet zu ver"andern, 
verwenden wir einige Zusammenhangsformeln in einer Gestalt, die das 
Herabsetzen der Parameter erm"oglicht. Es sind dies:

\begin{itemize}
\item zum Herabsetzten des zweiten und dritten Parameters mit besonderem Augenmerk 
auf den dritten Parameter
\eqnanf
{_2}F_1(a,b,c,z)& &= {{\left( -1 + c \right) \,
       {_2}F_1(a,-1 + b,-1 + c,z)}\over {-1 + b}} \schwups
& &  + {{\left( b - c \right) \,{_2}F_1(a,-1 + b,c,z)}\over {-1 + b}}. \ok
\eqnend

\item zum Herabsetzten des zweiten und dritten Parameters mit besonderem Augenmerk 
auf den zweiten Parameter

\eqnanf
{_2}F_1(a,b,c,z) &=& {{\left( 1 - c \right) \,
       {_2}F_1(a,-1 + b,-1 + c,z)}\over {\left( 1 + a - c \right) \,z}} 
       \schwups
& &  + {{\left( -1 + c + z - c\,z \right) \,
       {_2}F_1(a,b,-1 + c,z)}\over {\left( 1 + a - c \right) \,z}}. \ok
\eqnend

Diese und die vorige Regel m"ussen bei der Vereinfachung der Ausdr"ucke 
so eingesetzt werden, da"s nicht einer der Parameter zu schnell sinkt, da 
dann eine Vereinfachung im Sinne m"oglichst kleiner Werte f"ur beide 
Parameter nicht mehr m"oglich ist. 

\item zum Herabsetzten des ersten Parameters
\eqnanf
{_2}F_1(a,b,c,z) &=& {{\left( -1 + a - c \right) \,
       {_2}F_1(-2 + a,b,c,z)}\over 
     {\left( -1 + a \right) \,\left( -1 + z \right) }} \schwups
& &  + {{\left( 2 - 2\,a + c - z + a\,z - b\,z \right) \,
       {_2}F_1(-1 + a,b,c,z)}\over 
     {\left( -1 + a \right) \,\left( -1 + z \right) }}. \ok
\eqnend

\item zum Herabsetzten des zweiten Parameters
\eqnanf
{_2}F_1(a,b,c,z) &=& {{\left( -1 + b - c \right) \,
       {_2}F_1(a,-2 + b,c,z)}\over 
     {\left( -1 + b \right) \,\left( -1 + z \right) }} \schwups 
& &   + {{\left( 2 - 2\,b + c - z - a\,z + b\,z \right) \,
       {_2}F_1(a,-1 + b,c,z)}\over 
     {\left( -1 + b \right) \,\left( -1 + z \right) }}. \ok
\eqnend

\item zum Herabsetzten des dritten Parameters
\eqnanf
{_2}F_1(a,b,c,z) &=& {{\left( -2 + c \right) \,\left( -1 + c \right) \,
       \left( -1 + z \right) \,{_2}F_1(a,b,-2 + c,z)}\over 
     {\left( 1 + a - c \right) \,\left( -1 - b + c \right) \,z}} \schwups
& &    + {{\left( -1 + c \right) \,
       \left( -2 + c + 3\,z + a\,z + b\,z - 2\,c\,z \right) \,
       {_2}F_1(a,b,-1 + c,z)}\over 
     {\left( 1 + a - c \right) \,\left( -1 - b + c \right) \,z}}. \ok
\eqnend
\end{itemize}

\subsubsection{Abk"urzungen}

Da die folgenden Ausdr"ucke in dieser Arbeit sehr h"aufig vorkommen, 
wollen wir folgende Abk"urzungen vereinbaren:
\begin{equation}
s = \frac{1-t}{t+1}
\end{equation}
sowie 
\begin{equation}
F(t) = {_2}F_1(1, -2 \,t, 1 - 2 \, t, s^2)
\end{equation}
und
\begin{equation} 
F_2(t) = {_2}F_1(1, -2 \,t, 1 - 2 \, t, -s).
\end{equation}
Bei diesen Funktionen ist zur Verbesserung der Konvergenz der dritte 
Parameter gr"o"ser als der zweite .

%
%

\subsection{Der nichtrelativistische Dipol-Anteil}

Der nichtrelativistische Dipol-Anteil ist durch Gleichung \ref{Pnd} 
gegeben:
\mathanf
P_{nd} = \frac{1}{3 m} \bra{\phi} p^i 
\frac{1}{H_S - (E_S - \omega)}  
p^i \ket{\phi}.
\mathend
Zur Kontrolle wurde die Rechnung sowohl im Impuls- als auch im Ortsraum 
ausgef"uhrt. F"ur die Rechnung im Impulsraum wird die 
Fourier-Transformation der $2P$-Wellenfunktion ben"otigt, und es wird die 
Schwingersche Integraldarstellung des Schr"odinger-Coulomb-Propagators 
im Impulsraum verwandt.

F"ur die Rechnung im Ortsraum wird die ``"ubliche'' Darstellung des 
Propagators im Ortsraum mit Laguerre-Polynomen verwendet (siehe Abschnitt
\ref{lepmathe}). Die Summation 
bez"uglich $k$ wird ausgef"uhrt, und die entstehenden hypergeometrischen 
Funktionen werden mit Hilfe der Zusammenhangsformeln vereinfacht.

Au"serdem wird zum dimensionslosen Parameter $t$ [siehe Gl. (\ref{deft})]
"ubergegangen. Das Ergebnis f"ur $P_{nd}(t)$ lautet folgenderma"sen:
\eqnanf
P_{nd}(t) &=& {{256\,F(t)\,{t^7}\,\left( -3 + 11\,{t^2} \right) }\over 
    {9\,{{\left( 1 - t \right) }^5}\,{{\left( 1 + t \right) }^5}}} + 
    \schwups
& &  {{2\,{t^2}\,\left( -3 + 6\,t + 3\,{t^2} - 12\,{t^3} - 29\,{t^4} - 
       122\,{t^5} +  413\,{t^6} \right) }\over 
    {9\,{{\left( -1 + t \right) }^5}\,{{\left( 1 + t \right) }^3}}}. \ok
\eqnend
Die Integration bez"uglich $t$ mit der Formel \ref{FausP} liefert 
\begin{equation}
F_{nd} = 
  0.0400223 + {{2\,{{\alpha}^2}}\over {9\,\ep}}. 
\end{equation}  

%
%

\subsection{Korrekturen zum Strom}

\subsubsection{Der nichtrelativistische Quadrupol $F_{nq}$}

Im nichtrelativistischen Quadrupol werden diejenigen Korrekturen zum 
Strom zusammengefa"st, die sich aus den 
\mathanf
y_{1,FW}^j = \frac{p^j}{m} (i \scp{k}{r}) + \dots
\mathend
und  
\mathanf
y_{2,FW}^j = -\frac{1}{2} \frac{p^i}{m} (\scp{k}{r})^2 + \dots
\mathend
ergeben. Die Rechnung erfolgt im Impulsraum (eine vergleichbare Rechnung  
im Impulsraum f"ur $S$-Zust"ande
findet sich zum Beispiel in \cite{krpdiss}, und die Rechnung wird daher hier nicht 
n"aher beschrieben). Das Ergebnis f"ur $P_{nq}(t)$ 
lautet:
\eqnanf
P_{nq}(t) &=&
  {{64\,{{\alpha}^2}\,F(t)\,{t^5}\,\left( 25 - 106\,{t^2} + 29\,{t^4} \right) }\over
      {45\,{{\left( 1 - t \right) }^5}\,{{\left( 1 + t \right) }^5}}} + 
      \schwups
& &  {{\alpha}^2}\,\bigg( 15 - 30\,t - 58\,{t^2} + 146\,{t^3} + 812\,{t^4} + 
        4630\,{t^5} - 15174\,{t^6}  - \wups
& &   - 1418\,{t^7} + 4421\,{t^8} \bigg)  \bigg/ 
    \bigg( 360\,{{\left( -1 + t \right) }^5}\,{{\left( 1 + t \right) }^3} 
    \bigg). \ok
\eqnend
Nach der Integration bez"uglich $t$ ergibt sich folgender Ausdruck f"ur
$F_{nq}$:
\begin{equation}
F_{nq} =
  -1.20115\,{{\alpha}^2} - {{49\,{{\alpha}^2}\,\ln (\alpha)}\over {45}} + 
    {{49\,{{\alpha}^2}\,\ln (\ep)}\over {90}}.
\end{equation}

%
%

\subsubsection{Korrektur aufgrund des $p^i \psq$-Terms}

Die $p^i \vec{p}^2$-Korrektur kommt aufgrund des entsprechenden Terms 
(siehe Gl. \ref{y0}), 
\mathanf
y_{0,FW}^j = \dots - \frac{1}{2 m^3} p^j \psq + \dots, 
\mathend
der bei der FW-Transformation der reinen $\alpha^i$-Matrix entsteht, zustande. 
Der Ausdruck f"ur den $P$-Term lautet:
\eqnanf
P_{p^i p^2} &=& 
  2 \times \frac{m}{2} \delta^{T,ij} \bra{\phi} \left( -\frac{1}{2 m^3} p^i 
  \psq \right) \frac{1}{H_S - (E_S - \omega)} \frac{p^j}{m} 
  \ket{\phi} \schwups
&=& \frac{2}{3 m^3} \bra{\phi} \left( -\frac{1}{2} p^i 
  \psq \right) \frac{1}{H_S - (E_S - \omega)} p^i
  \ket{\phi}. \ok
\eqnend
Die Rechnung erfolgt im Ortsraum. Das Ergebnis f"ur $P_{p^i p^2}(t)$ lautet:
\eqnanf
P_{p^i p^2}(t) &=& 
  {{64\,{{\alpha}^2}\,F_2(t)\,{t^5}}\over 
    {9\,{{\left( -1 + t \right) }^3}\,{{\left( 1 + t \right) }^3}}}   
    \schwups
& &   + {{64\,{{\alpha}^2}\,F(t)\,{t^5}\,
      \left( -1 - 11\,{t^2} + 36\,{t^4} \right) }\over 
    {9\,{{\left( 1 - t \right) }^5}\,{{\left( 1 + t \right) }^5}}}  \wups
& & + {{{{\alpha}^2}\,{t^2}\,\left( 7 - 14\,t - 23\,{t^2} - 68\,{t^3} - 103\,{t^4} - 
        302\,{t^5} + 1271\,{t^6} \right) }\over 
    {18\,{{\left( -1 + t \right) }^5}\,{{\left( 1 + t \right) }^3}}} \ok
\eqnend
F"ur den $F$-Faktor wird schlie"slich erhalten:
\begin{equation}
F_{p^i p^2} = 0.618148\,{{\alpha}^2} + {{4\,{{\alpha}^2}\,\ln (\alpha)}\over 9} - 
  {{2\,{{\alpha}^2}\,\ln (\ep)}\over 9}  
\end{equation}

%
%

\subsubsection{Korrektur aufgrund des $\vec{r} \times \vec{\sigma}$-Terms}

Diese Korrektur kommt durch den Term 
\mathanf
y_{0,FW}^j = \dots - \frac{1}{2 m^2} \frac{\alpha}{r^3} \big( \vec{r}
  \times \vec{\sigma} \big)^j + \dots,
\mathend
welcher bei der FW-Transformation der reinen $\alpha^i$-Matrix entsteht,
zustande. Es ist das Matrix-Element
\eqnanf
P_{r \times \sigma} &=& 
  2 \times \frac{m}{2} \delta^{T,ij} \bra{\phi} \left( \frac{1}{2 m^2} 
  \frac{\alpha}{r^3} \big( - \vec{r} \times \vec{\sigma} \big)^i \right) \,
  \frac{1}{H_S - (E_S - \omega)} \, \frac{p^j}{m} \,
  \ket{\phi} \schwups
&=& \frac{2}{3 m^2} \bra{\phi} \left( \frac{1}{2}
  \frac{\alpha}{r^3} \big( - \vec{r} \times \vec{\sigma} \big)^i \right) \,
    \frac{1}{H_S - (E_S - \omega)} \, p^i \ket{\phi} \ok
\eqnend
auszuwerten. Mit einer Rechnung im Ortsraum erh"alt man f"ur 
$P_{r \times \sigma}(t)$:
\eqnanf
P_{r \times \sigma} &=& 
{{64\,{{\alpha}^2}\,F(t)\,{t^5}\,\left( -1 + 5\,{t^2} \right) }\over 
    {9\,{{\left( -1 + t \right) }^4}\,{{\left( 1 + t \right) }^4}}} + 
    \schwups
& &  {{2\,{{\alpha}^2}\,{t^2}\,\left( -1 + 2\,t + 14\,{t^3} - 47\,{t^4} \right)}\over
      {9\,{{\left( -1 + t \right) }^4}\,{{\left( 1 + t \right) }^2}}}. \ok
\eqnend      
Das Ergebnis f"ur den $F$-Faktor lautet:
\begin{equation}
F_{r \times \sigma} = 0.346690\,{{\alpha}^2}.
\end{equation}

%
%

\subsubsection{Korrektur aufgrund des $\scp{k}{r}$-Terms}

Dieser Beitrag wiederum kommt durch den Term
\mathanf
y_{1,FW}^j = \dots + \frac{1}{2 m} (\scp{k}{r}) 
  \big( \vec{k} \times \vec{\sigma} \big)^j + \dots,
\mathend
welcher bei der FW-Transformation des Ausdrucks $\alpha^i (i \scp{k}{r})^2$-Matrix 
entsteht, zustande. Es ist das Matrix-Element
\eqnanf
P_{k \cdot r} &=& 
  2 \times \frac{m}{2} \delta^{T,ij} \bra{\phi} \left( \frac{1}{2 m} (\scp{k}{r}) \,
  \big( \vec{k} \times \vec{\sigma} \big)^i \right) \, 
  \frac{1}{H_S - (E_S - \omega)} \, \frac{p^j}{m} 
  \ket{\phi} \wups
&=& \frac{2}{3 m} \bra{\phi} \left( \frac{1}{2} (\scp{k}{r}) \,  
  \big( \vec{k} \times \vec{\sigma} \big)^i \right) \,
    \frac{1}{H_S - (E_S - \omega)} \, p^i \ket{\phi} \glok
\eqnend
auszuwerten. Eine l"angere Rechnung ergibt, da"s dieser Beitrag 
gerade dem $-\frac{1}{2}$-fachen von $F_{r \times \sigma}(t)$ 
entspricht. Daher gilt
\begin{equation}
F_{k \cdot r} = -0.173345\,{{\alpha}^2}.
\end{equation}

%
%

\subsection{Korrektur zur Energie}

Bei der Korrektur zur Energie ist lediglich ein Matrix-Element zu 
berechnen, 
\begin{equation}
P_{\delta E} = 
  \frac{1}{3 m} \bra{\phi} p^i 
  \frac{1}{H_S - (E_S - \omega)} \, \delta E \, \frac{1}{H_S - (E_S - \omega)}
  p^i \ket{\phi}
\end{equation}
F"ur den $2P_{1/2}$-Zustand ist 
\begin{equation}
\delta E = - \frac{5}{128} \alpha^4 m.
\end{equation}
F"ur die Berechnung k"onnen wir die folgende Beziehung ausnutzen:
\begin{equation}
\left( \frac{1}{H_S - (E_S - \omega)} \right)^2 = 
  - \frac{\del}{\del \omega} \frac{1}{H_S - (E_S - \omega)}
\end{equation}  
Mit der Beobachtung, da"s sich der Selbstenergie-Beitrag $E_L$ aus dem 
$P$-Term durch Integration gem"a"s Gl. \ref{EausP}
\mathanf
E_L = - \frac{2 \alpha}{\pi \, m} \int_0^{\ep} d \omega \, \omega P(\omega)
\mathend
und der Beobachtung, da"s der $F$-Term direkt proportional zu $E_L$ ist,
l"a"st sich somit der $F$-Term vermittels einer partiellen 
Integration berechnen. Es entsteht ein Randterm und ein weiterer Term, 
der mit den "ublichen Methoden im Ortsraum berechnet wird. Das 
Endergebnis f"ur den $F$-Faktor lautet:

\begin{equation}
F = 0.0400946\,{{\alpha}^2} - {{5\,{{\alpha}^2}\,\ln (\alpha)}\over {48}} + 
  {{5\,{{\alpha}^2}\,\ln (\ep)}\over {96}}.
\end{equation}

%
%
%

\subsection{Korrektur zur Wellenfunktion}

Der Beitrag aufgrund der relativistischen Korrektur zur Wellenfunktion 
$\delta \phi$ ist gem"a"s Gl. \ref{Pdeltaphi} gegeben durch
\begin{equation}
P_{\delta \phi} = \frac{2}{3 m} \bra{\phi} p^i
\frac{1}{H_S - (E_S - \omega)}
p^i \ket{\delta \phi}.  
\end{equation}
Es ist die Differentialgleichung \ref{dgldeltaphi},
\mathanf
(H - \bra{\phi} \delta H \ket{\phi}) \, \ket{\phi} = (E_S - H_S) \, \ket{\delta \phi},
\mathend
zu l"osen. Deren L"osung ist durch die zus"atzlichen Bedingung 
\begin{displaymath}
\langle \phi | \delta \phi \rangle \demand 0
\end{displaymath}
eindeutig bestimmt. Wir 
erhalten folgenden Ausdruck f"ur $\ket{\delta \phi}$:
\eqnanf
\delta \phi(\vec{r}) &=&
  {{53\,{{\alpha}^{{9\over 2}}}\,{m^{{5\over 2}}}\,r}\over 
    {96\,{\sqrt{6}}}} \exp({{{-\alpha\,m\,r}\over 2}}) -  
    {{{{\alpha}^{{9\over 2}}}\,\gamma\,{m^{{5\over 2}}}\,r} \over 
    {4\,{\sqrt{6}}}} {\exp({{{-\alpha\,m\,r}\over 2}})} -\wups
& &  {{{{\alpha}^{{7\over 2}}}\,{m^{{3\over 2}}}}\over 
    {2\,{\sqrt{6}}}} {\exp({{{-\alpha\,m\,r}\over 2}})} +  
    {{{{\alpha}^{{{11}\over 2}}}\,{m^{{7\over 2}}}\,{r^2}}\over 
    {32\,{\sqrt{6}}}} {\exp({{{-\alpha\,m\,r}\over 2}})} - \wups
& & {{{{\alpha}^{{9\over 2}}}\,{m^{{5\over 2}}}\,r\,\ln (\alpha\,m\,r)}\over 
    {4\,{\sqrt{6}}}} {\exp({{{-\alpha\,m\,r}\over 2}})} \wups
&=& {\delta \phi}_1 + {\delta \phi}_2 + {\delta \phi}_3 + {\delta \phi}_4 + 
      {\delta \phi}_5.    
\eqnend
Um die Rechnung bei der praktischen Ausf"uhrung "ubersichtlich zu halten, 
behandeln wir $P_{\delta \phi}$ f"ur jeden der 5 Anteile von $\ket{\delta 
\phi}$ getrennt. Es entstehen also 5 einzelne Terme, $P_{\delta \phi,1}$ 
bis $P_{\delta \phi,5}$.

$F_{\delta \phi,1}$ bis $P_{\delta \phi,4}$ k"onnen im Ortsraum berechnet 
werden. Die jeweiligen $P$-Terme enthalten die Funktionen $F(t)$ und 
$F_2(t)$ und sind in ihrer Struktur den bereits aufgelisteten Ergebnissen 
f"ur die vorherigen Terme "ahnlich. Daher verzichten wir hier auf eine 
Darstellung dieser Ergebnisse und gehen gleich zu den $F$-Faktoren "uber.
Es gilt:
\eqnanf
F_{\delta \phi,1} &=& 0.0883825\,{{\alpha}^2} \schwups
F_{\delta \phi,2} &=& -0.0231015\,{{\alpha}^2} \schwups
F_{\delta \phi,3} &=& 
-0.608142\,{{\alpha}^2} - {{4\,{{\alpha}^2}\,\ln (\alpha)}\over 9} + 
  {{2\,{{\alpha}^2}\,\ln (\ep)}\over 9} \schwups
F_{\delta \phi,4} &=&
-0.0261919\,{{\alpha}^2} - {{{{\alpha}^2}\,\ln (\alpha)}\over {18}} + 
  {{{{\alpha}^2}\,\ln (\ep)}\over {36}} \glok
\eqnend
Aufgrund des Logarithmus im letzten Term, der zu $\ket{\delta \phi}$ 
beitr"agt, wird die Rechnung f"ur diesen Term sehr viel komplexer. Es 
m"ussen andere Summenformeln f"ur die Summation "uber $k$ bem"uht werden, 
und bereits das Ergebnis f"ur den $P$-Term wird um einiges l"anger:
\eqnanf
P_{\delta \phi,5} &=& {{-512\,{{\alpha}^2}\,{\tilde F}_{122}(t)\,{t^6}}\over 
    {27\,{{\left( -1 + t \right) }^2}\,{{\left( 1 + t \right) }^6}}} + 
    \schwups
& & {{4096\,{{\alpha}^2}\,{\tilde F}_{366}(t)\,\left( 1 - 2\,t \right) \,
      \left( -1 + t \right) \,{t^6}\,\left( 1 + 2\,t \right) }\over 
    {27\,{{\left( 1 + t \right) }^{11}}}} + \wups
& & {{16\,{{\alpha}^2}\,F_2(t)\,{t^5}}\over 
    {3\,{{\left( 1 - t \right) }^3}\,{{\left( 1 + t \right) }^3}}} + \wups
& & {{2\,{{\alpha}^2}\,\gamma\,{t^2}\,
      \left( -3 + 6\,t + 3\,{t^2} - 12\,{t^3} - 29\,{t^4} - 122\,{t^5} + 
        413\,{t^6} \right) }\over 
    {9\,{{\left( -1 + t \right) }^5}\,{{\left( 1 + t \right) }^3}}} + \wups
& & {{{{\alpha}^2}\,{t^2}\, p(t)}\over 
    {810\,\left( 2 - t \right) \,{{\left( -1 + t \right) }^5}\,
      {{\left( 1 + t \right) }^{13}}}} + \wups
& &  {{128\,{{\alpha}^2}\,{t^7}\,\left( -3 + 2\,t + 9\,{t^2} \right) \,
      \ln ({{2\,t}\over {1 + t}})}\over 
    {9\,{{\left( 1 - t \right) }^5}\,{{\left( 1 + t \right) }^4}}} +  
     {{16\,{{\alpha}^2}\,F(t)\,{t^5}\, q(t)}\over 
    {405\,{{\left( -1 + t \right) }^5}\,{{\left( 1 + t \right) }^5}}}. \ok
\eqnend
Die Funktionen $p(t)$ und $q(t)$ sind gegeben durch
\eqnanf
p(t) &=& 
       1710 + 12825\,t + 34920\,{t^2} + 25920\,{t^3} - 
        38160\,{t^4} + 391876\,{t^5}  \schwups
& &      + 2892232\,{t^6} + 6664816\,{t^7} + 
        4416444\,{t^8} - 15374274\,{t^9} - 44242888\,{t^{10}} \wups
& &       - 51115488\,{t^{11}} - 37744768\,{t^{12}} - 6464172\,{t^{13}} + 
        4472472\,{t^{14}} \wups
& &     + 4920240\,{t^{15}} + 1591846\,{t^{16}} + 
        186385\,{t^{17}}  \ok
\eqnend
sowie
\eqnanf
q(t) &=& 135 - 7594\,{t^2} + 2160\,\gamma\,{t^2} + 23251\,{t^4} - 
        7920\,\gamma\,{t^4} \schwups
& &        - 2160\,{t^2}\,\ln ({{2\,t}\over {1 + t}}) + 
        7920\,{t^4}\,\ln ({{2\,t}\over {1 + t}}). \ok
\eqnend        
Diejenigen Terme, welche die speziellen Funktionen ${\tilde F}_{122}(t)$ und
${\tilde F}_{366}(t)$ nicht enthalten, k"onnen mit den "ublichen Methoden 
nach $t$ integriert werden. Die restlichen Terme, welche die speziellen 
Funktionen enthalten, werden als numerischer 
Anteil $N_{\delta \phi,5}$ zusammengefa"st und sp"ater mit anderen 
numerischen Termen gemeinsam ausgewertet (siehe Abschnitt \ref{num}). Da 
es sich dabei um eine einfache numerische Integration handelt, kann das 
Ergebnis mit hoher Genauigkeit angegeben werden. Als Endergebnis f"ur
$F_{\delta \phi,5}$ erhalten wir:
\begin{equation} 
F_{\delta \phi,5} = 
-0.182817\,{{\alpha}^2} - {{2\,{{\alpha}^2}\,\ln (\alpha)}\over 9} + 
  {{{{\alpha}^2}\,\ln (\ep)}\over 9} + N_{\delta \phi,5}.
\end{equation}
Z"ahlt man alle Einzelbeitr"age zu $F_{\delta \phi}$ zusammen, so wird
\begin{equation}
F_{\delta \phi} =  
  -0.751870\,{{\alpha}^2} - {{13\,{{\alpha}^2}\,\ln (\alpha)}\over {18}} + 
    {{13\,{{\alpha}^2}\,\ln (\ep)}\over {36}} + N_{\delta \phi,5}.
\end{equation}    
  
%
%

\subsection{Korrekturen zum Hamiltonian}

Aufgrund der Form der relativistischen Korrektur zum Hamiltonian,
\mathanf
\delta H = - \frac{(\psq)^2}{8 \, m^3} + \frac{\pi \alpha}{2 m^2} 
  \delta(\vec{r}) + \frac{\alpha}{4 m^2 r^3} \vec{\sigma} \cdot \vec{L} 
\mathend
k"onnen wir den Beitrag aufgrund von $\delta H$ (siehe Gl. \ref{PdeltaH}),
\begin{equation} 
P_{\delta H}  = -\frac{1}{3 m} \bra{\phi} p^i \,
\frac{1}{H_S - (E_S - \omega)} \, \delta H \, \frac{1}{H_S - (E_S - \omega)} \,
p^i \ket{\phi} 
\end{equation}
aufspalten in einen $P_{\delta}$, $P_{p^4}$ und $P_{L \cdot S}$-Term, 
wobei die Namen den charakteristischen Elementen in den den 3 Summanden
in $\delta H$ entsprechen, das hei"st, es gilt:
\begin{equation} 
P_{\delta}  = -\frac{1}{3 m} \bra{\phi} p^i \,
\frac{1}{H_S - (E_S - \omega)} \, \left( \frac{\pi \alpha}{2 m^2} 
  \delta(\vec{r}) \right) \, \frac{1}{H_S - (E_S - \omega)} \,
p^i \ket{\phi} 
\end{equation}
\begin{equation} 
P_{p^4}  = -\frac{1}{3 m} \bra{\phi} p^i \,
\frac{1}{H_S - (E_S - \omega)} \, \left(  - \frac{(\psq)^2}{8 \, m^3} \right) \,  
  \frac{1}{H_S - (E_S - \omega)} \, p^i \ket{\phi} 
\end{equation}
\begin{equation} 
P_{L \cdot S}  = -\frac{1}{3 m} \bra{\phi} p^i \, 
\frac{1}{H_S - (E_S - \omega)} \, \left(  \frac{\alpha}{4 m^2 r^3} 
  \vec{\sigma} \cdot \vec{L} \right) \, \frac{1}{H_S - (E_S - \omega)} \, 
    p^i \ket{\phi} 
\end{equation}
Die Rechnungen zu diesen 3 Beitr"agen werden durch die Anwesenheit von 2 
Propagatoren in den Matrix-Elementen erheblich erschwert. F"ur 
$P_{\delta}$ nutzt man die Eigenschaft der $\delta$-Distribution aus und 
geht zur Darstellung des Schr"odinger-Coulomb-Propagators vermittels der 
Whittaker-Funktionen "uber. F"ur $P_{p^4}$ und $P_{L \cdot S}$ benutzt 
man folgenden Trick: Man schreibt die Operatoren in den Beitr"agen zu 
$\delta H$ als Kommutatoren anderer Operatoren mit $H_S - (E_s - \omega)$. 
Dadurch k"urzt sich einer der Propagatoren heraus.
 
\subsubsection{Beitrag aufgrund des $\delta(\vec{r})$-Terms in $\delta H$}

Das Matrix-Element

\begin{equation} 
P_{\delta}  = -\frac{1}{3 m} \bra{\phi} p^i
\frac{1}{H_S - (E_S - \omega)} \left( \frac{\pi \alpha}{2 m^2} 
  \delta(\vec{r}) \right) \frac{1}{H_S - (E_S - \omega)}
p^i \ket{\phi} 
\end{equation}

wird im Ortsraum unter Verwendung der Darstellung des 
Schr"odinger-Coulomb-Propagators mit Whittaker-Funktionen berechnet.
Das Ergebnis in Termen von $t$ ist:

\begin{equation} 
P_{\delta}(t) = 
{{-\left( {{\alpha}^2}\,{t^4}\,{{\left( -3 + 4\,t - 8\,F_2(t)\,t + 7\,{t^2}
             \right) }^2} \right) }\over 
   {27\,{{\left( -1 + t \right) }^4}\,{{\left( 1 + t \right) }^4}}}
\end{equation}

Nach der Integration bez"uglich $t$ ergibt sich als Endergebnis f"ur 
$F_{\delta}$

\begin{equation}
F_{\delta} = 0.00165032\,{{\alpha}^2} - {{{{\alpha}^2}\,\ln (\alpha)}\over 6} + 
  {{{{\alpha}^2}\,\ln (\ep)}\over {12}}
\end{equation}  

%
%
  
\subsubsection{Beitrag aufgrund des $p^4$-Terms in $\delta H$}

Dieser Beitrag 
\mathanf 
P_{p^4}  = -\frac{1}{3 m} \bra{\phi} p^i \,
\frac{1}{H_S - (E_S - \omega)} \, \left(  - \frac{(\psq)^2}{8 \, m^3} \right) \, 
  \frac{1}{H_S - (E_S - \omega)} \, p^i \ket{\phi} 
\mathend
oder
\mathanf 
P_{p^4}  = \frac{1}{24 m^4} \bra{\phi} p^i \,
\frac{1}{H_S - (E_S - \omega)} \, \left(  \psq \right)^2 
\, \frac{1}{H_S - (E_S - \omega)} \, p^i \ket{\phi} 
\mathend
wird unter Zuhilfenahme der zun"achst trivialen Beziehung
\mathanf
\vec{p}^2 = 2 m ((H_S - (E_S - \omega)) - V + (E_S - \omega)))
\mathend
berechnet. Diese Beziehung ergibt quadriert:
\eqnanf
\left( \psq \right)^2 &=& 
 4 m^2 ((H_S - (E_S - \omega))^2  + 
 2  (E_S - \omega) (H_S - (E_S - \omega)) \schwups 
& & - V  (H_S - (E_S - \omega))  
 - (H_S - (E_S - \omega)) V + 
 ((E_S - \omega) - V)^2 ). \ok
\eqnend
Setzt man diesen Ausdruck f"ur $p^4$ in das Matrix-Element ein,
so ergeben sich auf nat"urliche Weise 3 Terme, die zum 
$\ep^0$-Koeffizienten beitragen (der durch $(H_S - (E_S - \omega))^2$
erzeugte Beitrag ist zu vernachl"assigen, da er als Integral 
$\int_0^\ep d\omega \, \omega \, P(\omega) \to 0$ geht f"ur $\ep \to 0$).
Man erh"alt
\begin{enumerate}
\item durch den Term $2 \, (E_S - \omega) \, (H_S - (E_S - \omega))$
\begin{equation}
P_{p^4,1} = \frac{1}{6 m^2} 2 (E_S - \omega) 
  \bra{\phi} p^i \frac{1}{H_S - (E_S - \omega)}  p^i \ket{\phi},
\end{equation}
\item durch den Term $V \, (H_S - (E_S - \omega)) + 
 (H_S - (E_S - \omega)) \, V$
\begin{equation}
P_{p^4,2} = -2 \frac{1}{6 m^2}
  \bra{\phi} p^i \frac{1}{H_S - (E_S - \omega)}  V p^i \ket{\phi},
\end{equation}
\item durch den Term $((E_S - \omega) - V)^2$
\begin{equation}
P_{p^4,3} = -2 \frac{1}{6 m^2}
  \bra{\phi} p^i \frac{1}{H_S - (E_S - \omega)} ((E_S - \omega) - V)^2 
    \frac{1}{H_S - (E_S - \omega)} p^i \ket{\phi}.
\end{equation}
\end{enumerate}
Der Term $P_{p^4,3}$ mu"s nun noch weiter behandelt werden.
F"ur die zwei Propagatoren gilt, da $((E_S - \omega) - V)^2$ ein Ausdruck 
der  Ordnung $\alpha^4$ ist, in erster N"aherung
\eqnanf
& & \frac{1}{H_S - (E_S - \omega)} \, ((E_S - \omega) - V)^2 \,  
    \frac{1}{H_S - (E_S - \omega)} = \schwups
& & = \frac{1}{H_S - (E_S - \omega) - ((E_S - \omega) - V)^2} - 
  \frac{1}{H_S - (E_S - \omega)} \ok    
\eqnend
Wenn wir die tempor"are Abk"urzung $\Omega = E_S - \omega$ einf"uhren, so 
entspricht die Ersetzung
\eqnanf
& & H_S(l,\alpha) - \Omega =  H_S - (E_S - \omega) \schwups
& & \to H_S - (E_S - \omega) - ((E_S - \omega) - V)^2 = 
H_S(l',\alpha') - \Omega' \ok 
\eqnend
im Schr"odinger-Hamiltonian und damit im Schr"odinger-Coulomb-Propagator einer 
Verschiebung der Energie um $\Omega \to \Omega'$, der  
Drehimpuls-Quantenzahlen $l \to l'$ und des $\alpha \to \alpha'$ im 
Potential. Damit k"onnen alle diese Korrekturen st"orungstheoretisch 
behandelt werden, und es ergeben sich drei Beitr"age $P_{p^4,3a}$,
$P_{p^4,3b}$ und  $P_{p^4,3c}$, die respektive  $\Omega \to \Omega'$, 
$\alpha \to \alpha'$ und $l \to l'$ entsprechen.   

Die Beitr"age  $P_{p^4,1}$ und $P_{p^4,2}$ enthalten nur einen Propagator 
und sind von ihrer Struktur her einfach. Wir nehmen daher gleich das 
Endergebnis f"ur die $F$-Faktoren vorweg:
\eqnanf
F_{p^4,1} &=& -0.302463\,{{\alpha}^2} - {{4\,{{\alpha}^2}\,\ln (\alpha)}\over 9} + 
  {{2\,{{\alpha}^2}\,\ln (\ep)}\over 9}, \schwups
F_{p^4,2} &=& 
-0.00661108\,{{\alpha}^2} + {{2\,{{\alpha}^2}\,\ln (\alpha)}\over 9} - 
  {{{{\alpha}^2}\,\ln (\ep)}\over 9}. \glok
\eqnend
Im Matrix-Element $P_{p^4,3a}$ sind zwei Propagatoren vorhanden, der 
zwischen den Propagatoren stehende Ausdruck ist jedoch nur 
eine skalare Gr"o"se (eine Zahl) und 
kommutiert mit den Propagatoren. Daher k"onnen wir diesen Beitrag 
nach dem Muster von $P_{\delta E}$ berechnen (partielle Integration, 
Randterm, dann Integration des verbleibenden Beitrags). Das Ergebnis ist:
\begin{equation}
F_{p^4,3a} =  
  0.559288\,{{\alpha}^2} + {{11\,{{\alpha}^2}\,\ln (\alpha)}\over {16}} - 
    {{11\,{{\alpha}^2}\,\ln (\ep)}\over {32}}
\end{equation}
Schwieriger wird die Auswertung der verbleibenden Terme $P_{p^4, 3b}$ und
$P_{p^4, 3c}$. Zun"achst zu $P_{p^4, 3b}$. Bei der Summation bez"uglich 
$k$ treten Terme auf, die der Lerchschen Funktion $\Phi$ entsprechen. 
Das Ergebnis f"ur $P_{p^4, 3b}$ ist:
\eqnanf
P_{p^4, 3b} &=&
{{128\,{{\alpha}^2}\,{ \tilde \Phi }(s^2, 2, - 2t)\,{t^7}\,
      \left( -3 + 11\,{t^2} \right) }\over 
    {9\,{{\left( 1 - t \right) }^5}\,{{\left( 1 + t \right) }^5}}} \schwups
& &  +  {{32\,{{\alpha}^2}\,F(t)\,{t^5}\,\left( 5 - 46\,{t^2} + 57\,{t^4} \right) }\over
      {9\,{{\left( -1 + t \right) }^6}\,{{\left( 1 + t \right) }^6}}} + 
      \wups
& &  + {{{{\alpha}^2}\,{t^2}\,\left( -5 + 10\,t - 11\,{t^2} - 52\,{t^3} + 453\,{t^4} + 
        170\,{t^5} - 821\,{t^6} \right) }\over 
    {18\,{{\left( -1 + t \right) }^6}\,{{\left( 1 + t \right) }^4}}} \ok.
\eqnend
Die Integration bez"uglich $t$ liefert f"ur $P_{p^4, 3b}$:
\eqnanf
F_{p^4, 3b} = 
-0.0376386\,{{\alpha}^2} - {{7\,{{\alpha}^2}\,\ln (\alpha)}\over {12}} + 
  {{7\,{{\alpha}^2}\,\ln (\ep)}\over {24}}
\eqnend
Den verbleibenden Term $P_{p^4, 3c}$ berechnet man mit Hilfe einer 
Variation des $\alpha$-Faktors im 
Schr"odinger-Coulomb-Propagator. Das Ergebnis f"ur den $P$-Term ist recht 
komplex, in seiner Struktur jedoch dem Resultat f"ur $P_{\delta \phi,5}$
"ahnlich und kann hier in verk"urzter Form angegeben werden.
\eqnanf
P_{p^4, 3c} &=& {{1024\,{{\alpha}^2}\,{\tilde F}_{122}(t)\,{t^6}}\over 
    {27\,{{\left( -1 + t \right) }^2}\,{{\left( 1 + t \right) }^6}}} 
    \schwups
& &    + {{8192\,{{\alpha}^2}\,{\tilde F}_{366}(t)\,\left( -1 + t \right) \,{t^6}\,
      \left( -1 + 2\,t \right) \,\left( 1 + 2\,t \right) }\over 
    {135\,{{\left( 1 + t \right) }^{11}}}} \wups
& &  + {{512\,{{\alpha}^2}\,{\tilde \Phi}_{21}(t)\,{t^6}}\over 
      {27\,{{\left( -1 + t \right) }^2}\,{{\left( 1 + t \right) }^6}}} 
      \wups
& &   - {{4096\,{{\alpha}^2}\,{\tilde \Phi}_{23}(t)\,\left( -1 + t \right) \,{t^6}\,
        \left( -1 + 2\,t \right) \,\left( 1 + 2\,t \right) }\over 
           {135\,{{\left( 1 + t \right) }^{11}}}} \wups  
& &   + {{1024\,{{\alpha}^2}\,{\tilde \Psi}_{21}(t)\,{t^6}}\over 
        {27\,{{\left( -1 + t \right) }^2}\,{{\left( 1 + t \right) }^6}}} 
        \wups
& &  +{{8192\,{{\alpha}^2}\,{\tilde \Psi}_{63}(t)\,\left( 1 - 2\,t \right) \,
      \left( -1 + t \right) \,{t^6}\,\left( 1 + 2\,t \right) }\over 
    {135\,{{\left( 1 + t \right) }^{11}}}} \wups
& & +  f(\gamma , F_2(t), F(t), \ln(t)), \ok 
\eqnend
wobei $f(\gamma\, F_2(t), F(t), \ln(t))$  eine Funktion der aufgef"uhrten 
Gr"o"sen ist, deren Struktur den ensprechenden Termen aus $P_{\delta 
\phi, 5}$ "ahnelt und die aufgrund ihrer L"ange hier nicht weiter 
aufgef"uhrt wird. Wichtig ist, da"s im Ergebnis einige spezielle 
Funktionen enthalten sind, die einer getrennten numerischen 
Auswertung zugef"uhrt werden (siehe Abschnitt \ref{num}).

Als Ergebnis f"ur $F_{p^4,3c}$ erhalten wir
\begin{equation}
F_{p^4,3c} = 0.0895706\,{{\alpha}^2} + {{5\,{{\alpha}^2}\,\ln (\alpha)}\over 9} - 
  {{5\,{{\alpha}^2}\,\ln (\ep)}\over {18}} + N_{p^4,3c}.
\end{equation}
mit $N_{p^4,3c}$ als dem noch auszuwertenden numerischen Anteil, der die 
speziellen Funktionen enth"alt. Somit ergibt sich als Endergebnis f"ur 
$F_{p^4}$
\begin{equation}
F_{p^4} = 0.302146\,{{\alpha}^2} + {{7\,{{\alpha}^2}\,\ln (\alpha)}\over {16}} - 
  {{7\,{{\alpha}^2}\,\ln (\ep)}\over {32}} + N_{p^4,3c}.
\end{equation}  

%
%
  
\subsubsection{Beitrag aufgrund des $\scp{L}{S}$-Terms in $\delta H$}

Der Beitrag 
\begin{equation} 
P_{L \cdot S}  = -\frac{1}{3 m} \bra{\phi} p^i \,
\frac{1}{H_S - (E_S - \omega)} \, \left(  \frac{\alpha}{4 m^2 r^3} 
  \vec{\sigma} \cdot \vec{L} \right) \, \frac{1}{H_S - (E_S - \omega)}
    \, p^i \ket{\phi} 
\end{equation}
wird mit der Definition
\begin{equation}
D_r = \frac{1}{r} \frac{\del}{\del r} r
\end{equation}
unter Ausnutzung der Kommutatorbeziehung 
\begin{equation}
[H, D_r] = \frac{\vec{L}^2}{m r^3} - \frac{\alpha}{r^2}
\end{equation}
berechnet. Mit dieser Kommutatorbeziehung ergeben sich drei Beitr"age,
\begin{equation} 
P_{L \cdot S,1}  = -\frac{\alpha}{72 m^2} \bra{\phi} p^i
D_r \frac{1}{H_S - (E_S - \omega)} \, \left( \scp{\sigma}{L} \right) \,
    p^i \ket{\phi} 
\end{equation}
sowie
\begin{equation} 
P_{L \cdot S,2}  = \frac{\alpha}{72 m^2} \bra{\phi} p^i \, 
\frac{1}{H_S - (E_S - \omega)} \, D_r \, \left( \scp{\sigma}{L} \right) \,
    p^i \ket{\phi} 
\end{equation}
und
\begin{equation} 
P_{L \cdot S,3}  = -\frac{1}{72 m} \bra{\phi} p^i \,
\frac{1}{H_S - (E_S - \omega)} \, \left[ \frac{1}{m} V^2 \right]
\, \frac{1}{H_S - (E_S - \omega)} \, p^i \ket{\phi}. 
\end{equation}
Aufgrund der Anti-Hermitezit"at von $D_r$ folgt, da"s 
$P_{L \cdot S,1}$ und $P_{L \cdot S,2}$ gleich 
sind, und man erh"alt als Ergebnis f"ur den $P_{L \cdot S, 1}$-Term:
\eqnanf
P_{L \cdot S, 1} &=& {{8\,{{\alpha}^2}\,F_2(t)\,{t^5}}\over 
    {9\,{{\left( -1 + t \right) }^3}\,{{\left( 1 + t \right) }^3}}} 
    \schwups 
& & + {{8\,{{\alpha}^2}\,F(t)\,{t^5}\,
      \left( 3 - 74\,{t^2} + 167\,{t^4} \right) }\over 
    {27\,{{\left( 1 - t \right) }^5}\,{{\left( 1 + t \right) }^5}}} \wups
& & + {{{{\alpha}^2}\,{t^4}\,\left( -1 + 2\,t - 88\,{t^2} - 98\,{t^3} + 
        377\,{t^4} \right) }\over 
    {27\,{{\left( -1 + t \right) }^5}\,{{\left( 1 + t \right) }^3}}}. \ok
\eqnend
Nach Integration bez"uglich $t$ erh"alt man den $F$-Faktor
\begin{equation}
F_{L \cdot S, 1} = F_{L \cdot S, 2} = 
 0.0103947\,{{\alpha}^2} + {{{{\alpha}^2}\,\ln (\alpha)}\over {54}} - 
  {{{{\alpha}^2}\,\ln (\ep)}\over {108}}.
\end{equation}
Den verbleibenden $F_{L \cdot S}$-Term rechnet man mit einer Variation 
des Drehimpulses im Schr"odinger-Hamiltonian (wie bei $P_{p^4,3c}$)
und erh"alt wieder ein recht 
komplexes Ergebnis, das sehr verk"urzt folgenderma"sen lautet:
\eqnanf
P_{L \cdot S, 3} &=& {{2048\,{{\alpha}^2}\,{\tilde \Psi}_{63}(t)\,\left( 1 - 2\,t \right) \,
      \left( -1 + t \right) \,{t^6}\,\left( 1 + 2\,t \right) }\over 
    {135\,{{\left( 1 + t \right) }^{11}}}} \schwups
& &  + {{2048\,{{\alpha}^2}\,{\tilde F}_{366}(t)\,\left( -1 + t \right) \,{t^6}\,
      \left( -1 + 2\,t \right) \,\left( 1 + 2\,t \right) }\over 
    {135\,{{\left( 1 + t \right) }^{11}}}} \wups
& &  + {{1024\,{{\alpha}^2}\,{\tilde \Phi}_{23}(t)\,\left( -1 + t \right) \,{t^6}\,
      \left( -1 + 2\,t \right) \,\left( 1 + 2\,t \right) }\over 
    {135\,{{\left( 1 + t \right) }^{11}}}} \wups
& & +  f(\gamma , F_2(t), F(t), \ln(t)). \ok
\eqnend
Daraus ist ersichtlich, da"s auch $P_{L \cdot S, 3}$ wieder einen Beitrag 
zum numerischen Anteil beisteuert. F"ur $F_{L \cdot S, 3}$ erhalten wir
\begin{equation}
F_{L \cdot S,3} = -0.0018158\,{{\alpha}^2} + {{{{\alpha}^2}\,\ln (\alpha)}\over {54}} - 
  {{{{\alpha}^2}\,\ln (\ep)}\over {108}} + N_{L \cdot S, 3}.
\end{equation} 
Als Ergebnis f"ur $F_{L \cdot S}$ (Summe von $F_{L \cdot S,1}$,
$F_{L \cdot S,2}$ und $F_{L \cdot S,3}$) erhalten wir 
\begin{equation}
F_{L \cdot S} = 0.0189735\,{{\alpha}^2} + {{{{\alpha}^2}\,\ln (\alpha)}\over {18}} - 
  {{{{\alpha}^2}\,\ln (\ep)}\over {36}} + N_{L \cdot S, 3}.
\end{equation}

%
%

\subsection{Der numerisch auszuwertende Anteil}
\label{num}

Der verbleibende numerische Anteil aus $N_{\delta \phi, 5}$, $N_{L \cdot 
S, 3}$ und $N_{p^4,3c}$ ergibt sich zu
\eqnanf
N & = & N_{\delta \phi, 5} + N_{L \cdot S, 3} + N_{p^4,3c} \schwups
& = & \int_0^1 dt \bigg( {{512\,{{\alpha}^2}\,{\tilde \Psi}_{21}(t)\,t}\over 
    {27\,\left( 1 - t \right) \,{{\left( 1 + t \right) }^5}}} + 
  {{256\,{{\alpha}^2}\,{\tilde F}_{122}(t)\,t}\over 
    {27\,\left( -1 + t \right) \,{{\left( 1 + t \right) }^5}}} \wups
& &  {{256\,{{\alpha}^2}\,{\tilde \Phi}_{21}(t)\,t}\over 
    {27\,\left( -1 + t \right) \,{{\left( 1 + t \right) }^5}}} + 
  {{1024\,{{\alpha}^2}\,{\tilde F}_{366}(t)\,\left( 1 - 2\,t \right) \,
      {{\left( -1 + t \right) }^2}\,t\,\left( 1 + 2\,t \right) }\over 
    {27\,{{\left( 1 + t \right) }^{10}}}} \wups
& &   + {{1024\,{{\alpha}^2}\,{\tilde \Psi}_{63}(t)\,\left( 1 - 2\,t \right) \,
      {{\left( -1 + t \right) }^2}\,t\,\left( 1 + 2\,t \right) }\over 
    {27\,{{\left( 1 + t \right) }^{10}}}} \wups 
& &  + {{512\,{{\alpha}^2}\,{\tilde \Phi}_{23}(t)\,{{\left( -1 + t \right) }^2}\,t\,
      \left( -1 + 2\,t \right) \,\left( 1 + 2\,t \right) }\over 
    {27\,{{\left( 1 + t \right) }^{10}}}} \bigg)\,. \ok
\eqnend
Die numerische Integration wird mit der Gausschen Quadratur ausgef"uhrt. 
Dabei erweist es sich als sehr hilfreich, da"s der Integrand f"ur $t \to 
0$ und $t \to 1$ verschwindet. Dies wurde durch das Herausnehmen der 
$k=0$- und $k=1$-Terme in den speziellen Funktionen erreicht. 
Das Endergebnis f"ur $N$ lautet:
\begin{equation}
N = 0.0030113\,{{\alpha}^2}.
\end{equation}

%
%

\subsection{Ergebnis des Niedrigenergie-Anteils}

In allen Teilergebnissen zum Niedrigenergie-Anteil hatten wir der 
Einfachheit halber $Z \alpha$ durch $\alpha$ ersetzt. Wenn wir nun alle 
Teilergebisse aufaddieren und im Endergebnis $\alpha$ durch 
$Z \alpha$ ersetzen, so ergibt sich: 
\eqnanf
\label{Flep}
F_L &=& F_{nd} + F_{nq} + F_{p^i p^2} + F_{r \times \sigma} + F_{k \cdot r} \schwups
& &  + F_{\delta E} + F_{\delta \phi} + F_{\delta} + F_{p^4} + F_{L \cdot S} + N \wups
& = & 0.0400223(1) - 0.79565(1) \,(Z \alpha)^2 + 
  \frac{103}{180} \, (Z \alpha)^2 \, \ln ((Z \alpha)^{-2}) \wups 
& & + \frac{2\,(Z \alpha)^2}{9\,\ep} + 
        \frac{103}{180} \, (Z \alpha)^2 \, \ln (\ep). \ok
\eqnend
Der $A_{61}$-Koeffizient des $\ln((Z \alpha)^{-2})$-Terms stimmt gerade 
mit dem bekannten Ergebnis von $103/180$ "uberein \cite{kinoshita}.
Wir erinnern uns an das Ergebnis (\ref{Fhep}) f\"{u}r den Hochenergie--Anteil
\mathanf
F_H(2P_{1/2}) = -\frac{1}{6} + (Z \alpha)^2 
\left[ \frac{4177}{21600} - \frac{103}{180} \ln(2) -
\frac{103}{180} \ln{(\epsilon)} - \frac{2}{9 \epsilon} \right]
\mathend
und geben noch einmal die Formel f\"{u}r den Niedrigenergie--Anteil an,
\eqnanf
F_L(2P_{1/2}) &=&  0.0400223(1) - 0.79565(1)\,(Z \alpha)^2 + 
  \frac{103}{180} \, (Z \alpha)^2 \, \ln ((Z \alpha)^{-2}) \wups 
& & + \frac{2\,(Z \alpha)^2}{9\,\ep} + 
        \frac{103}{180} \, (Z \alpha)^2 \, \ln (\ep). \ok
\eqnend
In der Summe heben sich die Divergenzen in $\ep$ heraus, und wir erhalten 
als Summe der Hoch- und Niedrigenergie-Anteile
\begin{equation}
F(2P_{1/2}) =
  -0.126644(1) + \left(Z \alpha \right)^2 \left[ -0.99891(1) + 
     \frac{103}{180} \, \ln\left( (Z \alpha)^{-2} \right) \right].
\end{equation}
Die Koeffizienten sind somit bestimmt zu
\eqnanf
A_{40} &=&  A_{40}(2P_{1/2}) = -\frac{1}{6} - \frac{4}{3} \ln k_0(2P) = 
  -0.126644(1), \schwups
A_{60} &=&  -0.99891(1),  \wups
A_{61} &=& \frac{103}{180}. \ok
\eqnend
Die Werte f"ur $A_{40}$ und $A_{61}$ stimmen mit den bekannten Resultaten 
"uberein (siehe \cite{kinoshita}).
Der Wert des Bethe-Logarithmus $\ln k_0$ f"ur $2P$-Zust"ande ist
$\ln k_0(2P) = -0.0300167089(3)$.

Die numerischen Integrationen erfolgen jeweils mit einer Unsicherheit von 
$1$ in der letzten Dezimalstelle (d. h. in der 6. Nachkommastelle). 
Im ganzen werden 10 numerische Integrationen ausgef"uhrt.
Wir geben daher die Genauigkeit des 
$A_{60}$-Koeffizienten mit $A_{60} =  -0.99891(1)$ an (d. h. mit.
einer Unsicherheit von 1 in der 5. Nachkommastelle). 

Anmerkung: Es wird vermutet, da"s die Integration sehr viel besser konvergiert
als hier angegeben. Zum Beispiel wurden bei der Integration des 
nichtrelativistischen Dipol-Anteils alle 9 bekannten 
Nachkommastellen des Bethe-Logarithmus in "Ubereinstimmung mit
dem bekannten Resultat erhalten. Da aber beim $A_{60}$-Koeffizienten
die 5. Nachkommastelle bereits einer Genauigkeit von unter 
$1 \, {\rm Hz}$ entspricht, geben wir uns mit einer sehr 
vorsichtigen Absch"atzung zufrieden.

%
%

\chapter{Die Selbstenergie des 2P-Zustands (j=3/2) 
  zur sechsten Ordnung in Z\_alpha}

\section{Der Hoch- und Niedrigenergie-Anteil}

Die Rechnungen des vorangegangenen Kapitels k"onnen f"ur das
$2P_{3/2}$-Niveau in "ahnlicher Weise wiederholt werden.
F"ur den Hochenergie-Anteil wird erhalten
\begin{equation}
F_H(2P_{3/2}) = \frac{1}{12} + (Z \alpha)^2 
\left[ \frac{6577}{21600} - \frac{29}{90} \, \ln(2) -
\frac{29}{90} \, \ln{(\epsilon)} - \frac{2}{9 \epsilon} \right].
\end{equation}
Weil beim $2P_{3/2}$-Niveau der Spin des Elektrons mit seinem 
Bahndrehimpuls zum Gesamtdrehimpuls $j = 3/2$ koppelt, "andern sich
die Beitr"age $F_{L \cdot S}$, $F_{r \times \sigma}$ und
$F_{k \cdot r}$. Weil die relativistische Korrektur zur Energie 
f"ur $2P_{3/2}$ anders ist als f"ur $2P_{1/2}$, "andert sich
$F_{\delta E}$. Da der relativistische Hamiltonian $\delta H$ 
die $\scp{\sigma}{L}$-Kopplung enth"alt, "andert sich auch die
relativistische Korrektur zur Wellenfunktion $\delta \phi$. 
Damit erh"alt auch $F_{\delta \phi}$ einen anderen Wert.
Die anderen Beitr"age "andern sich nicht. 

Es soll hier nicht
auf die Rechnungen eingegangen werden. Im wesentlichen
wird dem bereits besprochenen $2P_{1/2}$-Zustand
gefolgt. Der Niedrigenergie-Anteil ergibt sich zu
\begin{equation}
F_L(2P_{3/2}) = - \frac{4}{3} \ln k_0 + (Z \alpha)^2 
\left[ -0.58451(1) + \frac{29}{90} \, \ln\left( (Z \alpha)^{-2} \right) +
\frac{29}{90} \, \ln{(\epsilon)} + \frac{2}{9 \epsilon} \right].
\end{equation}

\section{Endergebnis f"ur den 2P-Zustand (j = 3/2)}

In der Addition der Hoch- und Niedrigenergie-Anteile 
heben sich die $1/\ep$- und $\ln(\ep)$-Divergenzen
gerade heraus, und wir erhalten als Ergebnis f"ur $F$
\begin{equation}
F(2P_{3/2}) = 0.123356(1) + \left(Z \alpha \right)^2 \left[ -0.50337(1) + 
  \frac{29}{90} \, \ln\left( (Z \alpha)^{-2} \right) \right].
\end{equation}
mit $A_{40}= 0.123356(1) = 1/12 - 4/3 \ln k_0(2P)$ ($\ln k_0$ ist der
Bethe-Logarithmus). Durch die Rechnung ist der $A_{60}$-Koeffizient
bestimmt zu
\begin{equation}
A_{60}(2P_{3/2}) = -0.50337(1).
\end{equation}

%
%

\chapter{Ergebnisse und Diskussion}
\label{ergebnisse}

%
%

\section{Zusammenfassung der Ergebnisse}

Es wurde in dieser Arbeit die $\ep$fw-Methode entwickelt, welche
zur Berechnung der (Ein-Schleifen-) Selbstenergie eines gebundenen Elektrons 
in h"oherer Ordnung dient. Diese Methode wurde auf die $2P_{1/2}$- und
$2P_{3/2}$-Zust"ande angewandt.

F"ur die $F$-Faktoren der Selbstenergie 
(zur Definition von $F$ siehe Kapitel \ref{ggstd}) erhalten wir
\begin{equation}
F(2P_{1/2}) = 
  - \frac{1}{6} - \frac{4}{3} \ln k_0(2P) + 
    \left(Z \alpha \right)^2 \left[ -0.99891(1) + 
     \frac{103}{180} \ln\left( (Z \alpha)^{-2} \right) \right]
\end{equation}
und
\begin{equation}
F(2P_{3/2}) = 
  \frac{1}{12} - \frac{4}{3} \ln k_0(2P) + \left(Z \alpha \right)^2 \left[ -0.50337(1) + 
     \frac{29}{90} \ln\left( (Z \alpha)^{-2} \right) \right],
\end{equation}
wobei der Bethe-Logarithmus gegeben ist durch 
$\ln k_0 (2P) = -0.0300167089(3)$. Die $A_{40}$- und $A_{61}$-Koeffizienten
stimmen mit den bekannten Resultaten "uberein (diese Koeffizienten
sind in \cite{kinoshita} f"ur $P$ und $S$ Zust"ande aufgef"uhrt, f"ur die
$P$ Zust"ande siehe Kapitel \ref{ggstd}). 
Die $A_{60}$-Koeffizienten sind bestimmt zu
\begin{equation}
A_{60}(2P_{1/2}) = -0.99891(1)
\end{equation}
und
\begin{equation}
A_{60}(2P_{3/2}) = -0.50337(1).
\end{equation}
Diese Resultate sind konsistent mit den
aus der Extrapolation von numerischen Resultaten
gewonnenen Werte f"ur die $A_{60}$-Koeffizienten \cite{mohr},
\begin{equation}
A_{60}(2P_{1/2}) = - 0.8(3)
\end{equation}
und
\begin{equation}
A_{60}(2P_{3/2}) = - 0.5(3).
\end{equation}
Die Bestimmung der $A_{60}$-Koeffizienten
mit verbesserter Genauigkeit war neben der Entwicklung
der $\ep$fw-Methode das Hauptziel dieser Arbeit.
Mit der genauen Kenntnis der Koeffizienten wird es m"oglich, neue
(genauere) Werte f"ur die Lamb-Shift der beiden $2P$ Zust"ande 
anzugeben (Kapitel \ref{lambshift}).

%
%

\section{"Uberpr"ufung der Rechnung}

Lamb-Shift-Rechnungen allgemein 
sind sehr umfangreich und die Ausdr"ucke in den Zwischenschritten 
recht gro"s.
In dieser Arbeit "uberstiegen die Integranden im Hochenergie-Anteil
bei den Matrix-Elemente eine L"ange von 1000 Termen,
im Niedrigenergie-Anteil traten Ausdr"ucke vergleichbarer
L"ange auf, die dann vermittels der Zusammenhangsformeln
vereinfacht wurden.  
Ein einziger Rechenfehler kann das Ergebnis verf"alschen. 
Daher stellt sich die Frage nach den "Uberpr"ufungsm"oglichkeiten. Bei der 
vorliegenden Arbeit wurden folgende Pr"ufungen vorgenommen:

\begin{enumerate}
\item Die Matrix-Elemente f"ur den Hochenergie-Anteil wurden auf zwei 
verschiedene Arten (computergest"utzt und von Hand) ausgewertet.
\item Im Hochenergie-Anteil sind der Koeffizient der 
$\ln(\ep)$-Divergenz und der $1/\ep$-Divergenz bekannt und k"onnen 
"uberpr"uft werden.
\item Im Niedrigenergie-Anteil kann man f"ur jeden einzelnen Beitrag 
die $\ln(\ep)$-Divergenz berechnen und "uberpr"ufen. Dies wurde f"ur 
jeden einzelnen Beitrag getan. Die Summe der 
Divergenzen mu"s sich gerade mit dem Hochenergie-Anteil herausheben.
\item Man kann zeigen, da"s das Matrix-Element
\mathanf
P(\omega) = \frac{m}{2} \delta^{T,ij} {\tilde P}^{ij}(\omega)
\mathend
mit
\mathanf
{\tilde P}^{ij}(\omega) = \bra{\psi} \alpha^i e^{i \scp{k}{r}}
\frac{1}{H_D - (E_{\psi} - \omega)}
\alpha^j e^{-i \scp{k}{r}} \ket{\psi}
\mathend
f"ur $\omega \to 0$ verschwindet. Daher mu"s auch die Summe der einzelnen 
Beitr"age des Niedrigenergie-Anteils f"ur $\omega \to 0$ ($t \to 1$) 
verschwinden. Nach der FW-Transformation treten jedoch noch versteckte 
Beitr"age auf, da die $\alpha^i$-Matrizen obere und unteren Komponente
der Spinoren miteinander verbinden. Wenn man die versteckten Beitr"age jedoch 
miteinbezieht, so mu"s sich der konsistente Wert Null f"ur die Summe 
ergeben. Man kann den Wert der Beitr"age f"ur $t \to 1$ sogar einzeln 
berechnen, wodurch eine zus"atzliche "Uberpr"ufung des Resultats
m"oglich ist. Diese Pr"ufungsm"oglichkeit wurd ebenfalls in die Tat umgesetzt.
\item Alle im Niedrigenergie-Anteil verwandten Summations-Regeln wurden 
an geeigneten Beispielen gepr"uft.
\item Der nichtrelativistische Dipol-Anteil wurde auf zwei verschiedene 
Arten (im Impuls- und Ortsraum) berechnet.
\end{enumerate}
Diese Pr"ufungen beziehen sich auf die Korrektheit
der Rechenschritte.
Es ergibt sich noch ein weiteres Problem, denn
auch die verwandten Computer-Algebra-Systeme sind 
(wahrscheinlich) nicht
frei von Inkonsistenzen. Die "Uberpr"ufungsm"oglichkeiten
in Bezug auf diese Systeme sind eingeschr"ankt.
Man kann allenfalls die eigenen Programme so gestalten, da"s 
die einzelnen Rechenschritte sehr gut kontrolliert
werden. 

%
%

\section{Absch"atzung h"oherer Koeffizienten}

P. Mohr hat $F$ f"ur gr"o"se $Z$ aus numerischen Rechnungen gewonnen
\cite{mohr}. F"ur $Z = 5$ ist die Konvergenz bereits sehr langsam,
f"ur $Z < 5$ werden dort keine Resultate angegeben. 
Aufgrund der nichtanalytischen Terme in $\alpha$ ist eine
Extrapolation der Ergebnisse in den Bereich f"ur kleine $Z$
nur eingeschr"ankt m"oglich. Kennt man aber die Koeffizienten von
$F$ bis zu einer bestimmten Ordnung in $\alpha$, so lassen sich
durch Angleichen einer Funktion, die die n"achsth"oheren Koeffizienten
enth"alt, Absch"atzungen f"ur diese Koeffizienten gewinnen. Wir gleichen
dazu die Funktion
\eqnanf
G(Z) &=& \frac{F(Z) - A_{40} - (Z \alpha)^2 
\left( A_{60} + A_{61} \ln \left[ (Z \alpha)^{-2} \right] \right)}
{(Z \alpha)^3} = \schwups
& & A_{71} + (Z \alpha) \left[ A_{80}
+ A_{81} \ln \left[ (Z \alpha)^{-2} \right]  
+ A_{82} \left( \ln \left[ (Z \alpha)^{-2} \right] \right)^2 \right] +
O((Z \alpha)^2) \ok
\eqnend
an die numerischen Ergebnisse an (mit den in dieser Arbeit 
bestimmten, neuen Werten von $A_{60}$).
Daraus ergeben sich die folgenden Absch"atzungen f"ur die n"achsth"oheren
Koeffizienten.
\begin{equation}
A_{70}(2P_{1/2}) \approx \frac{49}{50} \pi,
\end{equation}
Die Angabe dieses Koeffizienten erfolgt in dieser Form,
da gezeigt werden kann, da\3 $A_{{\rm ungerade}, 0}$ 
gleich einer rationalen Zahl mal $\pi$) ist. 
\begin{equation}
A_{80}(2P_{1/2}) \approx 0.1,
\end{equation}
\begin{equation}
A_{81}(2P_{1/2}) \approx -1.5,
\end{equation}
\begin{equation}
A_{82}(2P_{1/2}) \approx 0.2.
\end{equation}
F"ur den $2P_{3/2}$-Zustand ist
\begin{equation}
A_{70}(2P_{3/2})  \approx  \frac{39}{50} \pi,
\end{equation}
\begin{equation}
A_{80}(2P_{3/2})  \approx  -1.7,
\end{equation}
\begin{equation}
A_{81}(2P_{3/2})  \approx  -0.6,
\end{equation}
\begin{equation}
A_{82}(2P_{3/2})  \approx  -0.2.
\end{equation}
Den Fehler dieser Absch"atzungen sollte man
aufgrund der systematischen Unsicherheit 
des Verfahrens mit mindestens der H"alfte des Absolutwertes
in jedem einzelnen Fall ansetzen.
 
F"ur die Werte von $G_1 = G(Z = 1)$ ergeben sich die
folgenden Absch"atzungen:
\begin{equation}
G_1(2P_{1/2}) = 3 \pm 2,
\end{equation}
und
\begin{equation}
G_1(2P_{3/2}) = 2 \pm 1.
\end{equation}
Diese Werte von $G$ werden f"ur die Angabe des theoretischen
Fehlers des Selbstenergiebeitrages bei der Berechnung der
Lamb-Verschiebung ben"otigt.

%
%

\section{Neue Werte f"ur die Lamb-Verschiebung}
\label{lambshift}

F"ur die Lamb-Shift $\delta E_{\rm Lamb}$ verwenden wir die 
Definition
\begin{equation}
\label{defElamb}
E = M + m_r \left[ f(n,j)-1 \right] - \frac{m_r^2}{2 M}
\left[ f(n,j) - 1 \right]^2 + \delta E_{\rm Lamb} +
E_{\rm hfs}.
\end{equation}
$E$ ist die Energie des Zweik"orpersystems und $f(n,j)$
die dimensionslose Dirac Energie.
F"ur die in dieser Arbeit
behandelte Ein-Schleifen-Selbstenergie mu"s zur Bestimmung der
Lamb-Verschiebung die Abh"angigkeit der einzelnen
Koeffizienten von der reduzierten Masse des Systems 
miteinbezogen werden. Diese Abh"angigkeit ist bekannt \cite{kinoshita}.
Au"ser der Ein-Schleifen-Selbstenergie tragen folgende
Effekte zur Lamb-Verschiebung bei :
\begin{enumerate}
\item Zwei-Schleifen-Korrekturen zur Selbstenergie. Der Hauptbeitrag
ist von der Ordnung $\alpha (Z \alpha)^4$. Es ist auch
ein logarithmischer Beitrag der Ordnung $\alpha^2 (Z \alpha)^6 
\ln((Z \alpha)^{-2})$ bekannt \cite{kinokino}.
\item Drei-Schleifen-Beitrag zur Selbstenergie. Der
Hauptbeitrag dieser Korrektur ergibt sich durch die
Drei-Schleifen-Beitr"age zum anomalen magnetischen 
Moment des Elektrons \cite{kinokino}. Der Effekt skaliert mit
$\alpha^3 (Z \alpha)^4$. 
\item Vakuumpolarisation. Die Vakuumpolarisation geht f"ur $P$-Zust"ande
nur in h"oherer Ordnung ($\alpha (Z \alpha)^6$) ein \cite{kinoshita}.
\item Eine mit $(m_r/M)^2 (Z \alpha)^4$ skalierende R"ucksto"skorrektur,
die aus dem Breit-Hamiltonian folgt und eine leichte $l$-Abh"angigkeit
der Energieniveaus bedingt \cite{kinoshita}. 
\item Die Salpeter Korrektur (relativistische R"ucksto"skorrektur)
in $(Z \alpha)^5$ \cite{kinoshita}.  
\item Relativistische R"ucksto"skorrekturen h"oherer Ordnung
($(Z \alpha)^6 \, m_r/M$) \cite{golosov}.
\end{enumerate}
Durch Absch"atzen h"oherer Terme in $\alpha$ kann f"ur die einzelnen
Beitr"age ein ``theoretischer Fehler'' angegeben werden. F"ur den
$2P_{1/2}$-Zustand ergibt sich folgende Tabelle,
\begin{center}
\begin{tabular}{|c|r|c|} \hline
\multicolumn{3}{|c|}{Lamb-Verschiebung des $2P_{1/2}$-Zustands} \\ \hline \hline
Beitrag & Wert in ${\rm kHz}$ & Fehler in ${\rm kHz}$ \\ \hline \hline
Ein-Schleifen-Selbstenergie & $-12846.9$ & $0.1$ \\ \hline
Zwei-Schleifen-Selbstenergie & $26.0$ & $0.1$ \\ \hline
Drei-Schleifen-Selbstenergie & $-0.2$ & vernachl"assigbar \\ \hline
Vakuumpolarisation & $-0.3$ & vernachl"assigbar \\ \hline
$(Z \alpha)^4$ R"ucksto"s & $2.2$ & vernachl"assigbar \\ \hline
$(Z \alpha)^5$ R"ucksto"s (Salpeter) & $-17.1$ & $0.1$ \\ \hline
$(Z \alpha)^6$ R"ucksto"s & $0.4$ & vernachl"assigbar \\ \hline \hline
Summe f"ur $2P_{1/2}$ & $-12836.0$ & $0.3$ \\ \hline \hline
\end{tabular}
\end{center}
und f"ur $2P_{3/2}$ haben wir
\begin{center}
\begin{tabular}{|c|r|c|} \hline
\multicolumn{3}{|c|}{Lamb-Verschiebung des $2P_{3/2}$-Zustands} \\ \hline \hline
Beitrag & Wert in ${\rm kHz}$ & Fehler in ${\rm kHz}$ \\ \hline \hline
Ein-Schleifen-Selbstenergie & $12548.0$ & $0.1$ \\ \hline
Zwei-Schleifen-Selbstenergie & $-12.8$ & $0.1$ \\ \hline
Drei-Schleifen-Selbstenergie & $0.1$ &  vernachl"assigbar\\ \hline
Vakuumpolarisation & $-0.1$ & vernachl"assigbar \\ \hline
$(Z \alpha)^4$ R"ucksto"s & $-1.1$ & vernachl"assigbar \\ \hline
$(Z \alpha)^5$ R"ucksto"s (Salpeter) & $-17.1$ & $0.1$ \\ \hline
$(Z \alpha)^6$ R"ucksto"s & $0.4$ & vernachl"assigbar \\ \hline \hline
Summe f"ur $2P_{3/2}$ & $12517.5$ & $0.3$ \\ \hline \hline
\end{tabular}
\end{center}
Bei den theoretischen Unsicherheiten addieren wir die Absolutwerte
der Fehler, da sich theoretische Unsicherheiten nicht wie statistisch 
verteilte Me\3werte verhalten. 
Die neuen Werte f"ur die Lamb-Verschiebungen der $2P$-Zust"ande sind
\begin{equation}
\delta E_{\rm Lamb}(2P_{1/2}) = -12836.0(3) \, {\rm kHz}
\end{equation}
sowie
\begin{equation}
\delta E_{\rm Lamb}(2P_{3/2}) = 12517.5(3) \, {\rm kHz}.
\end{equation}
Der theoretische Fehler f"ur beide Endergebnisse liegt bei unter 
$300 \, {\rm Hz}$. Der neue theoretische Wert f"ur die Feinstruktur ergibt 
sich aus Gl. \ref{defElamb} zu 
\begin{equation}
\delta E_{\rm fs} = E(2P_{3/2}) -  E(2P_{1/2}) = 10969043.2(6) \, {\rm kHz},
\end{equation}
wiederum mit der Ma\3gabe, die Absolutwerte der 
einzelnen theoretischen Fehler zu addieren.

%
%

\section{Diskussion der epsilon-fw-Methode}

Die in dieser Arbeit vorgestellte $\ep$fw-Methode dient zur Berechnung
der Selbstenergie eines gebundenen Zustands in h"oherer Ordnung. 
Die Rechnung wird in einen Hoch- und einen Niedrigenergie-Anteil aufgespalten.
Der Hochenergie-Anteil wird durch Einsatz von Computer-Algebra vereinheitlicht
behandelt. Im Niedrigenergie-Anteil hat die $\ep$fw-Methode
aufgrund der Foldy-Wouthuysen-Transformation folgende Vorteile:
\begin{enumerate}
\item Klare Trennung des Hauptbeitrages von den relativistischen Korrekturen,
\item Einsatz des nichtrelativistischen Propagators,
\item Aufteilung der Rechnung in Teilbeitr"age.
\end{enumerate}
Ultimatives Ziel der computergest"utzten Rechnungen ist automatisierte 
Berechnung von quantenelektrodynamischen Korrekturen. Nur durch
eine solche Automatisierung der Rechnung ist es m"oglich,
Korrekturen h"oherer Ordnung zuverl"assig zu berechnen
(wegen der L"ange der Terme). 
Von besonderer Wichtigkeit bleibt jedoch die Wahl 
einer ad"aquaten Rechenmethode.
Bei Wahl einer falschen Methode ``explodieren'' die 
Ausdr"ucke in mehrere zehntausend Terme, die auch mit schnellen 
Rechnern nicht mehr handhabbar sind. Daher ist es unter anderem
w"unschenswert, die Rechnung in Teilschritte aufzuspalten.

Wie wir am Beispiel der Integrationsroutinen zur Berechnung der 
Matrix-Elemente (siehe Abschnitt \ref{allghep}) sehen, kann durch 
geschicktes Programmieren h"aufig die Rechenzeit um mehrere 
Gr"o"senordnungen verringert werden. Die entwickelten 
Integrationsroutinen zur Berechnung von Matrix-Elementen kommen mit sehr 
wenigen Regeln aus und sind daher schnell. Die in $Mathematica$ 
eingebauten Routinen decken zwar eine gr"o"sere Klasse von Integralen ab, 
sind aber f"ur den speziellen Einsatz viel zu langsam. 

%
%

\chapter{Ausblick}

\section{Anmerkung zum Vergleich mit dem Experiment}

Zum direkten Vergleich von Theorie und Experiment sollte auf dem
${\rm kHz}$-Niveau noch folgender Effekt miteinbezogen werden, 
f"ur den hier der Name hfs-fs-Kopplung vorgeschlagen wird.
Der Effekt wird beispielsweise in \cite{muonic} f"ur myonischen
Wasserstoff behandelt. Er beeinflu"st 
die relative Lage von Zust"anden mit denselben
$n$, $l$, und $F$ Quantenzahlen, die aber verschiedene
$j$-Quantenzahlen aufweisen. Der Effekt wird bereits in dem klassischen
Buch von Bethe und Salpeter \cite{bookbethe} beschrieben.
Durch den Effekt wird der $2P_{3/2}$, $F=1$ Zustand energetisch
abgesenkt, der $2P_{1/2}$, $F=1$ Zustand jedoch energetisch angehoben um
jeweils $2.5 \, {\rm kHz}$. Der Effekt beeinflu"st nicht die relative 
Lage der $2P_{3/2}$, $F=2$ und $2P_{1/2}$, $F=0$ Zust"ande.  

%
%

\section{Hochpr"azisions-Tests der QED}

Durch den neuen theoretischen Wert f"ur den
$2P_{1/2}$ Zustand wird es m"oglich, von der $2P$ Lamb-Verschiebung, 
die als Referenzpunkt dient, die $2S$ Lamb-Verschiebung sehr genau zu 
bestimmen (Messung von Hagley und Pipkin, \cite{hagleypipkin}).
Andererseits wird die $2S$-Lamb-Verschiebung im Frequenzkettenlabor 
durch Vergleich der $nS$-"Ubergangsfrequenzen ermittelt.
Dadurch ist ein direkter Vergleich von zwei v"ollig 
unterschiedlichen experimentellen Methoden m"oglich.

Allgemein gilt f"ur die hier vorgestellte $\ep$fw-Methode,
da"s analytische Rechnungen zur Selbstenergie in h"oherer 
Ordnung sehr vereinfacht werden..
In leicht abgewandelter Form sollte die Methode zur Berechnung
von Strahlungskorrekturen zur Hyperfeinstrukter
in Myonium und Positronium sowie zur Selbstenergie in Helium
einsetzbar sein. 

Weitere Anwendungsm"oglichkeiten ergeben sich
f"ur $P$-Zust"ande mit h"oheren Hauptquantenzahlen ($n=3$ und $n=4$).
Die Computerunterst"utzung ist hier sehr
hilfreich, da die aufgrund h"oherer
Hauptquantenzahl l"angeren Ausdr"ucke f"ur die Wellenfunktionen
problemlos behandelt werden k"onnen.

Nach Wissen des Autors existieren f"ur Zust"ande mit $n > 3$ zur
Zeit weder numerische noch symbolische Rechnungen in h"oherer
Ordnung. Durch Rechnungen in h"oherer
Ordnung k"onnte die Lamb-Shift dieser Zust"ande und damit
die Feinstruktur sehr genau bestimmt werden. 

%
%

\chapter*{Danksagung}

Ich m"ochte mich ganz herzlich bei Herrn Dr. Krzysztof Pachucki f"ur die  
freundliche, unkomplizierte und stets offene Hilfe und Betreuung 
bedanken. Bei Herrn Prof. Theodor H"ansch
bedanke ich mich f"ur die Vergabe des Themas und die Inspiration,
die von den Hochpr"azisions-Messungen ausgeht.
Schlie"slich danke ich auch der Studienstiftung f"ur
Anregung und Unterst"utzung w"ahrend des Studiums.  

%
%


\newpage
\begin{thebibliography}{1000}
%
%
\bibitem{greiner7} W. Greiner, {\em Theoretische Physik--Band 7: 
Quantenelektrodynamik}, Verlag Harry Deutsch, Thun; Frankfurt
am Main, 2. Auflage 1995.
%
%
\bibitem{itzykson} C. Itzykson und J. B. Zuber, {\em Quantum Field Theory}, 
McGraw-Hill, New York, 1980.
%
%
\bibitem{bjorkendrell} Bjorken/Drell, {\em Relativistische 
Quantenmechanik}, Originalausgabe bei McGraw-Hill 1964, 
deutsche "Ubersetzung im B.-I. Wissenschaftsverlag, 1964. 
%
%
\bibitem{landau} L. D. Landau und E.M. Lifschitz, 
{\em Lehrbuch der Theoretischen 
Physik, Bd. 3: Quantenmechanik}, Akademie-Verlag, Berlin, 1979. 
%
%
\bibitem{kinokino} T. Kinoshita und D. Yennie, {\em ``High Precision Tests of
Quantum Electrodynamics-- An Overview''}, in {\em Quantum
Electrodynamics}, Herausgeber T. Kinoshita, World Scientific,
Singapur 1990.
%
%
\bibitem{kinoshita} J. Sapirstein und D. Yennie, 
{\em ``Theory of Hydrogenic Bound States''}, in {\em Quantum
Electrodynamics}, Herausgeber T. Kinoshita, World Scientific,
Singapur, 1990.
%
%
\bibitem{uehling} E. A. Uehling, Phys. Rev. {\bf 48}, 55 (1935).
%
%
\bibitem{bethe} H. A. Bethe, { Phys. Rev.} {\bf 72}, 339 (1947).
%
%
\bibitem{bookbethe} H. A. Bethe und E. E. Salpter, 
{\em Quantum Mechanics of One- and Two-Electron Atoms}, 
Springer-Verlag, Berlin/New York, 1957.
%
%
\bibitem{lamb} W. E. Lamb, Jr. und R. C. Retherford, { Phys. Rev. }
{\bf 72}, 241 (1947).
%
\bibitem{tdl} S. Triebwasser, E. S. Dayhoff und W.E. Lamb, { Phys. Rev. }
{\bf 89}, 98 (1953).
%
%
\bibitem{haensch1} M. Weitz, F. Schmidt-Kaler, und T. W. H"ansch,
Phys. Rev. Lett. {\bf 68}, 1120 (1992), F. Schmidt-Kaler, D. Leibfried,
M. Weitz, und T.W. H"ansch, Phys. Rev. Lett. {\bf 70}, 2261 (1993),
M. Weitz, A. Huber, F. Schmidt-Kaler, D. Leibfried,
und T.W. H"ansch, Phys. Rev. Lett. {\bf 72}, 328 (1994).
%
%
\bibitem{hagleypipkin} E. W. Hagley und F. M. Pipkin, 
Phys. Rev. Lett. {\bf 72}, 1172 (1994). 
%
%
\bibitem{mohr} 
P. J. Mohr, {Phys. Rev. A} {\bf 26}, 2338
(1982), P. Indelicato und P. J. Mohr, {Phys. Rev A} {\bf 46}, 172 (1992),
K. T. Cheng, W. R. Johnson, und J. Sapirstein,
{Phys. Rev. Lett.} {\bf 66}, 2960 (1991),
S. A. Blundell und N. J. Snyderman, {Phys. Rev. A}
{\bf 44}, 1427 (1991),
P. J. Mohr und Y. K. Kim, {Phys. Rev. A} {\bf 45}, 2727 (1992).
%
%
\bibitem{host} L. Hostler, { J. Math. Phys.} {\bf 5}, 591 (1964).
%
%
\bibitem{swainsondrake}
R. A. Swainson und G. W. F. Drake, { J. Phys. A Math. Gen.} {\bf
24}, 79, 95 (1991).
%
%
\bibitem{bateman} H. Bateman, {\em Higher Transcendental Functions},
McGraw--Hill, New York, 1953.
%
\bibitem{buchholz} H. Buchholz, {\em The Confluent Hypergeometric
Function}, Springer Verlag, Berlin/New York, 1969.
%
%
\bibitem{krpdiss} K. Pachucki, {\em Dissertation}, 
  in {\em Annals of Physics} (N.Y.) {\bf 226}, 1 (1993).
%
\bibitem{pac1} K. Pachucki, { Phys. Rev. Lett. } {\bf 72}, 3154 (1994). 
%
\bibitem{pac2} K. Pachucki und H. Grotch, { Phys. Rev. A } {\bf 51}, 1854 
(1995).
%
\bibitem{pac3} K. Pachucki, { Phys. Rev. A } {\bf 52}, 1079 (1995).
%
\bibitem{muonic} K. Pachucki, { Phys. Rev. A } {\bf 53}, 2092 (1996).
%
%
\bibitem{golosov} E. Golosov {\em et al.},
Zh. Eksp. Teor. Fiz. {\bf 107}, p. 393 (1995).
%
%
\bibitem{math} S. Wolfram, {\em Mathematica: A System for Doing Mathmatics
by Computer} (Addison-Wesley, Reading, MA, 1988). 
%
\bibitem{HIP} A. Hsieh und E. Yehudai, {\em H.I.P. - Symbolic High 
Energy Physics Calculations}, Computers in Physics, 1991. 
%
\end{thebibliography}
\end{document}